\RequirePackage{rotating}
\documentclass[a4paper,fleqn,usenatbib]{mnras}

\usepackage[T1]{fontenc}
\usepackage{ae,aecompl}

\usepackage{graphicx}	
\usepackage{amsmath}	
\usepackage{amssymb}	
\usepackage{subfig}
\usepackage{rotating}
\usepackage{float}
\usepackage{hyperref}
\usepackage{placeins}
\usepackage{color}

\def\gsim{\mathrel{\raise0.35ex\hbox{$\scriptstyle >$}\kern-0.6em\lower0.40ex\hbox{{$\scriptstyle \sim$}}}} 

\AtBeginDocument{%
 \abovedisplayskip=3pt plus 3pt minus 9pt
 \abovedisplayshortskip=0pt plus 3pt
 \belowdisplayskip=12pt plus 3pt minus 9pt
 \belowdisplayshortskip=7pt plus 3pt minus 4pt
}

\title[KROSS: Disk Turbulence in $z$\,$\sim$\,1 Galaxies]{The KMOS
  Redshift One Spectroscopic Survey (KROSS): the origin of disk
  turbulence in $z$\,$\approx$\,1 star-forming galaxies}

\author[H. L. Johnson et al.]{\parbox[h]{\textwidth}{
H. L. Johnson,$^{1\dagger}$ C. M. Harrison,$^{2,1}$ A. M. Swinbank,$^{1,3}$ A. L. Tiley,$^{1,4}$ J. P. Stott,$^{5,4}$}
\newauthor{R. G. Bower,$^{1,3}$ Ian Smail,$^{1,3}$ A. J. Bunker,$^{4,6}$ D. Sobral,$^{5,7}$ O. J. Turner,$^{8,2}$ P. Best,$^{8}$}
\newauthor{M. Bureau,$^{4}$ M. Cirasuolo,$^{2}$ M. J. Jarvis,$^{4,9}$ G. Magdis,$^{10,11}$ R. M. Sharples,$^{1,12}$ }
\newauthor{J. Bland-Hawthorn,$^{13}$ B. Catinella,$^{14}$ L. Cortese,$^{14}$ S. M. Croom,$^{13,15}$} 
\newauthor{C. Federrath,$^{16}$ K. Glazebrook,$^{17}$ S. M. Sweet,$^{17}$ J. J. Bryant,$^{13,18,15}$ M. Goodwin,$^{18}$}
\newauthor{I. S. Konstantopoulos,$^{18}$ J. S. Lawrence,$^{18}$ A. M. Medling,$^{16}$ M. S. Owers,$^{19,18}$}
\newauthor{S. Richards$^{20}$ \vspace{5pt}} \\
$^{1}$\,Center for Extragalactic Astronomy, Department of Physics, Durham University, South Road, Durham DH1 3LE, UK\\
$^{2}$\,European Southern Observatory, Karl-Schwarzschild-Str. 2, D-85748 Garching b. M{\"u}nchen, Germany \\
$^{3}$\,Institute for Computational Cosmology, Durham University, South Road, Durham DH1 3LE, UK \\
$^{4}$\,Astrophysics, Department of Physics, University of Oxford, Keble Road, Oxford OX1 3RH, UK \\
$^{5}$\,Department of Physics, Lancaster University, Lancaster LA1 4YB, UK \\
$^{6}$\,Kavli Institute for the Physics and Mathematics of the Universe, 5-1-5 Kashiwanoha, Kashiwa, 277-8583, Japan \\
$^{7}$\,Leiden Observatory, Leiden University, PO Box 9513, NL-2300 RA Leiden, The Netherlands \\
$^{8}$\,SUPA, Institute for Astronomy, Royal Observatory of Edinburgh, Blackford Hill, Edinburgh EH9 3HJ, UK \\
$^{9}$\,Department of Physics, University of Western Cape, Bellville 7535, South Africa \\
$^{10}$\,Dark Cosmology Centre, Niels Bohr Institute, University of Copenhagen, Juliane Mariesvej 30, DK-2100 Copenhagen, Denmark \\
$^{11}$\,Institute for Astronomy, Astrophysics, Space Applications and Remote Sensing, National Observatory of Athens, GR-15236 Greece \\
$^{12}$\,Centre for Advanced Instrumentation, Durham University, South Road, Durham DH1 3LE, UK \\
$^{13}$\,Sydney Institute for Astronomy, School of Physics, University of Sydney, NSW 2006, Australia \\
$^{14}$\,ICRAR, University of Western Australia Stirling Highway, Crawley, WA, 6009, Australia \\
$^{15}$\,ARC Centre of Excellence for All-sky Astrophysics (CAASTRO), 44-70 Rosehill Street, Redfern NSW 2016, Sydney, Australia \\
$^{16}$\,Research  School  for  Astronomy  \&  Astrophysics,  Australian National University Canberra, ACT 2611, Australia \\
$^{17}$\,Centre for Astrophysics and Supercomputing, Swinburne University of Technology, PO Box 218, Hawthorn, VIC 3122, Australia \\
$^{18}$\,Australian Astronomical Observatory, 105 Delhi Rd, North Ryde, NSW 2113, Australia \\
$^{19}$\,Department of Physics and Astronomy, Macquarie University, NSW 2109, Australia \\
$^{20}$\,SOFIA Operations Center, USRA, NASA Armstrong Flight Research Center, 2825 East Avenue P, Palmdale, CA 93550, USA \\
$\dagger$\,E-mail: h.l.johnson@dunelm.org.uk
\vspace{-0.5cm}
}

\date{\vspace{-0.1cm}Accepted XXX. Received YYY; in original form ZZZ}

\pubyear{2017}

\begin{document}
\label{firstpage}
\pagerange{\pageref{firstpage}--\pageref{lastpage}}
\maketitle

\begin{abstract}
We analyse the velocity dispersion properties of 472 $z$\,$\sim$\,0.9
star-forming galaxies observed as part of the KMOS Redshift One
Spectroscopic Survey (KROSS). The majority of this sample is
rotationally dominated (83\,$\pm$\,5\% with $v_{\rm
  C}$/$\sigma_0$\,>\,1) but also dynamically hot and highly
turbulent. After correcting for beam smearing effects, the median
intrinsic velocity dispersion for the final sample is
$\sigma_0$\,=\,43.2\,$\pm$\,0.8\,km\,s$^{-1}$ with a rotational
velocity to dispersion ratio of $v_{\rm
  C}$/$\sigma_0$\,=\,2.6\,$\pm$\,0.1. To explore the relationship
between velocity dispersion, stellar mass, star formation rate and
redshift we combine KROSS with data from the SAMI survey
($z$\,$\sim$\,0.05) and an intermediate redshift MUSE sample
($z$\,$\sim$\,0.5). While there is, at most, a weak trend between
velocity dispersion and stellar mass, at fixed mass there is a strong
increase with redshift. At all redshifts, galaxies appear to follow
the same weak trend of increasing velocity dispersion with star
formation rate. Our results are consistent with an evolution of galaxy
dynamics driven by disks that are more gas rich, and increasingly
gravitationally unstable, as a function of increasing
redshift. Finally, we test two analytic models that predict turbulence
is driven by either gravitational instabilities or stellar
feedback. Both provide an adequate description of the data, and
further observations are required to rule out either model.
\end{abstract}




\section{Introduction}
\label{introduction}

The past decade has seen significant advancements in our understanding
of the high-redshift Universe. The cosmic star formation rate density
peaks in the redshift range $z$\,$\sim$\,1--3
(e.g. \citealt{lilly1996,karim2011,burgarella2013,sobral2013}), and so
establishing the properties of galaxies at this epoch is key to
constraining models of galaxy formation and evolution. It is at this
crucial time that today's massive galaxies formed the bulk of their
stars. The increased activity is thought to be driven (at least in
part) by high molecular gas fractions
(e.g. \citealt{daddi2010,tacconi2010,tacconi2013,saintonge2013,
  genzel2015}), which may naturally explain the clumpy and irregular
morphologies prevalent in \textit{Hubble Space Telescope}
(\textit{HST}) images (e.g. \citealt{livermore2012,livermore2015}).

The introduction of integral field spectroscopy (e.g. see
\citealt{glazebrook2013} for review) has been pivotal in allowing us
to resolve the internal complexities of distant galaxies. Each spatial
pixel of an integral field unit (IFU) is associated with a spectrum
such that galaxy kinematics, star formation and metallicity can be
mapped. Early studies often involved the in-depth analysis of small
samples, since observations were time-consuming
(e.g. \citealt{forster2006,law2009,lemoine2010,swinbank2012b}). However
second-generation instruments such as the $K$-band Multi Object
Spectrograph (KMOS; \citealt{sharples2004,sharples2013}), now allow
for the simultaneous observation of multiple targets and as such we
can construct large and well-selected samples in reasonable exposure
times (e.g. \citealt{wisnioski2015,stott2016}).

A surprising discovery has been that while high-redshift samples are
kinematically diverse, with a higher incidence of mergers than
observed locally (e.g. \citealt{molina2017}), many galaxies appear to
be rotationally supported
(e.g. \citealt{forster2009,epinat2012,wisnioski2015,stott2016,harrison2017},
although see also \citealt{DiTeodoro16}). Often despite morphological
irregularity, the dynamical maps of these galaxies reveal a smooth,
continuous velocity gradient. Clumps visible in broad-band imaging
appear to be giant star-forming complexes
(e.g. \citealt{swinbank2012,genzel2011,livermore2012,wisnioski2012})
which are embedded within the disk and share the same underlying
dynamics.

The existence of settled disks supports the emerging consensus that a
galaxy's star-formation history is not dominated by mergers but by an
ongoing accretion of gas from the cosmic web
\citep{dekel2009,ceverino2010}. Observations of a tight relation
between stellar mass and star formation rate (the so-called galaxy
``main sequence''; \citealt{noeske2007,elbaz2011,karim2011}) are
considered further evidence of this. A gradual decrease in the
available gas supply would explain the evolution of this trend as a
function of redshift, whereas stochastic, merger-driven bursts would
introduce significantly more scatter.

Kinematic surveys have revealed that while typical rotation velocities
of high-redshift disks are similar to those seen locally, intrinsic
velocity dispersions are much higher
(e.g. \citealt{genzel2008,lehnert2009,gnerucci2011,epinat2012,newman2013,wisnioski2015,DiTeodoro16,turner2017}). These
dispersions are supersonic and most likely represent turbulence within
the interstellar medium (ISM). Measurements are consistently large,
both for natural seeing observations and those which exploit adaptive
optics (e.g. \citealt{law2009,wisnioski2011}) or gravitational lensing
(e.g. \citealt{stark2008,jones2010}). While most high-redshift studies
use emission lines such as H$\alpha$ or [O{\sc ii}] to trace the
ionised gas dynamics of galaxies, observations of spatially resolved
CO emission have been made
(e.g. \citealt{tacconi2010,tacconi2013,swinbank2011,genzel2013}). These
studies suggest that the molecular gas is also turbulent -- it is the
entire disk which is dynamically hot, and not just ``flotsam'' on the
surface that has been stirred up by star formation (see also
\citealt{bassett2014}).

Since turbulence in the ISM decays on timescales comparable to the
disk crossing time, a source of energy is required to maintain the
observed high velocity dispersions
(e.g. \citealt{maclow1998,stone1998}). Several potential mechanisms
have been suggested, including star formation feedback
(e.g. \citealt{lehnert2009,green2010,lehnert2013,letiran2011}),
accretion via cosmological cold flows \citep{klessen2010},
gravitational disk instabilities
(e.g. \citealt{bournaud2010,bournaud2014,ceverino2010,goldbaum2015}),
interactions between star-forming clumps \citep{dekel2009,aumer2010},
or some combination thereof. However there have been few observational
tests of these theories.

Recent advancements in instrumentation such as multi-IFU systems
(e.g. KMOS, SAMI; \citealt{sharples2013}, \citealt{croom2012}) and
panoramic IFUs (e.g. MUSE; \citealt{bacon2010}) allow for large,
un-biased samples to be subdivided into bins of redshift, star
formation rate, stellar mass and morphology. In this work we
investigate the velocity dispersion properties of high-redshift
galaxies using data from the KMOS Redshift One Spectroscopic Survey
(KROSS; \citealt{stott2016}). This mass-selected parent
  sample targetted with KMOS consists of $\sim$\,800
  H$\alpha$-detected, typical star-forming galaxies at $z$\,$\sim$\,1.
  Of these galaxies observed with KMOS, 586 are detected in H$\alpha$,
  and these are the sample analysed in this paper. We further
supplement these observations with data from SAMI ($z$\,$\sim$\,0.05)
and an intermediate redshift MUSE sample ($z$\,$\sim$\,0.5).

We organise the paper as follows. In \S\ref{section2} we describe the
KROSS survey, sample selection and observations. In \S\ref{section3}
we outline our analysis, the measurement of kinematic quantities and
corrections applied for beam smearing. In \S\ref{results} we present
our results. We discuss how velocity dispersion relates to star
formation rate and stellar mass, and explore how galaxy dynamics
evolve as a function of redshift. In \S\ref{theory} we investigate
which physical processes may drive turbulence in the ISM, using KROSS
to test the predictions of analytic models. Finally in
\S\ref{conclusions} we summarise our main conclusions. In this work,
we adopt a $H_0$\,=\,70\,km\,s$^{-1}$\,Mpc\,$^{-1}$, $\Omega_{\rm
  M}$\,=\,0.3, $\Omega_\Lambda$\,=\,0.7 cosmology. We assume a
Chabrier IMF \citep{chabrier2003}, and quote all magnitudes as
AB. Throughout, the errors associated with median values are estimated
from a bootstrap re-sampling of the data.

%
%
\section{Survey Properties, Sample Selection and Data Reduction}
\label{section2}

KROSS is an ESO Guaranteed Time survey (PI: R. Sharples) designed to
study the spatially resolved dynamics of typical $z$\,$\sim$\,1
star-forming galaxies using KMOS. With 24 individual near-infrared
IFUs, the high multiplexing capability of KMOS has allowed us to
efficiently construct a statistically significant sample at this
epoch. The programme is now complete, with a total of 795 galaxies
observed. Full details of the sample selection, observations and data
reduction can be found in \citet{stott2016} and \citet{harrison2017},
however in the following sub-sections we briefly summarise the key
aspects.

\begin{figure}
\includegraphics[width=0.48\textwidth]{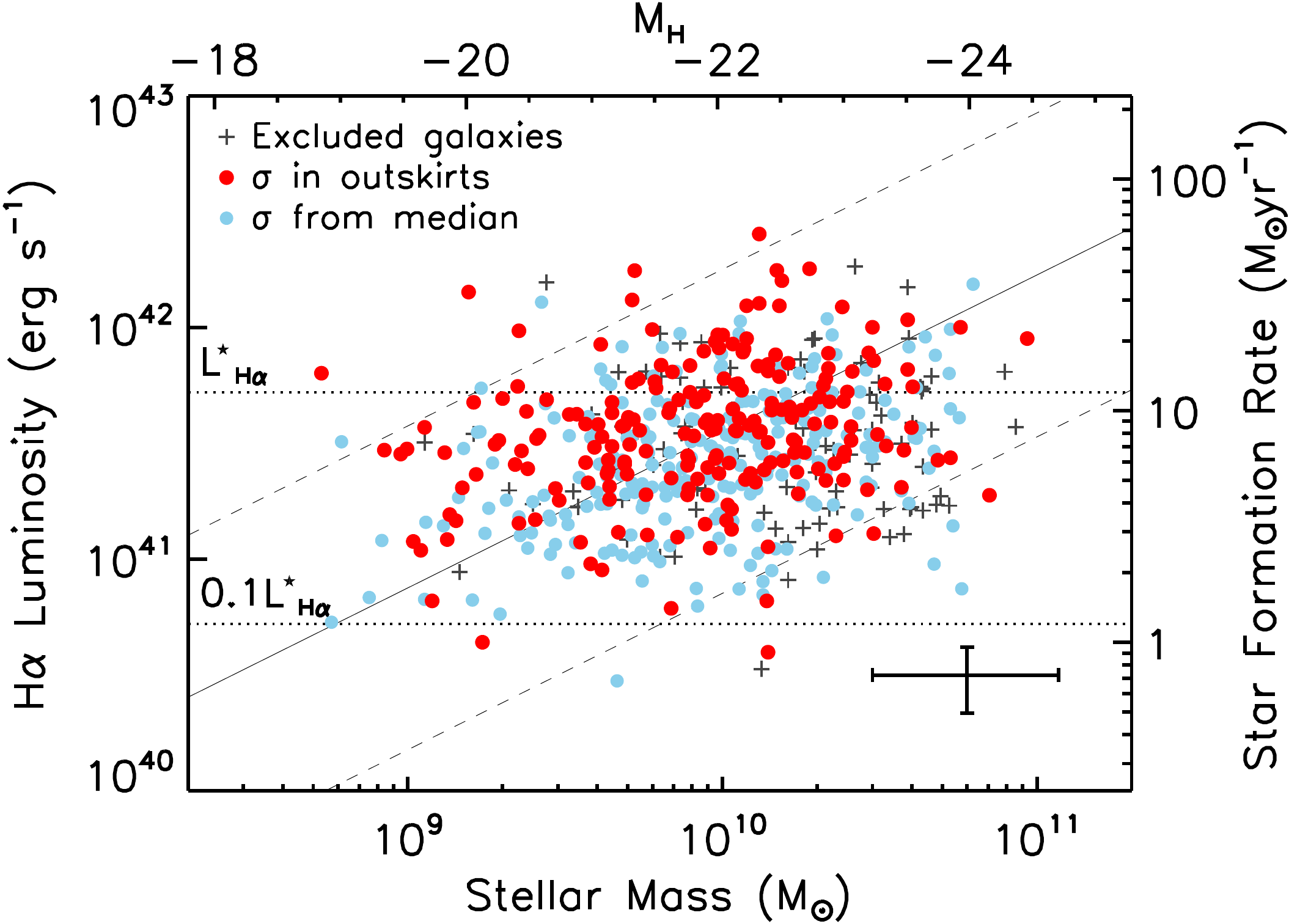}
\caption{Observed H$\alpha$ luminosity against stellar mass (scaled
  from $M_H$, top axis, assuming a constant mass-to-light ratio) for
  all 586 H$\alpha$--detected KROSS galaxies. Targets cut from the
  final kinematic sample (potential AGN or mergers, unresolved or low
  data quality sources; see \S\ref{final_sample}) are marked by
  crosses. We differentiate between galaxies for which the dispersion
  is measured in the outskirts of the disk, and those where it comes
  from the median of all available pixels (see
  \S\ref{measuring_sigma}). We find a median star formation rate of
  7\,M$_{\odot}$\,yr$^{-1}$ and a median stellar mass of
  10$^{10}$\,M$_{\odot}$, in line with the star forming ``main
  sequence'' at $z$\,=\,0.85 (\citealt{speagle2014}; solid line, with
  dashed lines a factor of five above or below). Dotted lines show
  0.1\,$\times$ and 1\,$\times$\,$L_{\rm H\alpha}$ at this redshift
  \citep{sobral2015}. A typical systematic error is shown in the
  bottom right.}
\label{fig:mass_sfr}
\end{figure}

\subsection{Sample selection}
\label{sample_selection}

The main aim of KROSS is to study the ionised gas kinematics of a
large and representative sample of star-forming galaxies at
$z$\,$\sim$\,1. We use KMOS to target the H$\alpha$ emission line,
which combined with the adjacent [N{\sc ii}] doublet allows us to
trace star formation, dynamics and chemical abundance
gradients. Targets were selected such that H$\alpha$ is redshifted
into the \textit{YJ} band and are located in the following
extragalactic fields: (1) Cosmological Evolution Survey (COSMOS); (2)
Extended \textit{Chandra} Deep Field South (ECDFS); (3) SA22 and (4)
UKIDSS Ultra-Deep Survey (UDS).

In addition to these redshift criteria we prioritised galaxies with an
observed $K$-band magnitude of $K_{\rm AB}$\,$<$\,22.5, which
translates to a stellar mass of
log(M$_\star$/M$_\odot$)\,$\gtrsim$\,9.5 at this redshift (see
\S\ref{stellar_masses}), and with colours of $r-z$\,$<$\,1.5. For
completeness, redder galaxies (more passive or potentially more dust
obscured) were also included but were assigned a lower priority for
observation. Our sample therefore favours star-forming and unobscured
galaxies which may have strong line emission.

From the original KROSS parent sample of 795 galaxies
  that were targetted with KMOS, we follow \citet{harrison2017} and
  first remove 52 galaxies which have unreliable photometry, or
  suffered from KMOS pointing errors.  The remaining sample consists
  of 743 galaxies between $z$\,=\,0.6\,--\,1.0, with a median redshift
  of $z$\,=\,0.85\,$^{+0.11}_{-0.04}$.  Of these, 586 are detected in
  H$\alpha$, with a total signal-to-noise in the one-dimensional
  spectrum (integrated over the FWHM of the H$\alpha$ emission line) of
  S\,/\,N\,$>$\,5.  These 586 galaxies are used for dynamical analysis
  described in this paper.

\subsection{Stellar masses}
\label{stellar_masses}

Since many our targets lie in deep extragalactic survey fields, a
wealth of archival photometry data (from X-ray to radio) exists.
Wherever possible, we use imaging from the $U$-band through IRAC
8.0\,$\mu$m to derive the best-fit SEDs and absolute magnitudes.
Briefly, we applied the SED fitting code {\sc hyperz}
\citep{bolzonella2000} to fit $U$-band through 4.5\,$\mu$m photometry
using spectral templates derived from the \citet{bruzual&charlot2003}
evolutionary code.  The model SEDs are characterised by
  star-formation histories which are parameterised by age and
  reddening.  We use the solar metallicity templates and consider
  seven star-formation histories (Burst, Constant, and six
  exponentially declining models with $\tau$\,=\,1, 2, 3, 5, 15 and
  30\,Gyr).  We allow stellar reddening ($A_{\rm V}$) of 0--5 magnitudes in steps
  of 0.1 and follow the \citet{calzetti2000} attenuation law.
  
Although individual estimates of stellar mass can be made from the
best-fit star formation histories, for consistency with
\citet{harrison2017}, and for the sake of reproducibility
  and homogeneity within our analysis, we apply a single mass-to-light
  ratio to ensure consistency across the four target fields.  This
  also allows homogeneity with the lower-redshift MUSE comparison
  sample, where there is often no long wavelength IRAC data (which is
  needed to break some of the degeneracies in the SED fitting between
  young/dusty versus old stellar populations). To derive stellar
masses we convert rest-frame $H$-band absolute magnitudes using the
median mass-to-light ratio returned by {\sc hyperz}
($\Upsilon_H$\,=\,0.2), as M$_\star$\,=\,$\Upsilon_H \times 10^{-0.4
  \times (M_H - 4.71)}$, resulting in a median stellar mass of
log(M$_\star$/M$_\odot$)\,=\,10.0\,$\pm$\,0.4. 

We note
  that adopting a single mass-to-light ratio or dust attenuation for
  the star formation rates (see \S\ref{sfrs}) may over- or under-
  estimate the stellar masses.  However, over the range of M$_\star$\,=\,5\,$\times$\,10$^9$ to
  5\,$\times$\,10$^{11}$\,M$_\odot$ the stellar masses as measured
  from the star formation histories (accounting for mass loss)
  and estimates made by adopting a single mass-to-light ratio
  typically agree to within 30\%.  Moreover, we do not identify a
  strong trend in dust attentuation, with A$_{\rm V}$ varying by less
  than 0.3\,magnitudes over this same stellar mass range.

\subsection{Star formation rates}
\label{sfrs}

We find a median H$\alpha$ luminosity for the KROSS sample of
log($L_{\rm H\alpha}$/erg\,s$^{-1}$)\,=\,41.5$\pm$\,0.3, which equates
to $\sim$\,0.6\,$\times$\,$L^\star_{\rm H\alpha}$ at $z$\,$\sim$\,1
\citep{sobral2015}. To convert to star formation rates we adopt a
simple approach and apply the \citet{kennicutt1998} calibration (using
a Chabrier IMF; \citealt{chabrier2003}), assuming a dust attenuation
of $A_{\rm H\alpha}$\,=\,1.73 (the median for the sample as returned
by {\sc hyperz}, converted from stellar to gas extinction using the
relation from \citealt{wuyts2013}). From this method, we derive a
median star formation rate of 7.0\,$\pm$\,0.3\,M$_\odot$yr$^{-1}$ (see
also \citealt{harrison2017}).

In Fig.\,\ref{fig:mass_sfr} we plot H$\alpha$ luminosity versus
estimated stellar mass for the 586 galaxies detected in H$\alpha$. We
overlay the star-forming ``main sequence'' (as described by
\citealt{speagle2014}) at the median redshift of KROSS and find the
properties of our sample to be consistent with this
trend. Approximately 95\% of galaxies have star formation rates within
a factor of five of the median for their mass. We therefore conclude
that our sample appears to be representative of typical star-forming
galaxies at this redshift.

\subsection{Observations and data reduction}
\label{observations}

Full details of the observations and data reduction can be found in
\citet{stott2016}, however the following is a brief
summary. Observations for KROSS were taken using KMOS, a near-infrared
integral field spectrograph on ESO/VLT. The instrument consists of 24
individual IFUs deployable within a 7.2 arcmin diameter patrol
field. Each covers a 2.8\,$\times$\,2.8 arcsec field of view with a
uniform spatial sampling of 0.2 arcsec. All targets were observed with
the \textit{YJ}-band filter which covers a wavelength range of
1.03\,--\,1.34\,$\mu$m, thus allowing us to measure the rest-frame
optical properties of our sample. The spectral resolution in this band
ranges between $R$\,$\sim$\,3000\,--\,4000.

Data was taken primarily between October 2013 and October 2015 using
guaranteed time, but was supplemented with some science verification
observations \citep{sobral2013b, stott2014}. Median seeing in the
$J$-band was 0.7 arcsec (which corresponds to a physical
  scale of 5.4\,kpc at the median redshift of our survey), with 92\%
of observations made during conditions of $<$\,1 arcsec, and
throughout the analysis we account for the seeing conditions of
individual observations. In Appendix \ref{appendix} we present a
detailed investigation into the impact of the seeing on our kinematic
measurements (so-called ``beam smearing''). Observations were made in
an ABAABAAB nod-to-sky sequence, where A represents time on target and
B time on sky. Total on-source integration time was an average of
9\,ks per galaxy.

Initial data reduction was performed using the standard {\sc
  esorex}/{\sc spark} pipeline which dark subtracts, flat-fields and
wavelength calibrates individual science frames, and applies an
additional illumination calibration. Each AB pair was reduced
individually, with the temporally closest sky subtracted from each
object frame. Further sky subtraction was then performed using
residual sky spectra extracted from a series of dedicated sky IFUs
(one for each of the three KMOS detectors). Finally, we combined all
observations of the same galaxy using a 3$\sigma$ clipped average and
re-sampled onto a pixel scale of 0.1 arcsec. This forms the final
datacube which we used to extract H$\alpha$ and continuum images, and
velocity and line of sight velocity dispersion maps discussed in the
following sections.
\vspace{-0.3cm}
\section{Analysis}
\label{section3}

In this work we explore the velocity dispersion properties of the
KROSS sample, investigating which processes may drive the high levels
of disk turbulence typically observed in galaxies at this redshift. We
first require measurements of galaxy size, inclination, position
angle, rotation velocity and velocity dispersion. \citet{harrison2017}
discussed how high resolution broad-band imaging can be combined with
KMOS data in order to make robust measurements of kinematic and
morphological properties. In the following section we summarise this
analysis. A catalogue of raw and derived properties for all 586
H$\alpha$ detected targets is available online (see Appendix
\ref{catalog}). With the release of this paper this has been updated
to include measurements and derived quantities relating to the
velocity dispersion, as also provided in Table\,\ref{table:Sigmas}. We
also discuss our method for mitigating the effects of beam smearing,
with a full, comprehensive analysis presented in Appendix
\ref{appendix}.

\subsection{Broad-band imaging}
\label{images}

We used the highest quality broad-band imaging available to measure
the half-light radius ($R_{1/2}$), inclination ($\theta$) and position
angle (PA$_{\rm im}$) of each galaxy. For 46\% of our sample there is
archival \textit{HST} imaging. All of our targets in ECDFS and COSMOS,
and a subset of those in UDS, have been observed with \textit{HST} in
the $H$, $I$ or $z'$-band. For all other targets we use $K$-band
ground-based imaging taken with the United Kingdom Infrared Telescope
as part of the UKIDSS survey \citep{lawrence2007}. These images have a
typical PSF of FWHM\,=\,0.65 arcsec in UDS and 0.85 arcsec in SA22.

In \citet{harrison2017} we discuss the implications of using imaging
of different rest-frame wavelengths and spatial resolutions, and
perform a series of tests to determine any systematics introduced. A
small ($\sim$\,10\%) correction is required such that the galaxy sizes
measured at different wavelengths are consistent. We also assign
greater uncertainties to position angles and inclinations derived from
ground-based images to account for the additional scatter introduced
to these measurements.

\subsection{Sizes, inclinations and position angles}
\label{image_measurements}

We first fit each image as a two dimensional Gaussian profile in order
to determine a morphological position angle and best-fit axis ratio
($b$/$a$). We deconvolve for the PSF of the image and convert this
axis ratio to an inclination angle as

\begin{equation}
cos^2\theta_{\rm im}\,=\,\frac{(b/a)^2-q_0^2}{1-q_0^2},
\end{equation}
where $q_0$ is the intrinsic axial ratio of an edge-on galaxy. This
parameter could have a wide range of values ($\approx$\,0.1\,--\,0.65;
see \citealt{law2012}), however we adopt the ratio for a thick disk,
$q_0$\,=\,0.2. Adjusting $q_0$ would not have a significant impact on
our results. For 7\% of galaxies we are unable to estimate
$\theta_{\rm im}$ due to poor resolution imaging. We therefore assume
the median axis ratio of the \textit{HST} observed sources and assign
these a ``quality 2'' flag (see \S\ref{final_sample}).

To estimate the half-light radius we measure the flux of each
broad-band image within a series of increasingly large elliptical
apertures. For each ellipse we use the continuum centre, and the
position angle and axis ratio derived above. We define $R_{1/2}$ as
the radius of the ellipse which contains half the total flux,
deconvolved for the PSF of the image.

For 14\% of the sample we are unable to measure the half-light radius
from the image, but instead infer an estimate using the turn-over
radius of the rotation curve ($R_{\rm d}$; see
\S\ref{measuring_vel}). We calibrate these radii using sources for
which both $R_{1/2}$ and $R_{\rm d}$ can be measured, and again assign
a ``quality 2'' flag. For an additional 6\% of sources neither of
these methods were suitable and we therefore place a conservative
upper-limit on $R_{1/2}$ of 1.8\,$\times$\,$\sigma_{\rm PSF}$. We
assign these a ``quality 3'' flag.  Quality 3 sources are
  not included in the results and discussion in Section 4 and 5 (see
  Section~\ref{final_sample}).

\subsection{Emission line fitting}
\label{making_maps}

A detailed description of how we extract two dimensional maps of
H$\alpha$ flux, velocity and velocity dispersion from the IFU data can
be found in \citet{stott2016}, however we include a brief summary
here. In each spatial pixel we fit the H$\alpha$ and [N{\sc
    ii}]\,6548,6583 emission lines via a $\chi^2$ minimisation
procedure, weighting against the positions of bright OH skylines
\citep{rousselot2000}. Each emission line is modelled as a single
Gaussian component within a linear local continuum. We fit the
H$\alpha$ and [N{\sc ii}] emission simultaneously, allowing the
centroid, intensity and width of the Gaussian profile to vary. The
FWHM of the lines are coupled, wavelength offsets fixed, and the flux
ratio of the [N{\sc ii}] doublet fixed to be 3.06
\citep{osterbrock2006}. During the fitting, we convolve the line
profile with the instrumental dispersion, as measured from the widths
of nearby skylines. As such, our dispersion measurements are corrected
for the instrumental resolution.

If the detection in a given pixel does not exceed a signal-to-noise of
$>$\,5 then we bin the data into successively larger regions, stopping
either when this criteria is met or an area of 0.7\,$\times$\,0.7
arcsec (the typical seeing of our observations) is reached. Using this
method, 552 (94\%) of the H$\alpha$ detected sample are spatially
resolved. We classify all unresolved sources as having a ``quality 4''
flag and these are not included in the results and
  discussion presented in Section 4 and 5. (see
  Section~\ref{final_sample}). In Fig.\,\ref{fig:examples} we show
example H$\alpha$ intensity, velocity and velocity dispersion maps for
eight KROSS galaxies.

\subsection{Rotation velocities}
\label{measuring_vel}

\begin{figure*}
\includegraphics[width=0.97\textwidth]{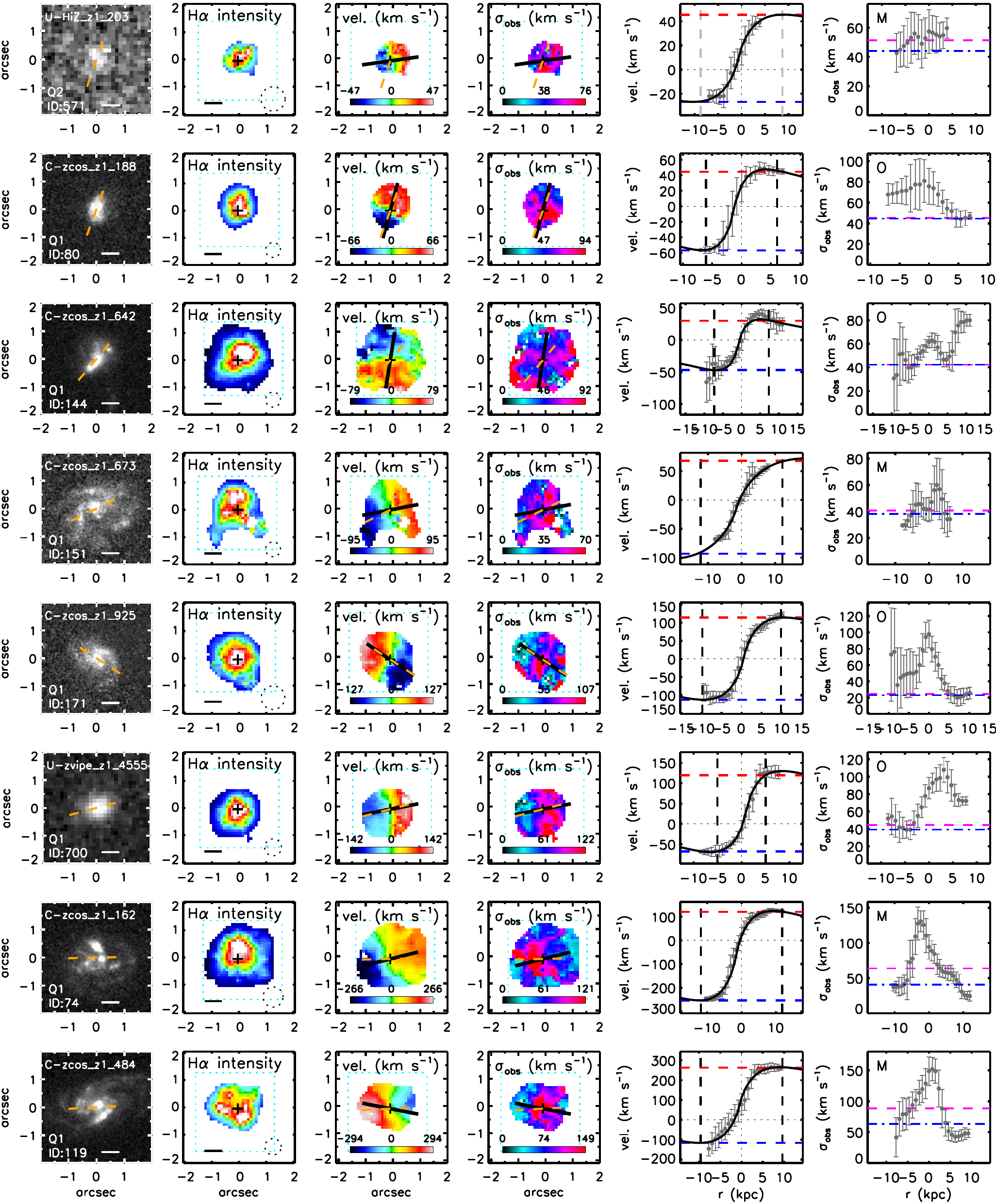}
\caption{Example data for eight galaxies in the KROSS sample (a
  complete set of figures is available in the online version of
  \citealt{harrison2017}), arranged by increasing stellar mass from top
  to bottom. \textit{Left to right:} (1) Broad-band image with orange
  dashed line to represent PA$_{\rm im}$. We also display the quality
  flag (see \S\ref{final_sample}) and a 5\,kpc scale bar. (2)
  H$\alpha$ intensity map with cross to mark the continuum centre and
  dashed circle to represent the seeing FWHM. (3) H$\alpha$ velocity
  map with dashed orange line to represent PA$_{\rm im}$ and solid
  black line to represent PA$_{\rm vel}$. (4) Observed H$\alpha$
  velocity dispersion map with lines as in panel 3. (5) Rotation curve
  extracted along a 0.7 arcsec wide `slit' of PA$_{\rm vel}$. The
  solid curve describes a disk model which we use to find the rotation
  velocity at $\pm$\,3.4\,$R_{\rm d}$ (dashed vertical lines). To
  estimate v$_{\rm obs}$ we take the average of these two values
  (horizontal dashed lines). (6) Observed velocity dispersion profile
  extracted along PA$_{\rm vel}$, with dashed line to represent
  $\sigma_{0, \rm obs}$ as measured in the outskirts of the disk (O)
  or from the median of all pixels (M). The dot-dashed line shows this
  same value corrected for beam smearing ($\sigma_0$). In general, as
  the stellar mass of the galaxy increases, we see a larger peak in
  the dispersion profile due to beam smearing.}
\label{fig:examples}
\end{figure*}

In order to measure a rotation velocity we must first establish the
position of the major kinematic axis (PA$_{\rm vel}$). We rotate the
H$\alpha$ velocity field around the continuum centre in 1 degree
increments and extract a velocity profile each time. We find the
profile with the largest velocity gradient and identify this position
angle as PA$_{\rm vel}$. To extract a rotation curve along this axis,
we calculate the median velocity at positions along a 0.7 arcsec
``slit'' through the continuum centre. Example rotation curves are
included in Fig.\,\ref{fig:examples}, where the error bar associated
with each point represents all variation within the ``slit''.

\begin{figure}
\includegraphics[width=0.465\textwidth]{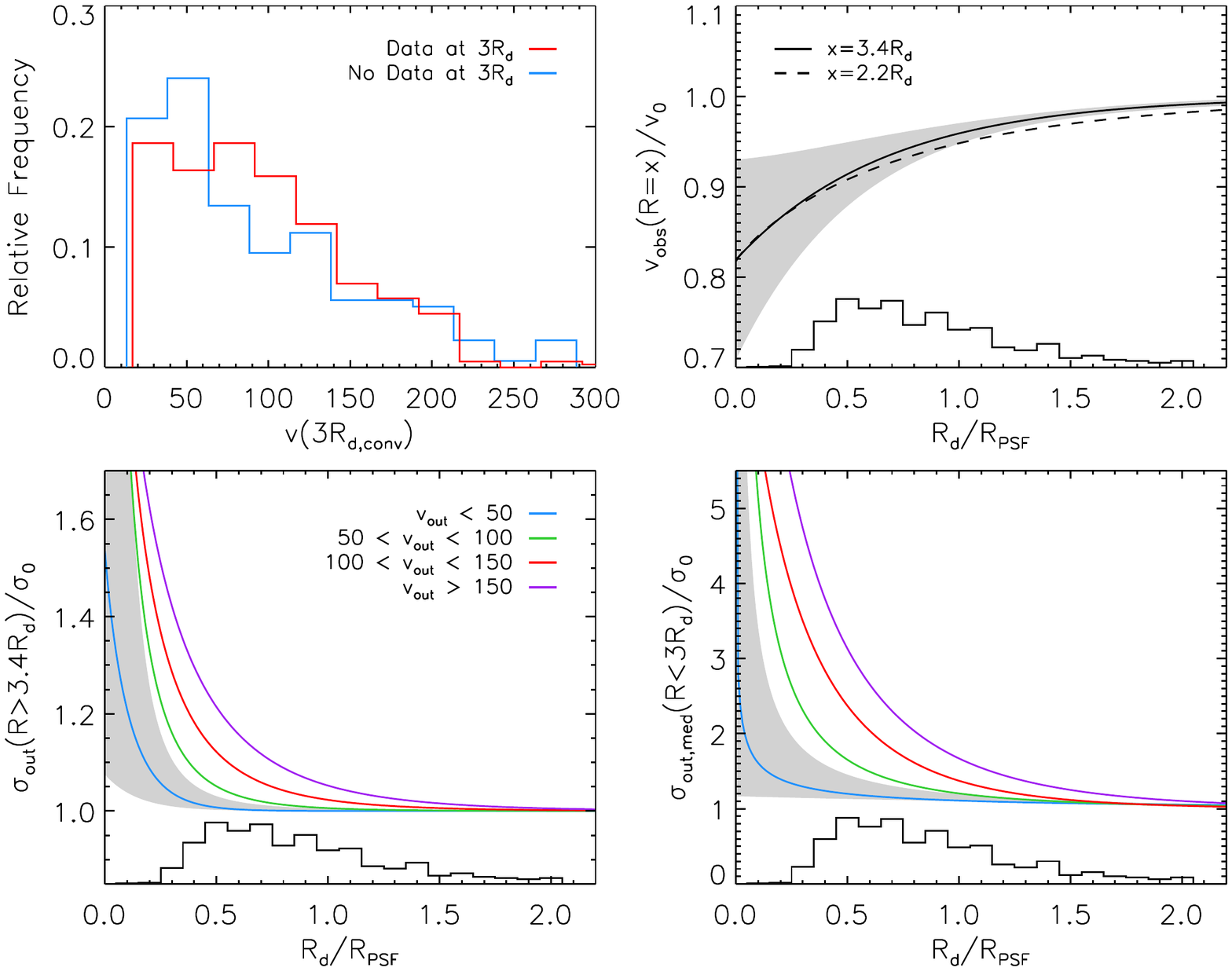}
\caption{Beam smearing correction applied to measurements of the
  rotation velocity at radii of 3.4 and 2.2$R_{\rm d}$ ($v_{\rm C}$
  and $v_{2.2}$, respectively), as a function of $R_{\rm d}$/$R_{\rm
    PSF}$. The shaded region represents the 1$\sigma$ scatter of
  outcomes for $\sim$\,10$^5$ mock galaxies. Tracks show the median of
  these outcomes and are defined by Eq.\,\ref{eq2} and the parameters
  listed in Table \ref{beam_smearing_params}. The histogram represents
  the $R_{\rm d}$/$R_{\rm PSF}$ distribution of the KROSS
  sample. Applying these beam smearing corrections to our data we find
  a modest median velocity correction of $\xi_v$\,=\,1.07\,$\pm$\,0.03
  and range of $\xi_v$\,=\,1.0\,--\,1.17.}
\label{fig:vel_tracks}%
\end{figure}

To minimise the impact of noise on our measurements, we fit each rotation curve as an exponential disk \citep{freeman1970} of the form:

\begin{equation}
v(r)^2 = \frac{r^2 \pi G \mu_0}{R_{\rm d}}(I_0K_0 - I_1K_1)+v_{\rm off},
\end{equation}
where $r$ is the radial distance, $\mu_0$ is the peak mass surface
density, $R_{\rm d}$ is the disk radius and $I_nK_n$ are the Bessel
functions evaluated at 0.5$r/R_{\rm d}$. The final parameter, $v_{\rm
  off}$, is the velocity measured at the centre of the galaxy and we
apply this offset to the rotation curve before making measurements. We
model each galaxy in this way with the intention of interpolating the
data to obtain a more robust measurement. However, for 13\% of
galaxies we must extrapolate ($>$\,0.4 arcsec; $\sim$\,3\,kpc) beyond
the data to evaluate the rotation velocity at the desired radius.

We measure the rotation velocities of our sample at two radii
frequently used within the literature, 1.3$R_{1/2}$ and 2$R_{1/2}$
($\approx$\,2.2$R_{\rm d}$ and 3.4$R_{\rm d}$ for an exponential
disk). The first of these coincides with the peak rotation velocity of
an ideal exponential disk, while the second probes outer regions of
the galaxy, where we expect the rotation curve to have flattened. We
refer to these measurements as $v_{2.2}$ and $v_{C}$,
respectively. For each galaxy we convolve $R_{1/2}$ with the PSF of
the KMOS observation\footnote{i.e. $R_{\rm
    1/2,conv}^2$\,=\,$R_{1/2}^2$\,+\,FWHM$_{\rm PSF}^2$} and extract
velocities from the model rotation curve. At a given radius, our final
measurement is half the difference between velocities on the blue and
red side of the rotation curve. We account for beam smearing using the
correction factors derived in \S\ref{beam_smearing}. Finally, we
correct for the inclination of the galaxy, as measured in
\S\ref{image_measurements}.

A small subset of our sample (11\%) are unresolved in the KMOS data
(``quality 4'') or the broad-band imaging (``quality 3''). As such we
are unable to extract rotation velocities for these galaxies from a
rotation curve. Instead we make estimates using the linewidth of the
galaxy integrated spectrum, and calibrate our results using galaxies
for which both methods are available. From a sample of 586 H$\alpha$
detected galaxies, 433 are flagged as ``quality 1'', 88 are ``quality
2'', 31 are ``quality 3'' and 34 are ``quality 4''. Only
  quality 1 and 2 sources are included in our results and discussion
  on the intrinsic velocity dispersions in Sections 4 and 5 (see
  Section~\ref{final_sample}).

\subsection{Velocity dispersions}
\label{measuring_sigma}

Throughout our analysis, we assume that the intrinsic velocity
dispersion is uniform across the disk (as in
e.g. \citealt{genzel2014,epinat2012,simons2016}). In the same way as
we extract a rotation curve from the velocity map, we also extract a
profile along the major kinematic axis of the velocity dispersion
map. We use this profile to measure the observed dispersion,
$\sigma_{0,{\rm obs}}$, by taking the median of values at either end
of the kinematic axis $|R|$\,$>$\,2$R_{1/2}$ and adopting whichever
value is smallest (see Fig.\,\ref{fig:examples}). We assume the
uncertainty on this measurement is the scatter of values included in
the median. Evaluating $\sigma_{0,{\rm obs}}$ at radii far from the
dynamical centre reduces any bias introduced by beam smearing (see
\S\ref{beam_smearing}), and measurements here should be close to the
intrinsic dispersion.

While this is our preferred method, 56\% of the resolved sample (307
galaxies) have insufficient signal-to-noise in the outer regions of
the galaxy ($\pm$\,2$R_{1/2}$) to be able to measure the dispersion in
this way. Instead we measure the median of all available pixels within
the dispersion map. Once we apply the relevant beam smearing
corrections derived in \S\ref{beam_smearing}, we find that the
$\sigma_{0,{\rm obs}}$ values from each method are in good
agreement. In cases where we can follow either approach the results
are (on average) consistent to within 4\%, with $\approx$\,50\%
scatter around this offset. We therefore assign an uncertainty of 50\%
to measurements made using this second method. We do not estimate
$\sigma_0$ for unresolved galaxies. 

We note that
  \citet{DiTeodoro16} derived intrinsic velocity dispersions using the
  KMOS data for 14 of the galaxies from our sample using their
  three-dimensional $^{3D}$BARLO technique.  12/14 of their derived
  values agree with 2$\sigma$ of our values but with $\sim$20\% lower
  velocity dispersions on average.  Our method, whilst less complex,
  has the advantage that is can be uniformly applied to a range of
  data quality across a wide range of redshifts, allowing us to
  explore trends with redshift, mass and star-formation rate in very
  large samples (e.g.\ see Section 4.3).

\subsection{Beam smearing corrections}
\label{beam_smearing}

Since our KMOS observations are seeing-limited, we must consider the
impact of the spatial PSF (the seeing) on our kinematic
measurements. As IFU observations are convolved with the PSF,
information from each spatial pixel is combined with that of
neighbouring regions -- a phenomenon known as ``beam smearing'' (see
e.g. \citealt{epinat2010,davies2011,burkert2016,federrath2017,zhou2017}). This
acts to increase the observed velocity dispersion (particularly
towards the dynamical centre) and to flatten the observed rotation
curve, thereby reducing the observed velocity. In order to calibrate
for these effects, we create a series of mock KMOS observations and
derive correction factors which can be applied to the kinematic
measurements. Our method for this correction is similar to that
adopted by other authors (e.g. \citealt{burkert2016,turner2017}) and
we derive similar results. In Appendix \ref{appendix} we present full
details of this investigation, however the following is a brief
summary.

\begin{figure*}%
  \centering
  \subfloat{{\includegraphics[width=0.45\textwidth]{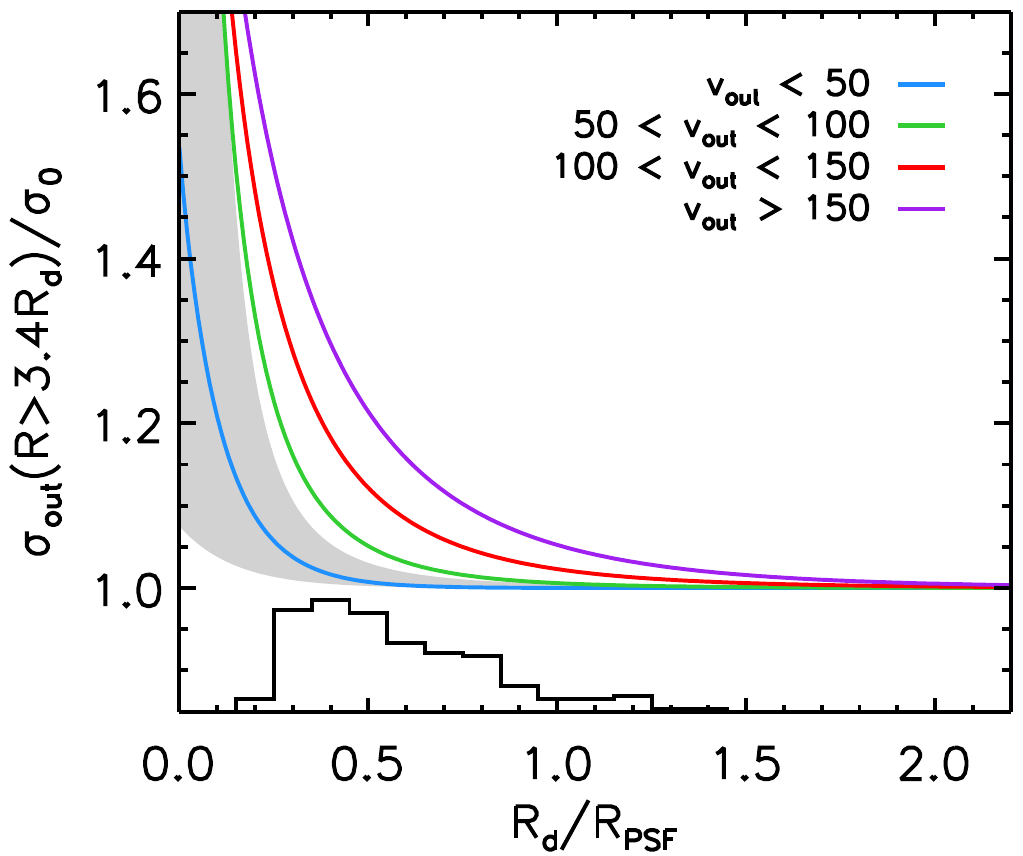} }}%
  \qquad
  \subfloat{{\includegraphics[width=0.43\textwidth]{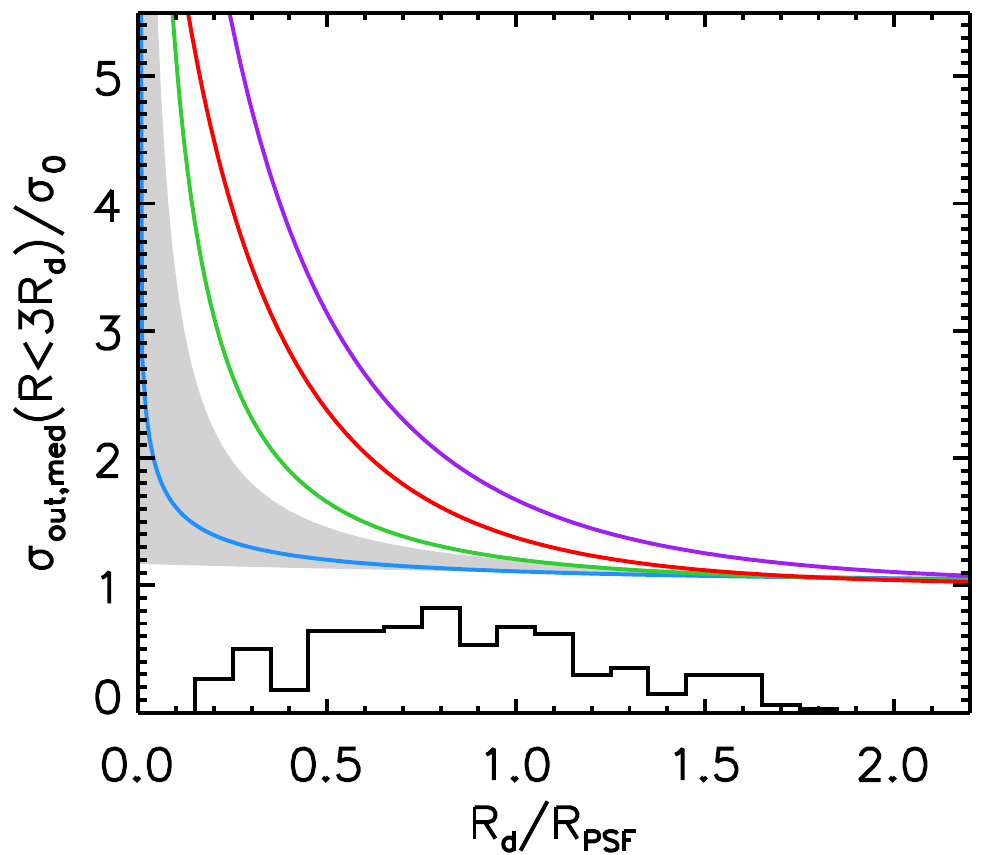} }}
  \caption{Beam smearing corrections for the velocity dispersion as a
    function of observed rotation velocity (v$_{\rm obs}$) and $R_{\rm
      d}$/$R_{\rm PSF}$. $R_{\rm PSF}$ is defined as half of the
    seeing FWHM and we assume an exponential disk such that $R_{\rm
      d}$\,=\,$R_{1/2}$\,/\,1.678. To derive these corrections we
    create $\sim$\,10$^5$ model galaxies of various masses, radii,
    inclinations, dark matter fractions and intrinsic dispersions
    ($\sigma_0$; uniform across the disk) and simulate the effects of
    beam smearing for a seeing of 0.5\,--\,0.9 arcsec (see
    \S\ref{beam_smearing} and Appendix \ref{appendix}). We fit a
    running median to the results of each velocity bin, with each
    track described by Eq.\,\ref{eq3} and the relevant parameters in
    Table\,\ref{beam_smearing_params}. Shaded regions demonstrate the
    typical 1$\sigma$ scatter of results in each bin, while the
    histograms represent the $R_{\rm d}$/$R_{\rm PSF}$ distribution of
    each subset. Note the different scales on the
    y-axes. \textit{Left:} Velocity dispersions measured in the
    outskirts ($R$\,$>$\,3.4$R_{\rm d}$) of the dispersion profile,
    relative to the intrinsic value. We apply an average correction of
    $\xi_\sigma$\,=\,0.97$^{+0.02}_{-0.06}$ to the KROSS
    sample. \textit{Right:} Dispersions measured as the median of all
    pixels. This method results in a greater overestimate of
    $\sigma_0$, with an average correction factor of
    $\xi_\sigma$\,=\,0.8$^{+0.1}_{-0.3}$.}
  \label{fig:beam_smearing_tracks}%
\end{figure*}

To begin this process we create a sample of $\sim$\,10$^5$ model disk
galaxies, with stellar masses and radii representative of the KROSS
sample. We assume an exponential light profile and model the galaxy
dynamics as the sum of a stellar disk plus a dark matter halo. An
appropriate range of dark matter fractions is determined using results
of the cosmological simulation suite ``Evolution and Assembly of
GaLaxies and their Environments'' ({\sc eagle};
\citealt{crain2015,schaye2015,schaller2015}). For simplicity, the
intrinsic velocity dispersion ($\sigma_0$) is assumed to be uniform
across the disk. From these properties we can predict the intensity,
linewidth and velocity of the H$\alpha$ emission at each position. We
use this information to create an ``intrinsic'' KMOS data cube for
each galaxy.

To simulate the effects of beam smearing we convolve each wavelength
slice of the cube with a given spatial PSF. We model a range of seeing
conditions to match our KMOS observations. This forms the ``observed''
data cube from which we extract dynamical maps (in the same way as for
the observations) and measure $v_{\rm C,obs}$, $v_{\rm 2.2,obs}$ and
$\sigma_{\rm 0,obs}$. Differences between the input values of the
model and these ``observed'' values then form the basis of our beam
smearing corrections. The amplitude of the beam smearing is most
sensitive to the size of the galaxy relative to the PSF. These
corrections are best parameterised as a function of $R_{\rm
  d}$/$R_{\rm PSF}$, where $R_{\rm PSF}$ is half of the FWHM of the
seeing PSF.

In Fig.\,\ref{fig:vel_tracks} we show the ratio of the observed and
intrinsic rotation velocity as a function of $R_{\rm d}$/$R_{\rm
  PSF}$. As expected, the larger the spatial PSF is compared to the
disk, the more we underestimate the intrinsic velocity. Averaging over
all stellar masses and inclinations, we find a median correction to
$v_{\rm C}$ of $\xi_v$\,=\,1.07\,$\pm$\,0.03, with a range of
$\xi_v$\,=\,1.0\,--\,1.17. Applying this correction acts to increase
the median rotation velocity measurement by 4\,km\,s$^{-1}$.

Similarly, the smaller the value $R_{\rm d}$/$R_{\rm PSF}$ the more we
overestimate the intrinsic velocity dispersion. However, the impact of
beam smearing on measurements of $\sigma_0$ also depends strongly on
the velocity gradient across the disk (which is a function of both
dynamical mass and inclination angle). In
Fig.\,\ref{fig:beam_smearing_tracks} we split corrections into four
separate tracks as a function of $v_{\rm obs}$. The majority of
galaxies in our sample (67\%) have observed rotation velocities of
$v_{\rm obs}$\,$\leq$\,100\,km\,s$^{-1}$, so most corrections are made
using the green and blue tracks of
Fig.\,\ref{fig:beam_smearing_tracks}. The required adjustments are
therefore relatively small. When using the velocity dispersions
extracted from outer regions of the disk, we apply a median beam
smearing correction of $\xi_\sigma$\,=\,0.97$^{+0.02}_{-0.06}$. If a
value is extracted from the median of the map, we apply a median
factor of $\xi_\sigma$\,=\,0.8$^{+0.1}_{-0.3}$. Applying these beam
smearing corrections to KROSS data reduces the median velocity
dispersion measurement by 9\,km\,s$^{-1}$.

\subsection{Definition of the final sample}
\label{final_sample}

In \S\ref{section2} we presented a mass- and colour-selected sample of
743 KROSS galaxies, 586 of which are detected in H$\alpha$. In
Fig.\,\ref{fig:mass_sfr} we show that this forms a representative
sample of star-forming galaxies at this redshift
($z$\,$\approx$\,0.85), in the context of the $M_\star$\,--\,SFR
``main sequence''. With kinematic and morphological properties of
these galaxies now established (e.g. Fig.\ref{fig:examples}), we make
a number of additional cuts to the sample.

Firstly, as in \citet{harrison2017} we exclude 20 galaxies with line
ratios of [N{\sc ii}]/H$\alpha$\,>\,0.8 and/or a broad-line component
to the H$\alpha$ emission of $\geq$\,1000\,km\,s$^{-1}$. These sources
may have a significant AGN component or kinematics which are
influenced by shocks (e.g. \citealt{kewley2013,harrison2016}). We also
remove 30 sources which have multiple components in their broad-band
imaging and/or IFU data. In doing so we hope to remove any potential
major mergers. Finally, we exclude ``quality 4'' and ``quality 3''
sources which are unresolved or without a half-light radius
measurement, respectively. This leaves a final sample of 472 galaxies.

Of this final sample, 18\% (84 galaxies) are classified as ``quality
2'', owing to a fixed inclination angle or half-light radius measured
from the rotation curve. For 49\% of the sample (231 galaxies) we are
able to measure the velocity dispersion ($\sigma_0$) using data in the
outer regions of the galaxy. For the remaining 51\% of cases (241
galaxies) we must measure the median of all IFU pixels and correct
this value appropriately. As discussed in \S\ref{measuring_sigma}
these two methods are consistent, however we attribute larger
uncertainities to measurements made using the latter approach. The
observed and beam smearing corrected velocity dispersions of each
galaxy are listed in Table\,\ref{table:Sigmas}, and a full catalogue
of galaxy properties is available online (see Appendix \ref{catalog}).

%
%
\section{Results}
\label{results}
In the previous section we summarised the morphological and kinematic
analysis of 586 H$\alpha$ detected galaxies in the KROSS sample.
After the removal of 114 sources which have either uncertain kinematic
measurements, or show signs of a significant AGN component or merger
event, we construct a final sample of 472 clean, well-resolved
galaxies. In the following subsections we present a detailed
discussion of the velocity dispersion properties of this sample.

\begin{figure}
\includegraphics[width=0.48\textwidth]{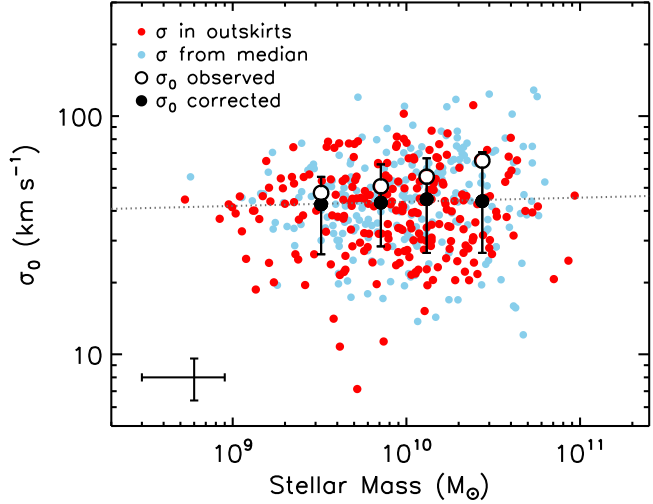}
\caption{Beam smearing corrected velocity dispersion against stellar
  mass, with points coloured by the technique used to measure
  $\sigma_0$. Large black symbols show the median dispersion (and
  standard deviation) in bins of stellar mass. If we consider only
  measurements made in the outskirts of the disk these average values
  are systematically a factor of 0.98\,$\pm$\,0.03 lower. Large open
  symbols show the median in each bin prior to the correction being
  applied. The large black points show that once we have accounted for
  the effects of beam smearing (see \S\ref{beam_smearing}) we find
  $\sigma_0$ to be independent of M$_\star$. The dotted line is a fit
  to this trend.}
\label{fig:sigma_mass}
\end{figure}

\subsection{Velocity dispersions}
We measure a median intrinsic velocity dispersion of
$\sigma_0$\,=\,43\,$\pm$\,1\,km\,s$^{-1}$ and a 16--84th
percentile range of 27\,--\,61\,km\,s$^{-1}$. This median dispersion
is lower than the 59\,$\pm$\,2\,km\,s$^{-1}$ previously reported for
KROSS in \citet{stott2016} due to a more rigorous beam smearing
analysis, different measurement techniques, and further refinement of
the kinematic sample (see \S\ref{section3} and
\citealt{harrison2017}). As discussed in \S\ref{measuring_sigma}, we
measure the dispersion of each galaxy using one of two different
methods. For approximately half of the sample we measure $\sigma_0$ in
outer regions of the disk ($|R|$\,$>$\,2$R_{1/2}$) while for the
remaining galaxies we calculate the median of all pixels. Since the
ability to resolve kinematics in the outskirts is dependent on galaxy
size and signal-to-noise, galaxies in the ``median'' sample tend to be
larger than those in the ``outskirts'' sample (median half-light radii
of 3.5\,$\pm$\,0.1\,kpc and 2.07\,$\pm$\,0.08\,kpc, respectively) and
also more passive (median star formation rates of
6.2\,$\pm$\,0.3\,M$_\odot$yr$^{-1}$ and
8.2\,$\pm$\,0.4\,M$_\odot$yr$^{-1}$). The velocity dispersions of this
subset are also slightly higher, with a median $\sigma_0$ of
45\,$\pm$\,1\,km\,s$^{-1}$ as opposed to 41\,$\pm$\,1\,km\,s$^{-1}$.

In Fig.\,\ref{fig:sigma_mass} we explore the relationship between
stellar mass and velocity dispersion. We may expect these quantities
to be related, since dispersions are important in measuring the
dynamical support of galaxies, regardless of morphological type. For
example, several authors have noted that the $S_{0.5}$ parameter
[$S_{0.5}$\,=\,(0.5\,$v^2$+$\sigma^2$)$^{1/2}$] correlates more
strongly with stellar mass than rotational velocity alone
(e.g. \citealt{kassin2007, vergani2012,
  cortese2014}). Fig.\,\ref{fig:sigma_mass} shows that before we
account for beam smearing, the average velocity dispersion increases
significantly with stellar mass. We measure a median $\sigma_{\rm
  obs}$ of 48\,$\pm$\,2\,km\,s$^{-1}$ in the lowest mass bin compared
to 64\,$\pm$\,5\,km\,s$^{-1}$ in the highest. However as discussed in
\S\ref{beam_smearing} (and extensively in Appendix \ref{appendix}), a
more massive galaxy is typically associated with a steeper velocity
gradient across the disk (e.g. \citealt{catinella2006}) and hence
stronger beam smearing. After we apply corrections as a function of
$R_{\rm d}$/$R_{\rm PSF}$ and $v_{\rm obs}$
(Fig.\,\ref{fig:beam_smearing_tracks}) we no longer observe this trend
and instead find the median $\sigma_0$ to be consistent across the
four mass bins, with values between 42$\pm$\,2\,km\,s$^{-1}$ and
45\,$\pm$\,3\,km\,s$^{-1}$. If we consider only dispersion
measurements made in the outskirts of the disk, values are almost
identical -- lower by a factor of 0.98\,$\pm$\,0.03. Our results are
consistent with $\sigma_0$ being almost independent of stellar mass
over the range log($M_{\star}$/M$_{\odot}$)\,=\,9.4\,--\,10.4.

\subsection{Rotational support}
To quantify the balance between rotational support and turbulence of
the gas, we calculate the ratio between rotation velocity and velocity
dispersion, $v_{\rm C}$/$\sigma_0$, for each of the KROSS galaxies. We
find a median value of $v_{\rm C}$/$\sigma_0$\,=\,2.6\,$\pm$\,0.1 and
a 16--84th percentile range of 0.9\,--\,5. We can use this ratio
between rotation velocity and intrinsic dispersion to achieve a crude
separation of ``dispersion dominated'' and ``rotationally dominated''
galaxies. Following e.g. \citet{genzel2006} we adopt $v_{\rm
  C}$/$\sigma_0$\,=\,1 as a boundary between the two. By this
definition we find a rotationally dominated fraction of
83\,$\pm$\,5\%, which suggests that the majority of star-forming
galaxies at this redshift are already settled disks. The KROSS sample
used for this work is slightly different to that presented in
\citet{harrison2017}, for example we include only ``quality 1'' or
``quality 2'' sources. However, our results are consistent, suggesting
that this does not introduce a bias. \citet{harrison2017} find a
median value of $v_{\rm C}$/$\sigma_0$\,=\,2.4\,$\pm$\,0.1 and a
rotationally dominated fraction of 81\,$\pm$\,5\%. Despite a more
detailed treatment of the beam smearing effects, our results are also
consistent with the initial KROSS values derived in \citet{stott2016}.
 
\begin{figure}
\includegraphics[width=0.478\textwidth]{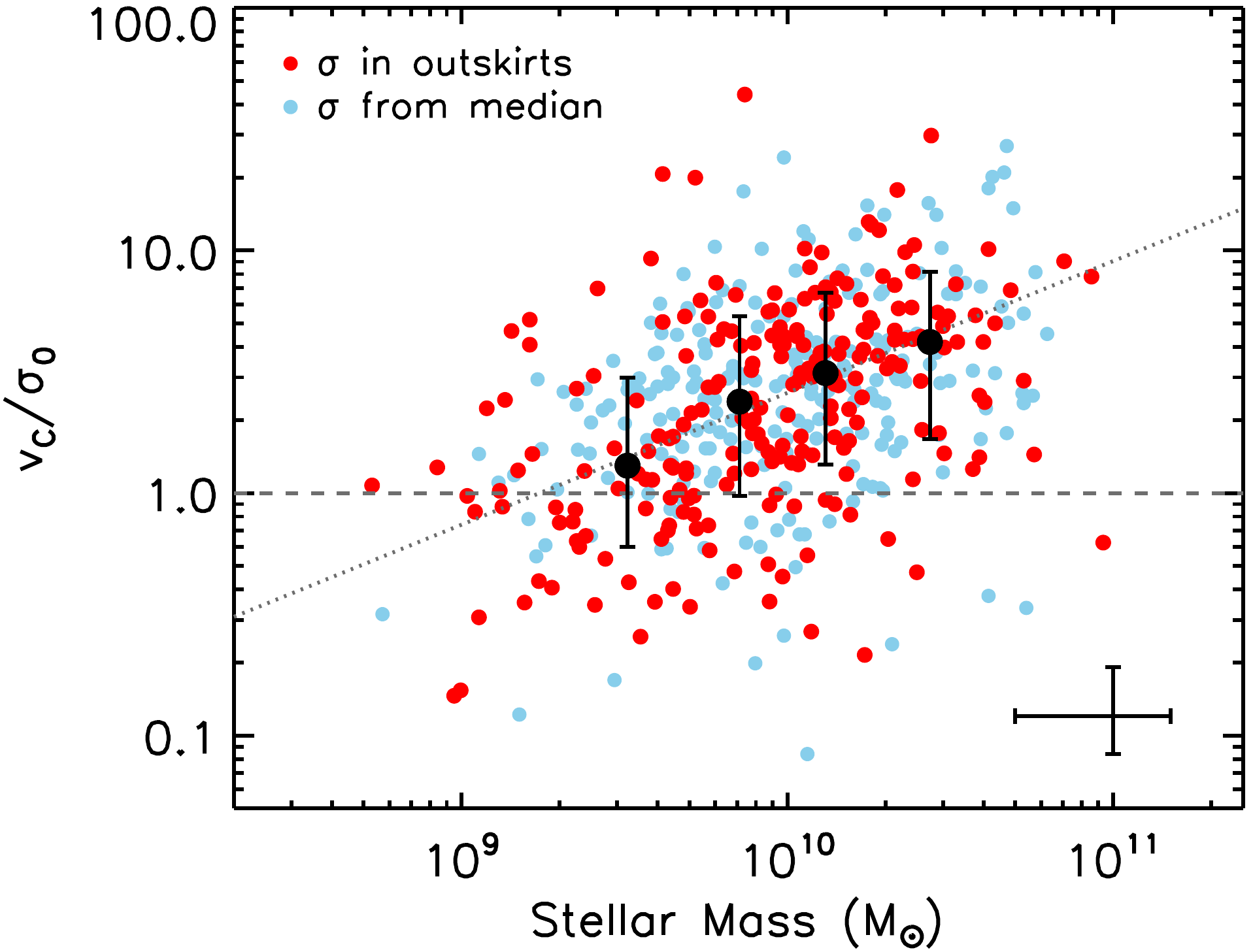}
\caption{Ratio between inclination corrected rotational velocity
  ($v_{\rm C}$) and intrinsic velocity dispersion ($\sigma_0$) against
  stellar mass. Fig.\,\ref{fig:sigma_mass} shows that the average
  $\sigma_0$ is roughly the same in each mass bin, however due to
  larger rotational velocities we see an increase in $v_{\rm
    C}$/$\sigma_0$ with increased stellar mass. We fit a trend to the
  median values in bins of increasing stellar mass (large black
  points) and plot this as a dotted line. The dashed line acts as a
  crude boundary between ``dispersion dominated'' (below) and
  ``rotationally dominated'' galaxies (above, $\sim$\,80\% of our
  sample). More massive galaxies appear to be more rotationally
  supported.}
\label{fig:vsigma_mass}
\end{figure}

In Fig.\,\ref{fig:vsigma_mass} we study how rotational support relates
to stellar mass. Observations suggest that galaxies evolve
hierarchically from disordered, dynamically hot systems to regularly
rotating disks, with the most massive galaxies settling first
(kinematic downsizing;
e.g. \citealt{kassin2012,vanderwel2014,simons2016,simons2017}). At a
given redshift it is expected that high mass galaxies are more stable
to disruptions due to gas accretion, winds or minor mergers
(e.g. \citealt{tacconi2013,genzel2014}). As such, we expect the most
massive galaxies to exhibit the largest $v_{\rm C}$/$\sigma_0$
values. Fig.\,\ref{fig:vsigma_mass} demonstrates that this is indeed
true for the KROSS sample, with median $v_{\rm C}$/$\sigma_0$ values
of 1.3\,$\pm$\,0.1 and 4.3\,$\pm$\,0.3 in the lowest and highest mass
bins, respectively, and ``dispersion dominated'' systems more
prevalent at low stellar mass. Since we observe no correlation between
velocity dispersion and stellar mass, this increase must be a result
of higher mass galaxies rotating more quickly. If
M$_{\star}$\,$\propto$\,$v_{\rm C}^2$ then we would expect $v_{\rm C}$
to increase by a factor of $\sim$\,3.2 over the mass range
log($M_{\star}$/M$_{\odot}$)\,=\,9.4\,--\,10.4.  Indeed,
  for KROSS galaxies, we find a slope of $\sim$\,2.1 in the $v_{\rm
  c}$ versus mass plane, consistent with previous multi-epoch galaxy
studies (e.g.\ \citealt{harrison2017}).  This is consistent with our
results in Fig.\,\ref{fig:vsigma_mass}.

\subsection{Trends between dispersion and stellar mass, star formation rate and redshift}
\label{sami_muse}

To analyse the kinematics of KROSS galaxies in an evolutionary
context, and to further explore how dispersion relates to other galaxy
properties, we introduce comparison samples. In the ``IFU era'' there
are a multitude of kinematic surveys to choose from, however it is
often difficult to make comparisons since the techniques used,
particularly for beam smearing corrections, can vary a great deal. In
this subsection we therefore consider only two additional samples, for
which we can measure (and correct) $\sigma_0$ in a consistent way. In
\S\ref{trends_with_redshift} we will study the average properties of a
further five comparison samples.

\subsubsection{SAMI sample}

Our first comparison sample consists of 824 galaxies from the
Sydney-AAO Multi-object Integral field (SAMI; \citealt{croom2012})
Galaxy Survey. The goal of this survey is to provide a complete census
of the spatially resolved properties of local galaxies
(0.004\,<\,$z$\,<\,0.095; \citealt{bryant2015,owers2017}). SAMI is a
front-end fibre feed system for the AAOmega spectrograph
\citep{sharp2006}. It uses a series of ``hexabundles''
\citep{bland2011,bryant2014}, each comprised of 61 optical fibres and
covering a $\sim$\,14.7\,arcsec field of view, to observe the stellar
and gas kinematics of up to 12 galaxies simultaneously. Reduced SAMI
datacubes have a 0.5 arcsec spatial sampling. A detailed description
of the data reduction technique is presented in \citet{sharp2015}. The
data used for this analysis was kindly provided by the SAMI team ahead
of its public release (Green et al. in prep), however an early data
release is presented in \citet{allen2015}.

In order to compare SAMI data to KROSS we first make a series of cuts
to the sample. In particular, the SAMI survey contains a number of
early-type and elliptical galaxies with high S$\'{e}$rsic indicies,
high stellar masses and low star formation rates (hence very low
specific star formation rates), which are not representative of the
KROSS sample selection, that is to select typical star-forming
galaxies for that epoch. We therefore remove galaxies from the SAMI
sample with masses greater than
$M_\star$\,=\,8$\times$10$^{10}$\,M$_{\odot}$ and a S$\'{e}$rsic index
of $n>$\,2 (since the derived $\sigma_0$ measurements for these
galaxies are likely to be measuring different physical processes). We
also remove sources which are unresolved at the SAMI resolution or
have kinematic uncertainities greater than 30\%. This leaves a total
of 274 galaxies with a median redshift $z$\,$\sim$\,0.04 and median
stellar mass log($M_{\star}$/M$_{\odot}$)\,=\,9.34\,$\pm$\,0.07.

In Fig.\,\ref{fig:sami_muse} we plot star formation rate versus
stellar mass for this sample. Stellar masses were estimated from $g-i$
colours and $i$-band magnitudes following \citet{taylor2011}, as
described in \citet{bryant2015}. Star formation rates were estimated
using H$\alpha$ fluxes corrected for dust attenuation. Most SAMI
galaxies are representative of the star-forming ``main sequence'' at
$z$\,=\,0 \citep{peng2010}, and hence at fixed stellar mass, star
formation rates are 30\,--\,50 times lower than for KROSS galaxies.

To measure rotation velocities and dispersions, we exploit the gas
velocity maps, which use 11 strong optical emission lines including
H$\alpha$ and [O{\sc ii}]. From these maps we make measurements using
the same methods as for the KROSS sample (for an independent study of
SAMI velocity dispersions see \citealt{zhou2017}). However, since the
angular sizes of galaxies at this redshift are much larger, the field
of view of SAMI often does not extend to 3.4$R_{\rm d}$. Instead, we
use a radius of 2$R_{\rm d}$ and correct the derived quantities
appropriately based on our modelling in Appendix \ref{appendix}.

\subsubsection{MUSE sample}

\begin{figure}
\includegraphics[width=0.48\textwidth]{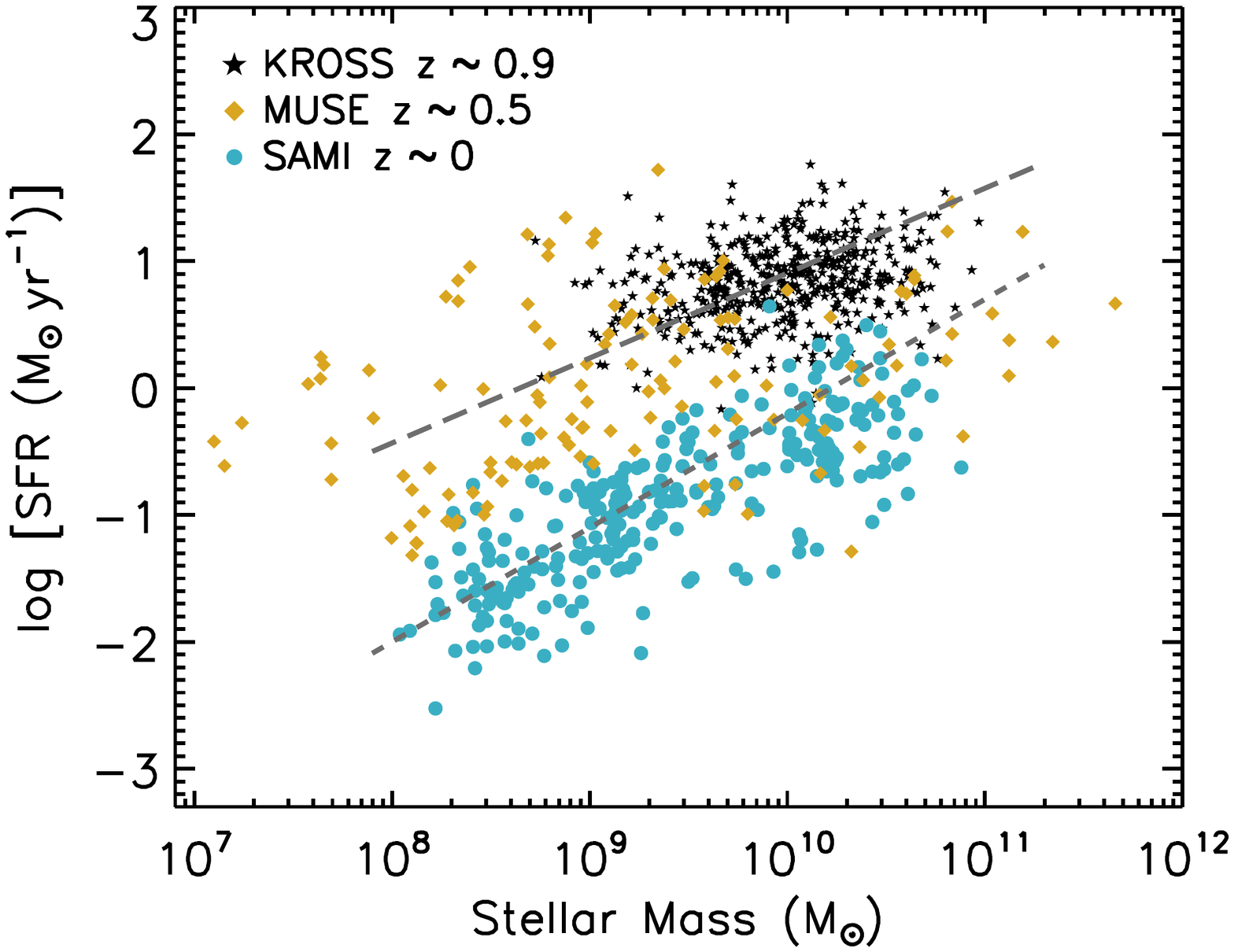}
\caption{Star formation rate versus stellar mass for the KROSS
  galaxies studied in this work (as in Fig.\,\ref{fig:mass_sfr}), and
  the MUSE and SAMI comparison samples discussed in
  \S\ref{sami_muse}. We overlay the star-forming ``main sequence'' at
  $z$\,=\,0 \citep{peng2010} and $z$\,=\,0.85 \citep{speagle2014},
  which illustrate that the KROSS and SAMI samples are representative
  of typical star-forming galaxies at their respective redshifts. The
  MUSE sample are [O{\sc ii}] emitters serendipitously detected within
  observations of other targets, hence these galaxies have a wide
  range of masses and star formation rates.}
\label{fig:sami_muse}
\end{figure}

For a second comparison we exploit the sample of \citet{swinbank2017},
who study the kinematics of 553 [O{\sc ii}] emitters serendipitously
detected in a series of commissioning and science verification
observations using MUSE (Multi-Unit Spectroscopic Explorer;
\citealt{bacon2010}), a panoramic IFU with 1\,$\times$\,1\,arcmin
field of view and 0.2\,arcsec spatial sampling. Science targets were
largely extragalactic ``blank'' fields or high-redshift galaxies and
quasars. Due to the nature of the sample, sources span a wide range of
redshifts, with 0.28\,<\,$z$\,<\,1.49. To provide an intermediate
between the redshifts of KROSS and SAMI, we restrict this sample to
galaxies between 0.3\,<\,$z$\,<\,0.7. In \citet{swinbank2017} sources
were classified as rotationally supported, merging, interacting or
compact, based on their dynamics and optical morphologies. We choose
to exclude major mergers and compact (unresolved) galaxies from our
analysis, and also those which have poorly defined masses or optical
radii. With the implementation of these cuts our comparison sample
consists of 133 galaxies with a median redshift of $z$\,$\sim$\,0.5
and median stellar mass
log($M_{\star}$/M$_{\odot}$)\,=\,9.1\,$\pm$\,0.1. Stellar masses were
derived from $M_{H}$ magnitudes, using the same method as for KROSS,
and the star formation rates calculated using dust-corrected [O{\sc
    ii}] fluxes. Fig.\,\ref{fig:sami_muse} shows that since the
selection is based only on [O{\sc ii}] flux, galaxies are scattered
within the $M_{\star}$\,--\,SFR plane and it is more difficult than
for SAMI and KROSS to identify a ``main sequence'', however star
formation rates are generally between those of the $z$\,$\sim$\,0 and
$z$\,$\sim$\,0.9 samples.

\begin{figure*}
\includegraphics[width=\textwidth]{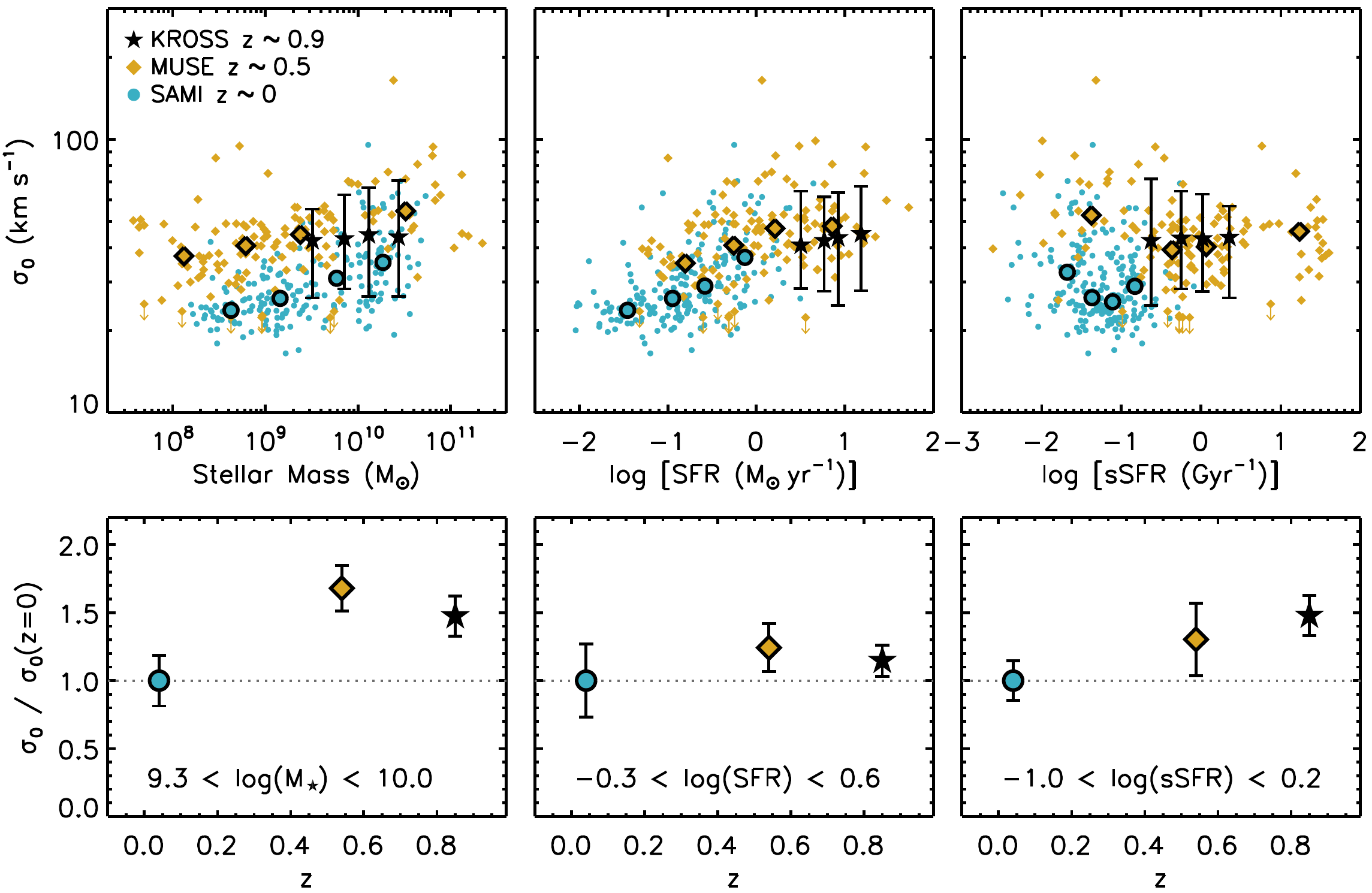}
\caption{\textit{Top:} Trends between velocity dispersion ($\sigma_0$)
  and a selection of non-kinematic properties, for KROSS galaxies
  (this study) and the two comparison samples outlined in
  \S\ref{sami_muse}. For the SAMI and MUSE samples we plot properties
  of individual galaxies and overlay medians within a series of x-axis
  bins (each containing 25\% of the sample). For clarity, for KROSS
  galaxies we show only the median values, with error bars to
  represent the 1$\sigma$ scatter. \textit{Top Left:} Velocity
  dispersion versus stellar mass. At any given redshift there is no
  strong correlation between dispersion and stellar mass, however
  higher redshift galaxies appear to have larger
  dispersions. \textit{Top Middle:} Velocity dispersion versus star
  formation rate. While there is little overlap in star formation rate
  between the three samples, we observe a weak trend of increasing
  dispersion with increasing star formation rate. \textit{Top Right:}
  Velocity dispersion versus specific star formation rate. For
  individual samples we see no significant trend between dispersion
  and specific star formation rate, but again there appears to be an
  increase with redshift. \textit{Bottom:} Velocity dispersion versus
  redshift, relative to the SAMI sample. We calculate the median
  dispersion of each sample over the same range in (left to right)
  M$_\star$, SFR or sSFR, and plot these values as a function of
  redshift.  In these plots, the error-bars denote the
    bootstrap error on the median. For fixed stellar mass or fixed
  sSFR we see a weak trend of increasing dispersion with redshift. For
  fixed SFR the values are consistent within the uncertainties.}
\label{fig:sami_muse_6panels}
\end{figure*}

\citet{swinbank2017} extract rotation velocities at radii of 3$R_{\rm
  d}$ and we apply the beam smearing corrections derived in
\S\ref{beam_smearing} to these values. Velocity dispersions are
calculated by first applying a pixel-by-pixel $\Delta v$/$\Delta R$
correction to the map (i.e. subtracting the average shear across the
pixel in quadrature), and then finding the median of all pixels
outside of the seeing PSF. This beam smearing method is very similar
to that for KROSS and so no additional corrections are applied in our
comparison.

\subsubsection{Dispersion properties}

In Fig.\,\ref{fig:sami_muse_6panels} we plots the relationships
between velocity dispersion and stellar mass, star formation rate and
specific star formation rate.  In the upper panels, the
  error-bars show the 1$\sigma$ scatter in the distributions.  In the
  lower panels, the errors show the uncertainty in the median values,
  measured from a bootstrap resample (with replacement) of the
  values.  At a given redshift, there appears to be at most only a
weak trend between stellar mass and gas velocity dispersion.  This is
consistent with the results of other high redshift kinematic studies
(e.g. \citealt{kassin2012,wisnioski2015,simons2017,turner2017}). We
observe a larger trend of increasing dispersion with stellar mass for
the SAMI sample than for KROSS (where any change is not significantly
detected) and MUSE, however this is still only a
12\,$\pm$\,5\,km\,s$^{-1}$ change associated with a factor $\sim$\,100
increase in stellar mass. Such a small trend can not be
  ruled out for the KROSS sample. What is more apparent is an
increase in $\sigma_0$ with redshift. In the lower left panel of
Fig.\,\ref{fig:sami_muse_6panels} we show that for a fixed stellar
mass the average velocity dispersions of KROSS and MUSE galaxies are
$\sim$50\,\% higher than for the SAMI sample at $z$\,$\sim$\,0 (see
also \citealt{zhou2017}). As we show below, the apparent
  decrease in $\sigma_{\rm 0}$ from $z\sim$\,0.5 to $z\sim$\,0.85 in
  the lower-left panel is not apparent when the samples are matched in
  mass-normalised star-formation rates.

In Fig.\,\ref{fig:sami_muse_6panels} we also investigate how
dispersion is affected by global star formation rate. While there is
little overlap between the three samples, the three samples combined
indicate a weak trend of increasing dispersion with increasing star
formation rate. Although we observe only a 20\,--\,25\,km\,s$^{-1}$
change (a factor of $\sim$\,2 increase) in $\sigma_0$ across three
orders of magnitude in star formation rate, this result is consistent
with a number of previous studies
(e.g. \citealt{lehnert2009,lehnert2013,green2010,green2014,letiran2011,moiseev2015}). Typically
a weak trend is observed below 10\,M$_{\odot}$yr$^{-1}$ and it is only
above this threshold that there is a strong increase in velocity
dispersion with star formation rate. Few KROSS galaxies fit this
criteria. Several authors have interpreted the relationship between
star formation and dispersion as evidence of feedback driven
turbulence, however \citet{krumholz2016} argue that turbulence driven
by disk instability would manifest in a similar way. In \S\ref{theory}
we investigate whether it is possible to distinguish between these two
different scenarios using our dataset.

One way to normalise for differences in star formation rate and mass
between samples is to plot the specific star formation rate (sSFR;
SFR/$M_{\star}$). In the top right panel of
Fig.\,\ref{fig:sami_muse_6panels} we plot velocity dispersion against
sSFR, and find that for all three samples $\sigma_0$ is remarkably
constant. There is a variation of less than 5\,km\,s$^{-1}$ across an
order of magnitude in sSFR for KROSS and SAMI, and of less than
10\,km\,s$^{-1}$ across three orders of magnitude for the MUSE
sample. In the panel below this we study the relationship between
velocity dispersion and redshift, calculating the median of each
sample for a fixed range in sSFR. It is difficult to make a robust
comparison since the SAMI galaxies tend to have a much lower specific
star formation rate, however there appears to be a systematic increase
in dispersion with redshift. We see an increase of $\sim$\,50\%
between $z$\,$\sim$\,0 and $z$\,$\sim$\,0.9.

\begin{figure*}%
  \centering
  \subfloat{{\includegraphics[width=0.478\textwidth]{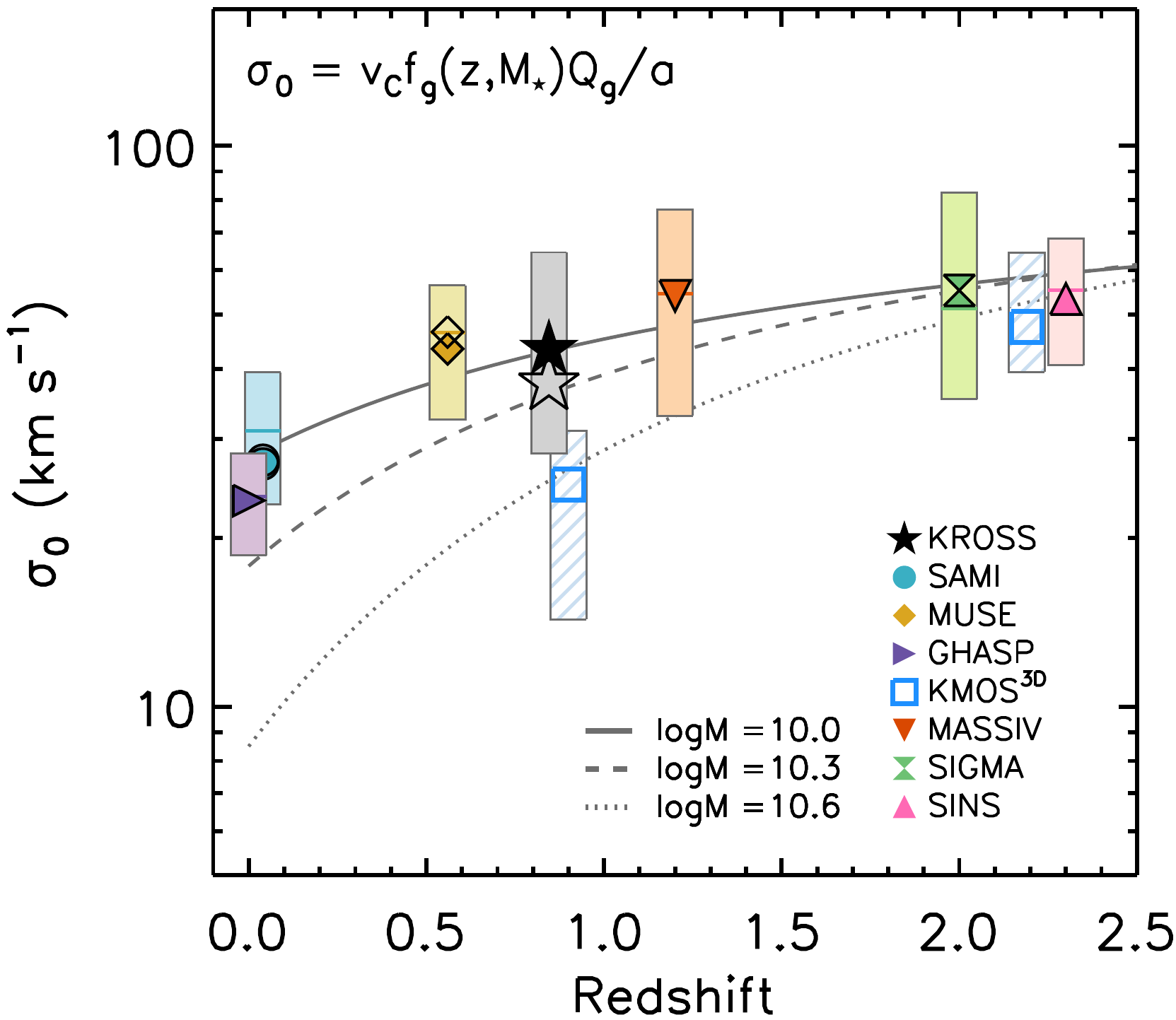} }}%
  \qquad
  \subfloat{{\includegraphics[width=0.463\textwidth]{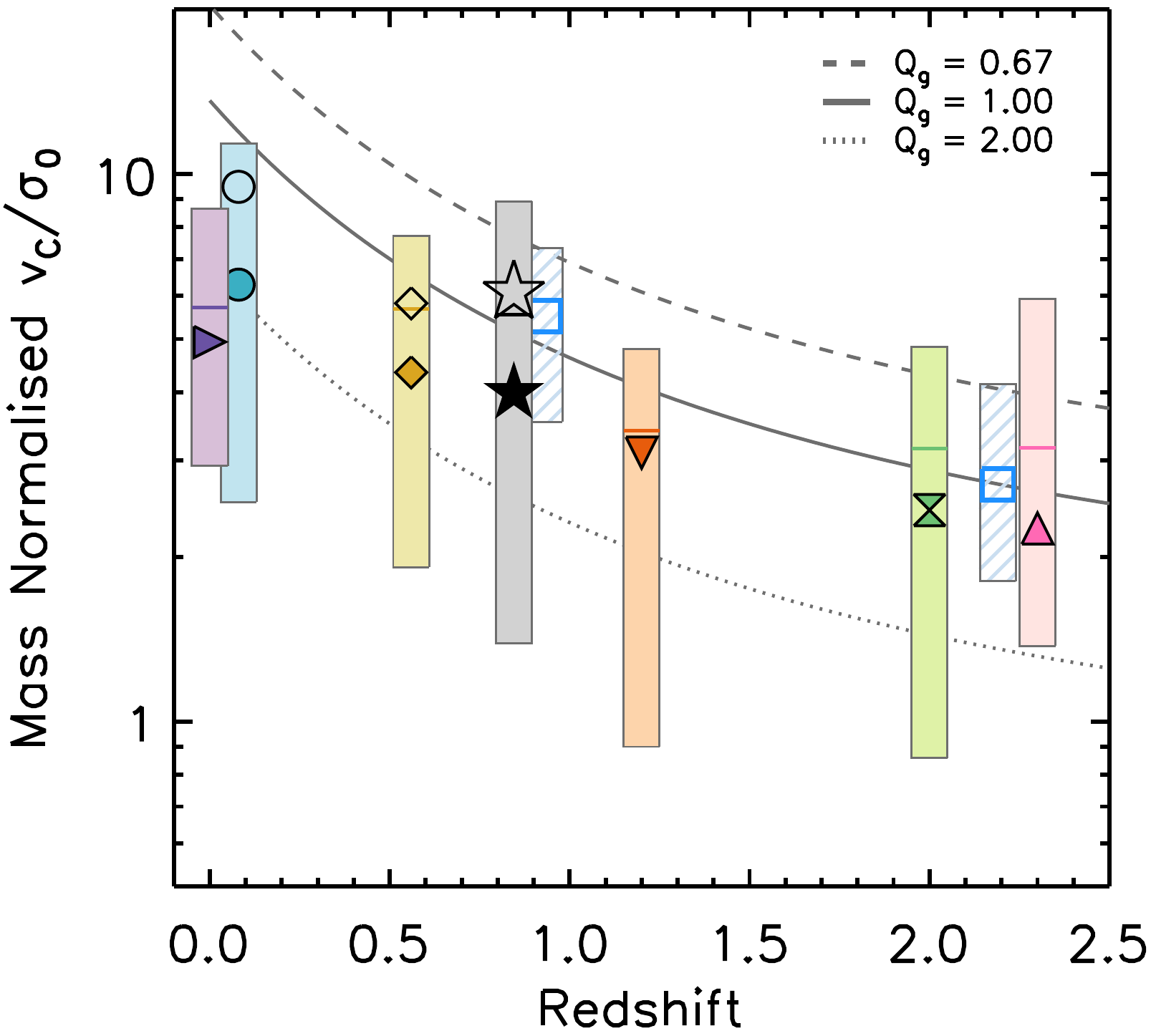} }}
  \caption{Velocity dispersion and (mass normalised) ratio between
    rotational velocity and velocity dispersion as a function of
    redshift. Alongside our results for KROSS we include the SAMI and
    MUSE samples described in \S\ref{sami_muse} and five samples from
    the literature, chosen such that our measurements and beam
    smearing corrections are comparable. Filled symbols represent the
    median, horizontal lines the mean, and vertical bars the 16-84th
    percentile range. Symbols for KMOS$^{\rm 3D}$ represent the median
    of ``rotationally dominated'' galaxies only, and the shaded bars
    represent the central 50\% of the distribution. We plot open
    symbols for the KROSS, SAMI and MUSE samples for comparison,
    showing the median of galaxies with $v_{\rm C}$/$\sigma_0$\,>\,2.
    \textit{Left:} Intrinsic velocity dispersion as a function of
    redshift, with a simple Toomre disk instability model
    (Eq.\,\ref{eqn:fgas}--\ref{eqn:toomre}) plotted for
    log(M$_{\star}$)\,=\,10.0\,--\,10.6. The model appears to provide
    a good description of the data. \textit{Right:} $v_{\rm
      C}$/$\sigma_0$ as a function of redshift, with a simple disk
    model overlaid for $Q_{\rm g}$\,=\,0.67\,--\,2. Values have been
    normalised to a stellar mass of log(M$_{\star}$)\,=\,10.5. The
    data is broadly consistent with the model, and we observe a
    decrease in $v_{\rm C}$/$\sigma_0$ with redshift. For KMOS$^{\rm
      3D}$ data was only available for ``rotationally dominated''
    galaxies. If we consider the same subsample of KROSS our results
    are similar.}
  \label{fig:redshift_trends}%
\end{figure*}

\subsection{Dynamics in the context of galaxy evolution}
\label{trends_with_redshift}

Kinematic studies at high-redshift suggest that star-forming galaxies
at early times were dynamically ``hot'', with velocity dispersions
much larger than those observed for disks in the local Universe. In
this section we examine how the KROSS galaxies fit within a wider
evolutionary context, comparing their dynamics to those of the SAMI
and MUSE samples discussed in \S\ref{sami_muse} and five additional
comparison samples between 0\,<\,$z$\,<\,2.5. For this comparison we
include data from the GHASP (\citealt{epinat2010}; log($M_{\star}^{\rm
  avg}$/M$_{\odot}$)\,=\,10.6), KMOS$^{\rm 3D}$
(\citealt{wisnioski2015}; log($M_{\star}^{\rm
  avg}$/M$_{\odot}$)\,=\,10.7 and 10.9 for the $z$\,$\sim$\,1 and 2
samples respectively), MASSIV (\citealt{epinat2012};
log($M_{\star}^{\rm avg}$/M$_{\odot}$)\,=\,10.5), SIGMA
(\citealt{simons2016}; log($M_{\star}^{\rm
  avg}$/M$_{\odot}$)\,=\,10.0) and SINS
(\citealt{cresci2009,newman2013}; log($M_{\star}^{\rm
  avg}$/M$_{\odot}$)\,=\,10.6) surveys. These are all large samples
($\gtrsim$\,50 galaxies) of ``typical'' star-forming galaxies, with
star formation rates representative of the main sequence at a
particular redshift. Beam smearing of the intrinsic velocity
dispersion has been accounted for in each sample, either through disk
modelling or post-measurement corrections. With the exception of GHASP
(Fabry-P$\'e$rot) and SIGMA (MOSFIRE), these are IFU-based studies.

In calculating average dispersion and $v_{\rm C}$/$\sigma_0$ values,
we note that different authors adopt different approaches. For example
\citet{wisnioski2015} consider only ``disky'' galaxies within the
KMOS$^{\rm 3D}$ sample, selected based on five criteria including
$v_{\rm C}$/$\sigma_0$\,>\,1, a smooth gradient within the velocity
map (``spider diagram''; \citealt{vanderkruit1978}), and a dispersion
which peaks at the position of the steepest velocity gradient. However
it is difficult to isolate a similar subset for each of the samples
discussed here. For example, \citet{epinat2010} have shown that up to
30\% of rotators may be misclassified if a velocity dispersion central
peak is required. Low spatial resolution may also lead kinematically
irregular galaxies to be misidentified as rotators
(e.g. \citealt{leethochawalit2016}).

In the left panel of Fig.\,\ref{fig:redshift_trends} we plot the
median, mean and distribution of velocity dispersion measurements for
each of the eight samples, as a function of redshift. As has been
noted before, there is a gradual increase in the average velocity
dispersion from $\sim$\,25\,km\,s$^{-1}$ at $z$\,=\,0 to
$\sim$\,50\,km\,s$^{-1}$ at $z$\,=\,2. At $z$\,$\sim$\,1
\citet{wisnioski2015} report an average of
$\sigma_0$\,=\,25\,$\pm$\,5\,km\,s$^{-1}$ for the KMOS$^{\rm 3D}$
sample, whereas for KROSS we measure a median of
$\sigma_0$\,=\,43.2\,$\pm$\,0.8\,km\,s$^{-1}$.  We attribute this
difference to the samples used to calculate the median (also see
\citealt{DiTeodoro16}), with stricter criteria applied to isolate
'disky' galaxies.  Applying a series of criteria that directly or
indirectly isolate 'disky' galaxies has the effect of removing the
higher velocity dispersion sources and selecting the most dynamically
mature galaxies, with such samples being increasingly less
representative of the overall star-forming population with increasing
redshift (e.g.\ see \citealt{turner2018} for a discussion).  We
restrict the KROSS, SAMI and MUSE samples to ``rotationally
dominated'' galaxies, towards being more consistent with their
sample, and plot the medians as open symbols. For KROSS we find a
reduced median of $\sigma_0$\,=\,36\,$\pm$\,2\,km\,s$^{-1}$, which is
in better agreement.

There has been much discussion as to which physical processes drive
the observed evolution of velocity dispersion with redshift. We
explore the theoretical arguments in \S\ref{theory}. However in this
subsection we follow the analysis of \citet{wisnioski2015},
interpreting the results of Fig.\,\ref{fig:redshift_trends} in the
context of a rotating disk with a gas fraction and specific star
formation rate that evolve as a function of redshift. In this simple
model the gas fraction of the disk is defined as \citet{tacconi2013}:

\begin{equation}
\label{eqn:fgas}
f_{\rm gas}\,=\,\frac{1}{1+(t_{\rm dep}{\rm sSFR})^{-1}},
\end{equation}
where the depletion time evolves as $t_{\rm dep}({\rm
  Gyr})\,=\,1.5\,\times\,(1+z)^\alpha$. From molecular gas
observations of $z$\,=\,1\,--\,3 galaxies, \citet{tacconi2013} measure
$\alpha$\,=\,$-$0.7 to $-$1.0, however the analytic models of
\citet{dave2012} predict $\alpha$\,=\,$-$1.5. Here $\alpha$\,=\,$-$1.0
is adopted as a compromise. The cosmic specific star formation rate is
assumed to follow the evolution described in \citet{whitaker2014},
where

\begin{equation}
\label{eqn:ssfr}
{\rm sSFR}(M_{\star},z)\,=\,10^{A(M_{\star})}(1+z)^{B(M_{\star})}.
\end{equation}
This sSFR relation was derived to fit UV$+$IR star formation rates of
$\sim$\,39,000 galaxies in the redshift range 0.5\,<\,$z$\,<\,2.5
(3D-\textit{HST} survey; \citealt{momcheva2016}).
Finally, the Toomre disk stability criterion for a gas disk ($Q_{\rm
  g}$; \citealt{toomre1964}) can be rewritten in terms of $f_{\rm
  gas}$ (see \citealt{glazebrook2013}) as

\begin{equation}
\label{eqn:toomre}
\frac{v_{\rm C}}{\sigma_0}\,=\,\frac{a}{f_{\rm gas}(z,M_{\star})Q_{\rm g}}
\end{equation}
where $a=\sqrt{2}$ for a disk of constant rotational velocity. In the
left panel of Fig.\,\ref{fig:redshift_trends} we overplot the
relationship between velocity dispersion and redshift derived for a
range of stellar masses. Following the approach of
\citealt{turner2017}, the value of $Q_{\rm g}v_{\rm C}$ is set such
that the log($M_{\star}$/M$_{\odot}$)\,=\,10.0 track is normalised to
fit the median dispersion of the KROSS sample. For a marginally stable
thin gas disk ($Q_{\rm g}$\,=\,1), this requires a model rotation
velocity of $v_{\rm C}$\,=\,150\,km\,s$^{-1}$. By comparison, we
measure a median velocity for the KROSS sample of $v_{\rm
  C}$\,=118\,$\pm$\,4\,km\,s$^{-1}$. This simple model appears to
provide a reasonable description of the data.

While these tracks provide useful guidance, we would typically expect
the average rotation velocity, and not just the gas fraction, to vary
as a function of mass. To eliminate this dependency, in the right
panel of Fig.\,\ref{fig:redshift_trends} we show how $v_{\rm
  C}$/$\sigma_0$ is expected to evolve for a galaxy of stellar mass
log($M_{\star}$/M$_{\odot}$)\,=\,10.5 and Toomre parameter of $Q_{\rm
  g}$\,=\,0.67, 1 or 2. These are the critical values for a thick gas
disk, thin gas disk and stellar-plus-gas disk, respectively (see
e.g. \citealt{kim2007}). We then plot the median, mean and
distribution of $v_{\rm C}$/$\sigma_0$ values for the eight samples,
normalising each to the stellar mass of the models. These correction
factors range between 0.7\,--\,3.0 with a median of 1.1\,$\pm$\,0.2,
with the largest applied to the SAMI sample (median
log($M_{\star}$/M$_{\odot}$)\,=\,9.3).

The data follow the general trend described by the model, with a
decrease from $v_{\rm C}$/$\sigma_0$\,$\sim$\,6 at $z$\,=\,0 to
$v_{\rm C}$/$\sigma_0$\,$\sim$\,2 at $z$\,=\,2. This general trend is
consistent with the results of \citet{turner2017}\footnote{We note
  that for the GHASP sample, \citet{turner2017} use the results from
  \citet{epinat2008}, whereas we use the results presented in
  \citet{epinat2010}.}. The model predicts that this is due to higher
gas fractions in galaxies at early times. Again we highlight the
effect of restricting the KROSS, SAMI and MUSE samples to
``rotationally dominated'' galaxies, with open symbols. For KROSS, the
median \textit{mass-weighted} $v_{\rm C}$/$\sigma_0$ increases from
3.9\,$\pm$\,0.2 to 6.1\,$\pm$\,0.2. This result is consistent with
KMOS$^{\rm 3D}$ at $z$\,$\sim$\,1, who find $v_{\rm
  C}$/$\sigma_0$\,=\,5.5 \citep{wisnioski2015}.

\begin{figure}
\includegraphics[width=0.45\textwidth]{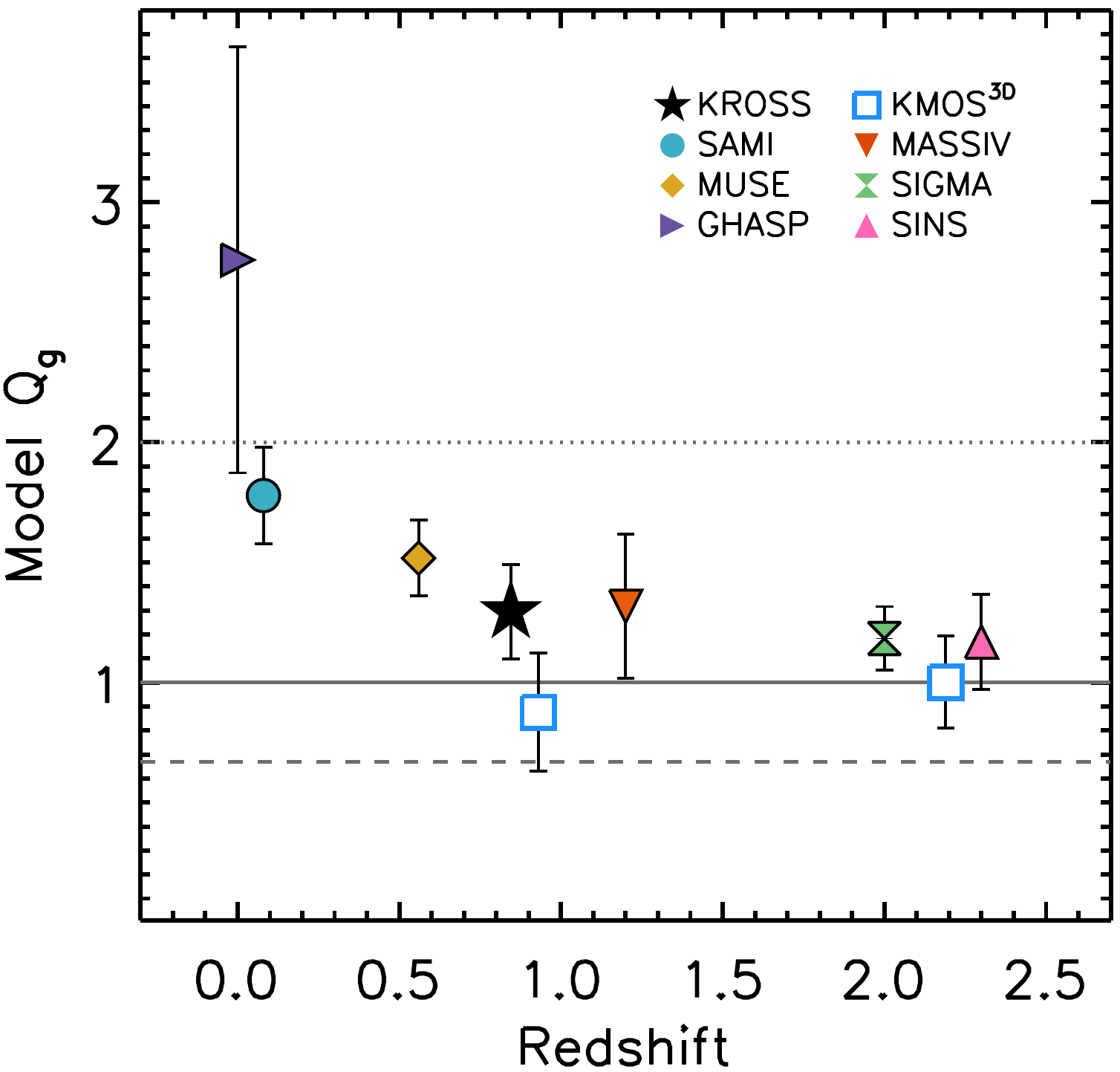}
\caption{Inferred Toomre $Q$ versus redshift for KROSS and comparison
  samples. Assuming a simple disk instability model (see
  \S\ref{trends_with_redshift} and
  Eq.\,\ref{eqn:fgas}--\ref{eqn:toomre}), we calculate the Toomre $Q$
  parameter required to fit the average $v_{\rm C}$/$\sigma_0$
  observed for KROSS and various comparison samples, given their
  respective redshifts and stellar masses. Error bars reflect typical
  uncertainties associated with measurements of the stellar mass and
  dynamics. The lines overplotted at $Q_{\rm g}$\,=\,0.67, 1 and 2
  represent the critical values for a thick gas disk, thin gas disk
  and stellar-plus-gas disk, respectively. We find that higher
  redshift samples are best fit by lower values of $Q$, which would
  suggest that these galaxies are more unstable disks.}
\label{fig:Qmodel}
\end{figure}

The right panel of Fig.\,\ref{fig:redshift_trends} appears to suggest
a weak trend between Toomre $Q_{\rm g}$ and redshift. We caution that
$Q_{\rm g}$ is a galaxy averaged value, sensitive to systematics, and
is therefore only a crude measure of disk stability. However to
explore this potential trend, in Fig.\,\ref{fig:Qmodel} we plot the
``best fit'' Toomre parameter required to fit the observed median
$v_{\rm C}$/$\sigma_0$ for each of the samples, given their respective
redshifts and stellar masses. To calculate error bars we propagate the
typical uncertainties associated with measurements of the dynamics and
stellar mass. Within the framework of this model, we find that lower
redshift samples are associated with higher average values of $Q_{\rm
  g}$.

This increase in $Q_{\rm g}$ is consistent with recent numerical
simulations \citep{danovich2015,lagos2016} and observational studies
\citep{obreschkow2015,burkert2016,harrison2017,swinbank2017} which
suggest that the specific angular momentum of galaxy disks ($j_{\rm
  s}$) increases with time. An increase in angular momentum would act
to increase the global $Q$ parameter, stabilising disks against
fragmentation. \citet{obreschkow2015} and \citet{swinbank2017} suggest
this is likely what drives the morphological transition between
clumpy, irregular disks at high redshift, and the bulge-dominated
galaxies with thin spiral disks we see today. \citet{obreschkow2015}
propose that $Q\,\propto\,(1-f_{\rm g})j_{\rm s}\sigma_0$. Hence if
the gas fraction decreases by a factor of four between $z$\,$\sim$\,2
and $z$\,$\sim$\,0 (e.g. \citealt{tacconi2010,saintonge2013,
  genzel2015}), and the dispersion by a factor of two
(Fig.\,\ref{fig:redshift_trends}), then a factor of $\sim$\,2.6
increase in specific angular momentum would achieve the increase in
disk stability suggested by Fig.\,\ref{fig:Qmodel}.

The ``toy model'' described in this section is a useful tool, allowing
us to interpret the evolution of galaxy dynamics in terms of gas
fraction and disk instability. However it provides little information
about the physical mechanisms involved. For a deeper understanding we
must combine our observations with theoretical predictions.

\section{The origin of disk turbulence: Star formation feedback versus gravitational instability}
\label{theory}

Although the simple framework in \S\ref{trends_with_redshift} provides
an adequate description of the data, other -- more detailed --
physical models have been proposed to explain the origin of these high
turbulent motions. Turbulence in the interstellar medium (ISM) decays
strongly within the disk crossing time ($\sim$\,15\,Myr;
\citealt{maclow1998, maclow1999, stone1998}), so a source of energy is
clearly required to maintain disorder in the system. What this might
be is the subject of active discussion, however a large and
well-selected sample such as KROSS may be able to provide useful
constraints. In this section we consider whether our data can be used
to distinguish between two potential disk turbulence mechanisms.

One model is that the high level of turbulence is driven by stellar
feedback. Supernovae and winds inject energy into the ISM, and several
authors have identified a correlation between velocity dispersion and
star formation rate, either on global or spatially resolved scales
(e.g. \citealt{lehnert2009,lehnert2013,green2010,green2014,letiran2011};
but see \citealt{genzel2011}). However simulations including only
stellar feedback struggle to reproduce these large observed
dispersions \citep{joung2009,shetty2012,kim2013,kim2014} without high
rates of momentum injection (e.g. \citealt{hopkins2011,hopkins2014}).

An alternative framework is a clumpy, gas-rich disk fed by rapid
accretion from the intergalactic medium (IGM). While accretion of
material \textit{onto} the disk appears in itself insufficient to
drive large velocity dispersions
(e.g. \citealt{elmegreen2010,hopkins2013}, though see
\citealt{klessen2010}), simulations suggest that gravitational
instabilities may induce high levels of turbulence
(e.g. \citealt{aumer2010,bournaud2010,bournaud2014,ceverino2010,goldbaum2015,goldbaum2016})
which can be sustained by the accretion of mass \textit{through} the
disk. As mass is transported inwards, the dispersion, and hence $Q$,
is increased. Decay of this turbulence then acts to reduce $Q$, and
eventually the disk saturates at a state of marginal
stability. Several authors have considered whether gravitational
interactions between clumps (formed via fragmentation of the disk) may
also help to generate turbulence \citep{dekel2009,aumer2010}.

\citet{krumholz2016} noted that while the origins of disk turbulence
have been explored in detail from a theoretical point of view, there
had previously been few direct observational tests. To address this,
the authors formulated two simple models -- describing gravity-driven
turbulence and feedback-driven turbulence -- which could be used to
make observational predictions. We outline each of these below.

\subsection{Gravity-driven model}
\label{gravity_driven}

For a model in which turbulence is driven by gravitational
instabilities in the gas, \citet{krumholz2016} adopt expressions for
gas surface density ($\Sigma$) and velocity dispersion ($\sigma$)
derived for the ``steady state configuration'' described in
\citet{krumholz2010}. Within this framework the gas surface density
depends on the total Toomre $Q$ parameter (as opposed to that of the
gas or stars alone; i.e. $Q_{\rm g}$ or $Q_{\star}$), since the
turbulence is driven by a global instability of the disk. The
\citet{wang1994} approximation is adopted such that
$Q^{-1}$\,=\,$Q^{-1}_{\rm g}$\,+\,$Q^{-1}_{\star}$ and

\begin{equation}
\label{eqn:toomre2}
Q \approx \frac{av_{\rm C}\sigma f_{\rm g}}{\pi Gr\Sigma},
\end{equation}
with $a$\,=\,$\sqrt{2}$. Here $v_{\rm C}$ is the rotational velocity
measured at a radius of $r$, $\sigma$ is the velocity dispersion and
$\Sigma$ is the gas surface density. It is expected that the disk
self-regulates at $Q$\,$\approx$\,1. Star formation is then added to
the model assuming a so-called ``Toomre regime'' \citep{krumholz2012},
in which the ``entire ISM is a single star-forming structure''. This
is a key distinction between this model and the feedback-driven model
discussed below. Together, these assumptions lead to a star formation
rate which depends on the velocity dispersion as

\begin{equation}
\label{eqn:gravity_driven}
SFR = \frac{16}{\pi}\sqrt{\frac{\phi_{\rm P}}{3}}\left(\frac{\epsilon_{\rm ff}v_{\rm C}^2}{G}{\rm ln}\frac{r_1}{r_0}\right)f_{\rm g}^2\sigma,
\end{equation}
where $\epsilon_{\rm ff}$ is the star formation rate per freefall
time, $f_{\rm g}$ the gas fraction, $\phi_{\rm P}$ a constant to
account for the presence of stars, and ln($r_1$/$r_0$) relates to the
radial extent of the disk.

\subsection{Feedback-driven model}
\label{feedback_driven}

One way for analytic models to achieve large velocity dispersions via
stellar feedback, is to assume that the star formation efficiency
within giant molecular clouds (GMCs) is closely coupled to the Toomre
parameter of the gas disk ($Q_{\rm g}$). Activity on the scale of GMCs
is driven by self-gravity of the gas clouds and hence feedback-driven
models do not require a global $Q$\,$\approx$\,1 provided $Q_{\rm
  g}$\,$\approx$\,1. The expression for the gas Toomre parameter is
similar to Eq.\,\ref{eqn:toomre2},

\begin{equation}
\label{eqn:toomre3}
Q_{\rm g} \approx \frac{av_{\rm C}\sigma}{\pi Gr\Sigma}.
\end{equation}
In their model \citet{krumholz2016} adopt the star-forming relation of
\citet{faucher2013}, which balances the momentum per unit mass
($p_{\star}$/$m_{\star}$) injected by feedback against the gas surface
density squared. This results in a relationship between star formation
rate and velocity dispersion of

\begin{equation}
\label{eqn:feedback_driven}
SFR = \frac{8\sqrt{2} \phi v_{\rm C}^2}{\pi GQ_{\rm
    g}\mathcal{F}}\left({\rm
  ln}\frac{r_1}{r_0}\right)\left(\frac{p_{\star}}{m_{\star}}\right)^{-1}\sigma^2,
\end{equation}
where $\phi$ and $\mathcal{F}$ are constants associated with various
model uncertainties. There are two key differences between this and
the gravity-driven model. Firstly, since stellar feedback depends on
the amount of gas unstable to gravitational collapse, we assume
$Q_{\rm g}$\,$\approx$\,1 and not $Q$\,$\approx$\,1. As a consequence,
Eq.\,\ref{eqn:feedback_driven} does not depend on $f_{\rm
  g}$. Secondly, the star formation rate is more strongly dependent on
the velocity dispersion than for a gravity-driven model. For
turbulence to balance gravity in the ISM the star formation rate
density must be proportional to the gas surface density squared. Since
$\Sigma$\,$\propto$\,$\sigma$ for constant $Q_{\rm g}$, we therefore
obtain SFR\,$\propto$\,$\sigma^2$, as opposed to
SFR\,$\propto$\,$\sigma$ for the gravity-driven model
(Eq.\ref{eqn:gravity_driven}).

\subsection{Comparison of models to observations}

KROSS offers a large and representative sample of $\sim$\,500
star-forming galaxies, with velocity dispersions measured and
corrected for beam smearing in a consistent way.  This is an ideal
opportunity to test predictions of the aforementioned analytic
models. \citet{krumholz2016} compared observational data to their
models of feedback-driven and gravity-driven turbulence. 
  As shown by \citet{krumholz2016}, most of the diagnostic power in
  differentiating between the gravity-driven versus feedback-driven
  turbulence models is from galaxies with the highest star formation
  rate (above star formation rates of
  $\gsim$\,50\,M$_\odot$\,yr$^{-1}$ the predictions for the amount of
  turbulence from the two models most rapidly diverge). However,
while this data covers many orders of magnitude in star formation
rate, it consists of samples of differing selection criteria, redshift
and data quality.  In this section, we attempt to use our
  sample to test these predictions, although we note that the majority
  of our sample have star formation rates in the range
  2--30\,M$_\odot$\,yr$^{-1}$.

\subsubsection{Model tracks}
\begin{figure*}%
  \centering
\includegraphics[width=0.9\textwidth]{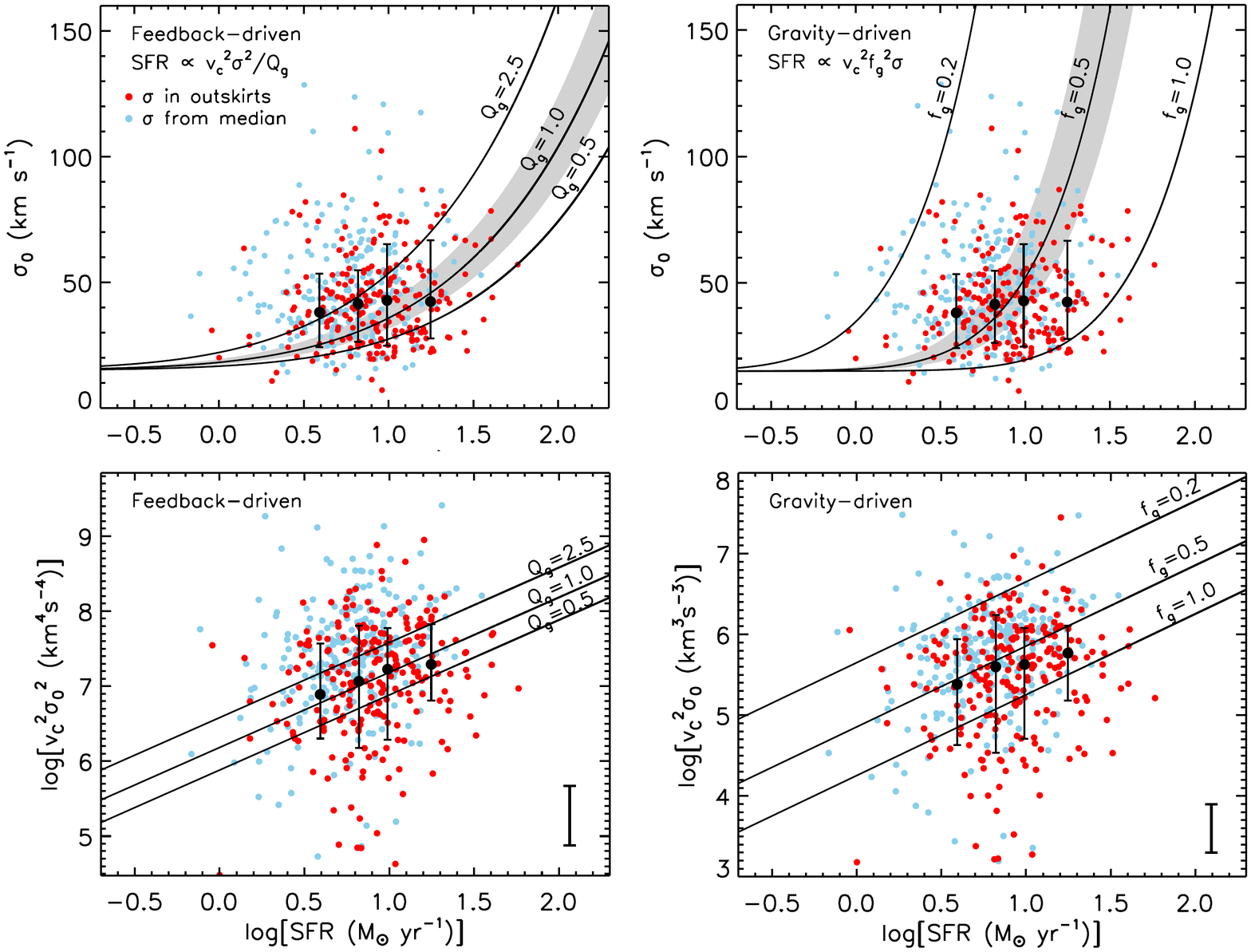}
  \caption{Properties of KROSS galaxies compared to predictions of the
    analytic models discussed in \S\ref{theory}. Points are coloured
    by the technique used to measure the velocity dispersion,
    $\sigma_0$. Large black symbols show the median dispersion (and
    standard deviation) in bins of star formation rate. \textit{Top
      Left:} Predictions of a model in which turbulence is driven by
    star formation feedback (see \S\ref{feedback_driven},
    Eq.\,\ref{eqn:feedback_driven}) assuming the median rotation
    velocity of the sample ($v_{\rm C}$\,$\approx$\,120\,km\,s$^{-1}$)
    and a gas Toomre parameter of $Q_{\rm g}$\,=\,0.5, 1.0, 2.0. The
    shaded region shows the impact of increasing/decreasing the
    rotation velocity by 20\,km\,s$^{-1}$; the 68th percentile range
    of our data is $v_{\rm C}$\,=\,44\,--\,204\,km\,s$^{-1}$ so we
    would expect a large amount of scatter even if only one value of
    $Q_{\rm g}$ was valid.  \textit{Top Right:} Predictions of a model
    in which turbulence is driven by gravitational instabilities (see
    \S\ref{gravity_driven}, Eq.\,\ref{eqn:gravity_driven}), assuming
    the median $v_{\rm C}$ and gas fractions of $f_{\rm g}$\,=\,0.25,
    0.5, 1.0. This model results in a much steeper increase in
    $\sigma_0$ as a function of star formation rate. We measure
    $\sigma_0$\,$\geq$\,100\,km\,s$^{-1}$ for only a handful of KROSS
    galaxies, and do not observe a strong trend with star formation
    rate. However this model could still be valid if the galaxies have
    a wide range of rotation velocities and/or gas
    fractions. \textit{Bottom:} To eliminate dependency on the
    rotation velocity, we also plot log($v_{\rm C}^2\sigma_0^2$) and
    log($v_{\rm C}\sigma_0^2$) as a function of star formation rate
    for the feedback- and gravity-driven model, respectively
    (in both panels, we plot a representative error-bar
      to highlight the uncertainty on individual measurements). Both
    models provide an adequate description of the data, however there
    is a large amount of residual scatter. This could be due to
    measurement uncertainties, an intrinsic variation of $Q_{\rm g}$
    and $f_{\rm g}$, or (most likely) a combination of these two
    factors.}
  \label{fig:model_tracks}
\end{figure*}

In Fig.\,\ref{fig:model_tracks} we plot velocity dispersion against
star formation rate for KROSS and overlay the models of
\citet{krumholz2016}\footnote{We adopt the same fiducial values of
  $\epsilon_{\rm ff}$\,=\,0.01, $\phi_{\rm P}$\,=\,3, $r_1$\,=\,10,
  $r_0$\,=\,0.1, $\phi$\,=\,1, $\mathcal{F}$\,=\,2 and
  $p_{\star}$/$m_{\star}$\,=\,3000}. In the top left panel we plot
trends for a feedback-driven model, adopting the median rotation
velocity of the sample ($\sim$\,120\,km\,s$^{-1}$) and Toomre $Q_{\rm
  g}$\,=\,0.5, 1.0 and 2.5. These tracks show only a moderate increase
in velocity dispersion with star formation rate, which is consistent
with our data. For KROSS, galaxies in the lower quartile of star
formation rate have a median dispersion of 42\,$\pm$\,2\,km\,s$^{-1}$
and those in the upper quartile a median of
45\,$\pm$\,2\,km\,s$^{-1}$. The dispersion predicted by the model is
much more sensitive to rotation velocity than $Q_{\rm g}$. The shaded
region around the $Q_{\rm g}$ \,=\,1.0 track shows the effect of
adjusting the rotation velocity of the model by 20\,km\,s$^{-1}$, with
larger values of $v_{\rm C}$ corresponding to smaller values of
$\sigma_0$. The 68th percentile range for our sample is
44\,--\,204\,km\,s$^{-1}$, so it is possible that datapoints are
consistent with a narrow range of $Q_{\rm g}$ if this effect dominates
the scatter.

In the top right panel of Fig.\,\ref{fig:model_tracks} we show trends
for a gravity-driven model, with the same rotation velocity and gas
fractions of $f_{\rm g}$\,=\,0.2, 0.5 and 1.0. These models predict a
sharp increase in velocity dispersion with star formation rate,
however this is not something seen in the data -- we measure
$\sigma_0$\,$\geq$\,100\,km\,s$^{-1}$ for only a handful of KROSS
galaxies. The expected velocity dispersion is very sensitive to the
input rotation velocity and gas fraction. Despite our data being
predominately low dispersion, this model may still be valid if the
galaxies have a wide range in these other properties.

To eliminate this dependency on rotation velocity, in the lower panels
of Fig.\,\ref{fig:model_tracks} we plot log($v_{\rm C}^2\sigma_0^2$)
and log($v_{\rm C}\sigma_0^2$) as a function of star formation rate
for the feedback- and gravity-driven model, respectively. We note that
low dispersion galaxies ($\sigma_0$\,$\leq$\,20\,kms$^{-1}$, scattered
below the model trends) tend to be smaller compared to the seeing, and
as such have larger beam smearing corrections (see
\S\ref{beam_smearing}). This sample has a median of $R_{\rm
  d}$/$R_{\rm PSF}$\,=\,0.35\,$\pm$\,0.08, as opposed to $R_{\rm
  d}$/$R_{\rm PSF}$\,=\,0.61\,$\pm$\,0.02 for all KROSS galaxies.  We
caution that in this situation it is more difficult to recover the
intrinsic velocity dispersion. Galaxies that lie \textit{above} the
model trends tend to be those where the dispersion comes from the
median of all available pixels.  As discussed in
\S\ref{measuring_sigma}, these measurements are associated with larger
uncertainties.  Nevertheless, the median properties of the KROSS
sample follow trends similar to those predicted by both models.

Over the star formation rate range measured by our
  observations, each model appears provides an adequate description of the
data.
For example, adopting a single track for models, with
  $Q$\,=\,1.0 in the feedback model and $f_{\rm g}$\,=\,0.5 in the
  gravity driven model, the reduced $\chi^2$ (accounting for
  measurement errors) agree within $\Delta\chi^2$\,=\,1 (the
  individual reduced $\chi^2$ are 3.4 and 4.1 for the gravity driven
  and feedback driven model respectively).  Although some of the
  scatter will be due to measurement errors, there will be intrinsic
  variations within the sample in $Q_{\rm g}$ and $f_{\rm g}$ and we
  will investigate this further in the next section.
  
\subsubsection{Best-fit model Toomre Q and gas fractions}

Directly comparing the observed velocity dispersions to those
predicted by the analytical models is not an efficient test of gravity-driven
versus feedback-driven turbulence. Offsets for the feedback-driven
model tend to be smaller than for the gravity-driven model, since the
latter has a much steeper relationship between star formation rate and
velocity dispersion. An alternative approach is to calculate the
Toomre parameter and gas fraction required for each galaxy to be fit
by the models. These are properties which we can also estimate
directly from the observations, independent of any turbulence model,
with a few simple assumptions.  By comparing these two sets of
parameters, we can test which model provides a better fit to the data.

\begin{figure*}%
  \centering
  \subfloat{{\includegraphics[width=0.47\textwidth]{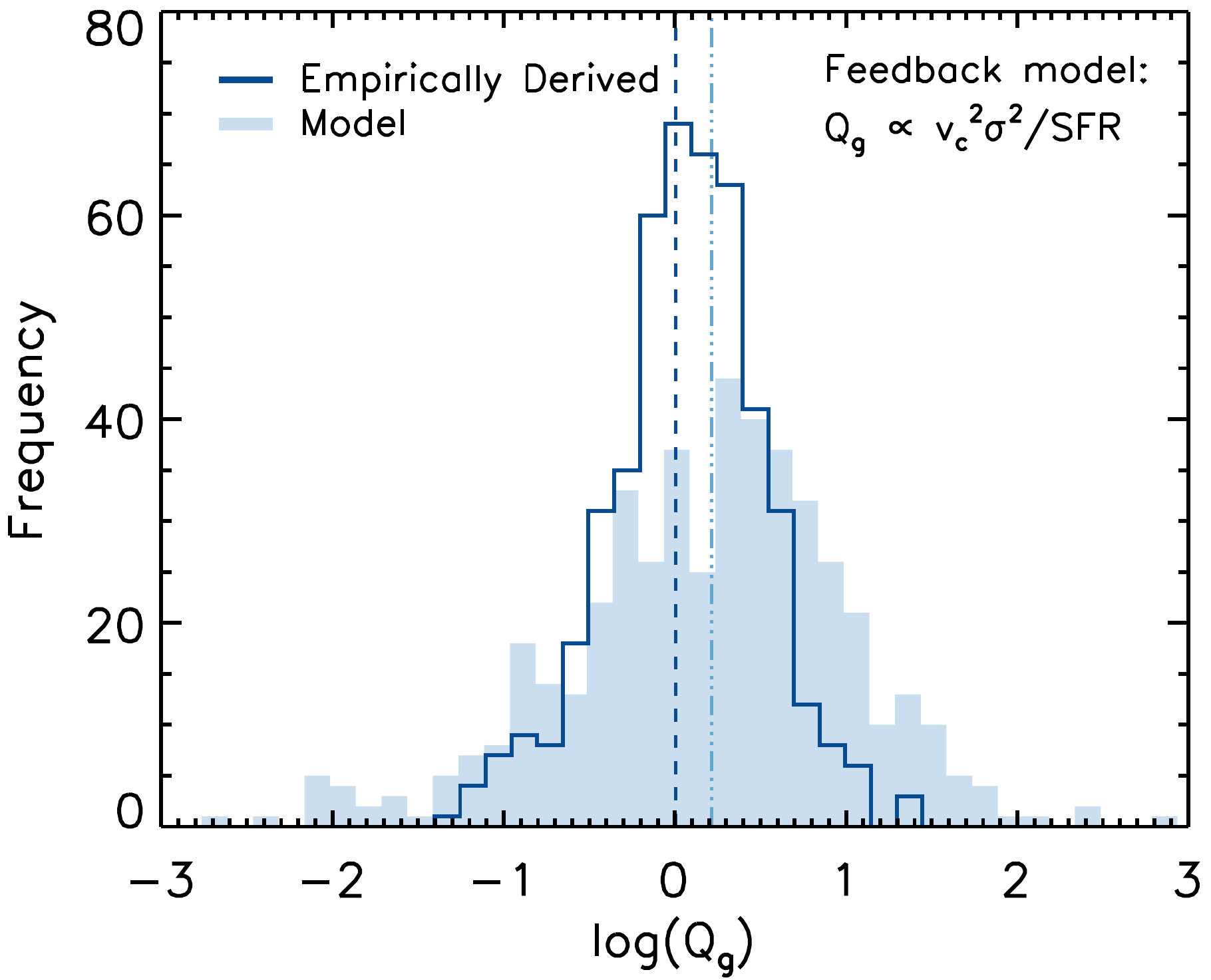} }}%
  \qquad
  \subfloat{{\includegraphics[width=0.484\textwidth]{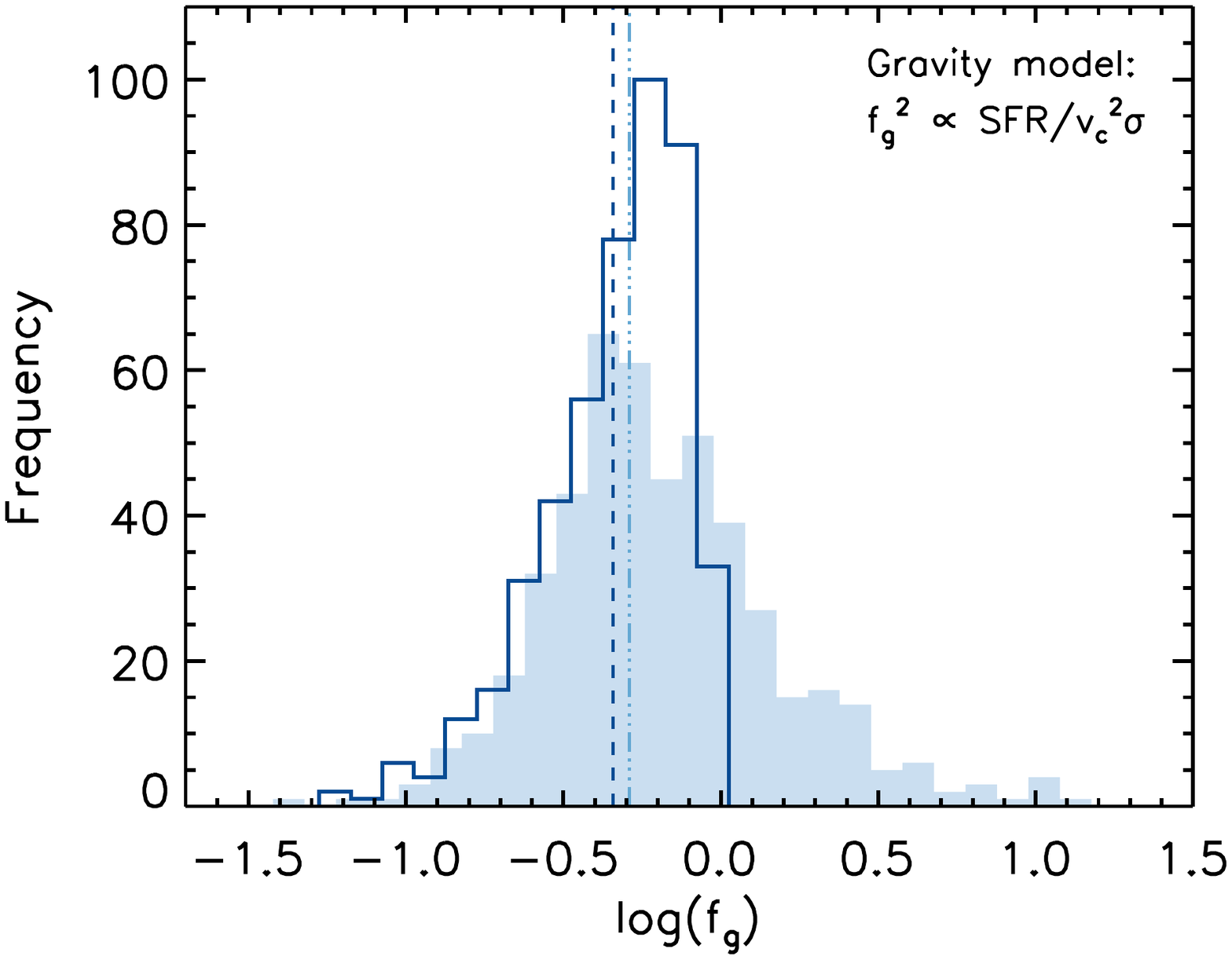} }}
  \caption{Toomre parameters and gas fractions required for the
    turbulence models in \S\ref{theory} to fit KROSS galaxies,
    compared to estimates of these properties made from
    observations. Dashed vertical lines show the median of each
    distribution. \textit{Left:} Model
    (Eq.\,\ref{eqn:feedback_driven}, filled histogram) versus
    empirically derived (outlined histogram) Toomre $Q_{\rm g}$. The
    model distribution is broader and has a slightly larger median --
    $Q_{\rm g,med}$\,=\,1.6$\pm$\,0.2 as opposed to $Q_{\rm
      g,med}$\,=\,1.01$\pm$\,0.06. \textit{Right:} Model
    (Eq.\,\ref{eqn:gravity_driven}, filled histogram) versus
    empirically derived (outlined histogram) gas fractions. Model
    values have a similar median -- $f_{\rm
      g,med}$\,=\,0.52$\pm$\,0.02 versus $f_{\rm
      g,med}$\,=\,0.45$\pm$\,0.01. In both panels the offset between
    distributions is small, and could be accounted for with a minor
    adjustment in model parameters. We are unable to definitively rule
    out either turbulence model. Uncertainties associated with the
    best-fit model parameters are approximately twice as large, which
    may explain why these distributions are broader.}
  \label{fig:model_histograms}
\end{figure*}

In Fig.\,\ref{fig:model_histograms} we compare the distribution of
Toomre $Q_{\rm g}$ values inferred from the feedback-driven turbulence
model (rearranging Eq.\,\ref{eqn:feedback_driven}) to those estimated
using Eq.\,\ref{eqn:toomre3}. To estimate the gas surface density we
calculate the star formation rate surface density and then invert the
Kennicutt-Schmidt relation, $\Sigma_{\rm SFR} = A\,\Sigma_{\rm
  gas}^n$, where
$A$\,=\,1.5\,$\times$\,10$^{-4}$\,M$_{\odot}$yr$^{-1}$pc$^{-2}$ and
$n$\,=\,1.4 (\citealt{kennicutt1998}; for a Chabrier IMF). We note
that an alternative approach would be to estimate $\Sigma_{\rm gas}$
by inverting the multi-freefall star formation relation
\citep{federrath2017b}, however this is not something we explore
here.  We find a median of $Q_{\rm g,med}$\,=\,1.6$\pm$\,0.2 for the
model and $Q_{\rm g,med}$\,=\,1.01$\pm$\,0.06 for the empirically
derived values (close to the $Q_{\rm g}$\,$\sim$\,1 expected for a
marginally unstable disk). The model distribution is noticeably
broader, with a 68th percentile range of 7.0 as opposed to 2.1 for the
empirically derived values.

In Fig.\,\ref{fig:model_histograms} we also compare gas fractions
inferred from the gravity-driven turbulence model (rearranging
Eq.\ref{eqn:gravity_driven}) to those calculated using the inverse
Kennicutt-Schmidt relation. We estimate the gas mass within twice the
half-light radius, and then express this as a fraction of the total
baryonic mass $f_{\rm g}$\,=\,$M_{\rm g}$/($M_{\rm
  g}$\,+\,$M_{\star}$). We find a median of $f_{\rm
  g,med}$\,=\,0.52$\pm$\,0.02 for the model and $f_{\rm
  g,med}$\,=\,0.45$\pm$\,0.01 for the empirically derived values. In
comparison, the relations described in \S\ref{trends_with_redshift}
predict an average gas fraction of $f_{\rm
  g}$\,=\,0.41$\pm$\,0.02. The model distribution is again the
broadest, with a 68th percentile range of 0.95 compared to 0.44 for
the observations, and this additional scatter results in unphysical
values of $f_{\rm g}$\,$>$\,1 for $\sim$\,25\% of galaxies.

Thus, over the mass and redshift range probed by our
  observations, both models can provide a reasonable match to the
observations and as such, we are unable to definitively rule out
either turbulence mechanism at this mass and star formation rate.
  First, we note that the medians of the
  distributions for model and empirically derived quantities are very
  similar. For the gravity-driven model, increasing the star formation
  rate per freefall time from $\epsilon_{\rm ff}$\,=\,0.01 to 0.013 in
  Eq.\,\ref{eqn:gravity_driven} would eliminate the offset completely
  (note \citealt{federrath2013,federrath2015} suggest values between
  $\epsilon_{\rm ff}$\,=\,0.01--0.02). For the feedback-driven model
  this could be achieved by adjusting $\mathcal{F}$\,=\,2 to
  $\mathcal{F}$\,=\,3 in Eq.\,\ref{eqn:feedback_driven}. This
  dimensionless normalisation parameter ensures that the model fits
  observations of the relationship between gas surface density and
  star formation rate surface density (Fig.\,4 of
  \citealt{faucher2013}).  Such an increase would be inconsequential
  in this regard.

Second, although the distributions of the best-fit model parameters
are much broader and include unphysical or implausible values
(e.g. $f_{\rm g}$\,>\,1 or $Q_{\rm g}$\,>\,100), this is likely due to
measurement uncertainties.  Estimates of the Toomre $Q_{\rm g}$
parameter for the model have a stronger dependence on rotation
velocity and velocity dispersion than the observational
estimates. Similarly, best-fit model gas fractions depend on $v_{\rm
  C}$ and $\sigma$ but the observational estimates do not. These
dynamical parameters are the largest source of uncertainty and as
such, fractional errors associated with model $Q_{\rm g}$ and $f_{\rm
  g}$ values are approximately twice as large as for the empirically
derived values.

Direct observation of the molecular gas component would help to
provide further constraints.  If turbulence in the ISM is
gravity-driven, we would expect the velocity dispersion to be strongly
dependent on the gas fraction ($\sigma\,\propto\,1/f_{\rm
  g}^2$). However in a feedback-driven scenario, the two properties
should not be related.

We note that these conclusions do not contradict the
  previous analysis by \citet{krumholz2016} who concluded that the
  better match to observations is from the gravity-driven turbulence
  model.  Their conclusion is mainly driven by the match to the
  galaxies with the highest star formation rates, for which we have
  few galaxies in our sample.  However, it is interesting to also note
  that \citet{krumholz2017}
predict a transition from mostly gravity-driven turbulence at high
redshift, to feedback-driven turbulence at low redshift.  They argue
that this evolution would explain why bulges form at high redshift and
disks form at lower redshift. Galaxies at $z$\,$\sim$\,0.9 (of a
similar mass to the KROSS sample) would be expected to have a ratio
between star formation-supported dispersion ($\sigma_{\rm SF}$) and
total gas velocity dispersion ($\sigma_{\rm g}$) of $\sigma_{\rm
  SF}$/$\sigma_{\rm g}$\,$\sim$\,0.3\,--\,0.4. In this context, it
would be unsurprising that we are unable to definitively rule out
either model.

\vspace{-0.5cm}
\section{Conclusions}
\label{conclusions}

In this work we have analysed the velocity dispersion properties of
472 H$\alpha$-detected star-forming galaxies observed as part of KROSS
\citep{stott2016, harrison2017}. KROSS is the largest near-infrared
IFU survey of $z$\,$\sim$\,1 galaxies to date, and consists of a mass-
and colour-selected sample which is typical of the star-forming ``main
sequence'' at this redshift. Mitigating the effects of beam smearing
is essential to understanding the dynamics of high redshift galaxies,
and in Appendix\,\ref{appendix} we present a detailed analysis of this
phenomenon. We derive correction factors as a function of $R_{\rm
  d}$/$R_{\rm PSF}$ (the ratio between the galaxy radius and half of
the spatial PSF) which we apply to our measurements of rotation
velocity and velocity dispersion. Our key results are as follows:

\begin{itemize}
\item Galaxies at this epoch are highly turbulent with large intrinsic
  velocity dispersions. We measure a median dispersion of
  $\sigma_0$\,=\,43.2\,$\pm$\,0.8\,km\,s$^{-1}$ and rotational
  velocity to dispersion ratio of $v_{\rm
    C}$/$\sigma_0$\,=\,2.6\,$\pm$\,0.1 for galaxies with stellar
  masses of log(M$_\star$/M$_\odot$)\,=\,8.7\,--\,11.0. Although
  dynamically hotter than their local counterparts, the majority of
  our sample are rotationally dominated (83\,$\pm$\,5\,\%). We observe
  a strong increase in $v_{\rm C}$/$\sigma_0$ with increasing stellar
  mass: evidence of ``kinematic downsizing''.

\item We combine KROSS with data from SAMI ($z$\,$\sim$\,0.05;
  \citealt{croom2012}) and an intermediate redshift MUSE survey
  ($z$\,$\sim$\,0.5; \citealt{swinbank2017}) to explore the
  relationship between intrinsic velocity dispersion, stellar mass and
  star formation rate. At a given redshift we see, at most, a
  $\sim$\,15\,km\,s$^{-1}$ increase in dispersion for a factor
  $\sim$\,100 increase in stellar mass.

\item All three samples (SAMI, MUSE and KROSS) are consistent with a
  weak increase in velocity dispersion with increasing star formation
  rate. We see an increase of 20\,--\,25\,km\,s$^{-1}$ across three
  orders of magnitude in star formation rate. This trend appears to be
  independent of redshift.

\item At a given redshift the average velocity dispersion is
  consistent across several orders of magnitude in specific star
  formation rate. Normalising for the effects of star formation rate
  and stellar mass, we see a $\sim$\,50\% increase in velocity
  dispersion between $z$\,$\sim$\,0 and $z$\,$\sim$\,0.9.

\item To understand the dynamics of KROSS in a wider evolutionary
  context, we consider five additional samples between
  0\,<\,$z$\,<\,2.5. We find an increase in the average velocity
  dispersion with redshift, from $\sigma_0$\,$\sim$\,25\,km\,s$^{-1}$
  at $z$\,=\,0 to $\sigma_0$\,$\sim$\,50\,km\,s$^{-1}$ at
  $z$\,=\,2. After normalising for the effects of stellar mass, we
  also find a decrease in the average $v_{\rm C}$/$\sigma_0$ ratio for
  a log($M_{\star}$/M$_{\odot}$)\,=\,10.5 galaxy, from $v_{\rm
    C}$/$\sigma_0$\,$\sim$\,6 at $z$\,=\,0 to $v_{\rm
    C}$/$\sigma_0$\,$\sim$\,2 at $z$\,=\,2.

\item We show that the observed evolution in galaxy dynamics can be
  reasonably well described by a simple ``toy model'', in which
  galaxies are asummed to be thin disks of constant rotational
  velocity with higher gas fractions at early times. To provide the
  best possible fit to the data, this model would require lower
  redshift samples to be associated with higher average values of
  Toomre $Q_{\rm g}$ (a more stable gas disk).

\item Finally, we test the predictions of two different analytical
  models -- one which assumes turbulence is driven by stellar feedback
  and another which assumes it is driven by gravitational
  instabilities. Each predicts a different relationship between star
  formation rate and velocity dispersion, with tracks parameterised by
  Toomre $Q_{\rm g}$ or gas fraction, respectively. We find that both
  models provide an adequate description of the data, with best-fit
  parameters close to what we derive independently from the
  observations, using a different set of assumptions. Direct
  measurement of the gas fraction, $f_{\rm g}$, would help to provide
  further constraints.
\end{itemize}

\section*{Acknowledgements}

We thank the anonymous referee for their comments and
  suggestions, which improved the content and clarity of the paper.
This work is based on observations obtained at the Very Large
Telescope of the European Southern Observatory. Programme IDs:
60.A-9460; 092.B-0538; 093.B-0106; 094.B-0061; 095.B-0035. HLJ, AMS,
CMH, RGB, IRS and RMS gratefully acknowledge support from the Science
and Technology Facilities Council (grant codes ST/K501979 and
ST/L00075X/1). AMS acknowledges the Leverhulme Foundation. JPS
acknowledges support from a Hintze Research Fellowship. IRS
acknowledges support from an ERC Advanced Investigator programme {\sc
  dustygal} 321334 and a Royal Society Wolfson Merit Award. AJB
gratefully acknowledges the hospitality of the Australian National
University. DS acknowledges financial support from the Netherlands
Organisation for Scientific research (NWO) through a Veni fellowship
and Lancaster University through an Early Career Internal Grant
(A100679). GEM acknowledges support from the ERC Consolidator Grant
funding scheme (project ConTExt, grant number 648179) and a research
grant (13160) from Villum Fonden. This work is based on observations
taken by the CANDELS Multi-Cycle Treasury Program with the NASA/ESA
\textit{HST}, which is operated by the Association of Universities for
Research in Astronomy, inc., under NASA contract
NAS5-26555. \textit{HST} data were also obtained from the data archive
at Space Telescope Science Institute. We thank Holly Elbert, Tim Green
and Laura Pritchard for carrying out some of the KMOS observations,
and the staff at Paranal.

Finally, we would like to thank the SAMI team for sharing their data
with us and Ivy Wong for useful discussion. BC acknowledges support
from the Australian Research Council's Future Fellowship (FT120100660)
funding scheme. CF gratefully acknowledges funding provided by the
Australian Research Council's Discovery Projects (grants DP150104329
and DP170100603). M.S.O. acknowledges the funding support from the
Australian Research Council through a Future Fellowship
(FT140100255). The SAMI Galaxy Survey is based on observations made at
the Anglo-Australian Telescope. The Sydney-AAO Multi-object Integral
field spectrograph (SAMI) was developed jointly by the University of
Sydney and the Australian Astronomical Observatory. The SAMI input
catalogue is based on data taken from the Sloan Digital Sky Survey,
the GAMA Survey and the VST ATLAS Survey. The SAMI Galaxy Survey is
funded by the Australian Research Council Centre of Excellence for
All-sky Astrophysics (CAASTRO), through project number CE110001020,
and other participating institutions. The SAMI Galaxy Survey website
is \url{http://sami-survey.org/}.




\bibliographystyle{mnras}
\bibliography{thesis_refs} 



\FloatBarrier 

\appendix

\renewcommand{\arraystretch}{1.5}
\begin{sidewaystable*}
\begin{center}{
\caption{Intrinsic velocity dispersion and related quantities}
\label{TableSigma}
\vspace{2mm}
\begin{tabular*}{\textwidth}{@{\extracolsep{\fill}} l | cccccccc}
\hline
Name & RA & Dec & $z$ & $\sigma_{\rm 0, obs}$ & $\sigma_{\rm 0}$ & Flag & $R_{\rm d}$\,/ \,$R_{\rm PSF}$ & $Q_{\rm g}$\\
 & (J2000) & (J2000) &  & (km\,s$^{-1}$) &  (km\,s$^{-1}$) &  &  & \\
\hline
\hline
C-HiZ\_z1\_111  &  +49:55:07  &  00:08:27.2  &  0.8498  &  79\,$\pm$\,40  &  53\,$\pm$\,27  &  M  &  0.8\,$\pm$\,0.2  &  2.5$^{+1.8}_{-1.6}$ \\
C-HiZ\_z1\_112  &  +49:55:13  &  00:09:08.0  &  0.8539  &  33\,$\pm$\,13  &  33\,$\pm$\,13  &  O  &  0.4\,$\pm$\,0.1  &  0.1$^{+0.1}_{-0.1}$ \\
C-HiZ\_z1\_186  &  +50:08:04  &  00:09:05.5  &  0.8445  &  46\,$\pm$\,3  &  45\,$\pm$\,3  &  O  &  0.3\,$\pm$\,0.1  &  0.2$^{+0.1}_{-0.1}$ \\
C-HiZ\_z1\_195  &  +50:08:40  &  00:08:58.0  &  0.8454  &  22\,$\pm$\,5  &  22\,$\pm$\,5  &  O  &  0.5\,$\pm$\,0.2  &  0.5$^{+0.3}_{-0.2}$ \\
C-HiZ\_z1\_215  &  +50:11:50  &  00:09:08.7  &  0.8441  &  12\,$\pm$\,3  &  11\,$\pm$\,3  &  O  &  0.6\,$\pm$\,0.2  &  1.3$^{+2.7}_{-0.9}$ \\
C-HiZ\_z1\_224  &  +50:13:05  &  00:08:27.7  &  1.0137  &  46\,$\pm$\,23  &  40\,$\pm$\,20  &  M  &  0.7\,$\pm$\,0.2  &  0.9$^{+0.7}_{-0.6}$ \\
C-HiZ\_z1\_230  &  +50:13:39  &  00:09:02.1  &  0.8445  &  48\,$\pm$\,3  &  46\,$\pm$\,3  &  O  &  0.2\,$\pm$\,0.1  &  0.6$^{+0.3}_{-0.3}$ \\
C-HiZ\_z1\_231  &  +50:13:40  &  00:08:38.6  &  0.8377  &  58\,$\pm$\,2  &  55\,$\pm$\,2  &  O  &  0.3\,$\pm$\,0.1  &  0.1$^{+0.1}_{-0.1}$ \\
C-HiZ\_z1\_235  &  +50:14:02  &  00:08:30.2  &  0.8378  &  85\,$\pm$\,42  &  37\,$\pm$\,18  &  M  &  0.3\,$\pm$\,0.1  &  0.8$^{+0.9}_{-0.6}$ \\
C-HiZ\_z1\_245  &  +50:15:26  &  00:07:27.4  &  0.8334  &  66\,$\pm$\,33  &  52\,$\pm$\,26  &  M  &  1.0\,$\pm$\,0.3  &  1.7$^{+1.2}_{-1.1}$ \\
C-HiZ\_z1\_246  &  +50:15:33  &  00:09:17.8  &  0.8422  &  51\,$\pm$\,26  &  32\,$\pm$\,16  &  M  &  0.5\,$\pm$\,0.1  &  0.8$^{+0.6}_{-0.6}$ \\
C-HiZ\_z1\_251  &  +50:15:57  &  00:09:20.7  &  0.8544  &  57\,$\pm$\,10  &  45\,$\pm$\,8  &  O  &  0.2\,$\pm$\,0.1  &  0.9$^{+0.6}_{-0.5}$ \\
C-HiZ\_z1\_255  &  +50:16:17  &  00:09:19.6  &  0.8502  &  41\,$\pm$\,21  &  34\,$\pm$\,17  &  M  &  0.9\,$\pm$\,0.3  &  1.4$^{+1.0}_{-0.9}$ \\
C-HiZ\_z1\_257  &  +50:16:24  &  00:09:05.6  &  0.8501  &  73\,$\pm$\,14  &  66\,$\pm$\,13  &  O  &  0.3\,$\pm$\,0.1  &  0.9$^{+0.6}_{-0.5}$ \\
C-HiZ\_z1\_258  &  +50:16:25  &  00:07:31.8  &  0.8376  &  43\,$\pm$\,21  &  38\,$\pm$\,19  &  M  &  1.3\,$\pm$\,0.4  &  2.1$^{+1.5}_{-1.4}$ \\
C-HiZ\_z1\_263  &  +50:17:11  &  00:08:42.1  &  0.8370  &  41\,$\pm$\,10  &  40\,$\pm$\,10  &  O  &  0.3\,$\pm$\,0.1  &  0.3$^{+0.2}_{-0.2}$ \\
...  &  ...  &  ...  &  ...  &  ...  &  ...  &  ...  &  ...  &  ... \\
\hline
\end{tabular*}
\label{table:Sigmas}
}\end{center}
\vspace{-2mm}
{\sc notes:} A catalog of all 586 H$\alpha$ detected galaxies in the KROSS sample is available online at \url{http://astro.dur.ac.uk/KROSS}. Columns $\sigma_{\rm 0,obs}$ and $\sigma_0$ are the observed and beam smearing corrected velocity dipersions, respectively. Corrections were applied as a function of $R_{\rm d}$/$R_{\rm PSF}$ -- the ratio between the disk radius (in arcsec) and half of the seeing FWHM (see \S\ref{beam_smearing}). We flag whether the dispersion was measured in the outskirts of the disk (O), or from the median of all available pixels (M). We also provide the global Toomre $Q_{\rm g}$ parameter for each galaxy, which we derived by inverting the Kennicutt-Schmidt relation to estimate $\Sigma_{\rm gas}$.
\end{sidewaystable*}

\section{Catalogue}
\label{catalog}
With \citet{harrison2017} we released a catalogue of all 586 H$\alpha$
detected galaxies in the KROSS sample. This is available online at
\url{http://astro.dur.ac.uk/KROSS}. We have updated the catalogue to
include all velocity dispersion measurements discussed in this
paper. Examples of this additional data are provided in
Table\,\ref{table:Sigmas}.

\section{Beam smearing analysis}
\label{appendix}
\subsection{Motivation}
Integral field spectroscopy has allowed us to study the spatially
resolved gas dynamics, star formation and ISM properties of distant
galaxies in unprecendented detail. However as with any other technique
it is not immune to systematics; in particular observations of galaxy
dynamics can be biased as a result of ground-based seeing. Each of the
24 deployable IFUs on KMOS has a spatial sampling of 0.2 arcsec,
however the observations are seeing-limited, and as such we must
consider the impact of the spatial PSF (the seeing) on our
measurements.

As the observations are convolved with the PSF, information from each
spatial pixel is combined with that of neighbouring regions -- a
phenomenon known as ``beam smearing''. Effects of this on the observed
gas kinematics are two-fold. Firstly, the spectrum at each pixel is
contaminated by components of slightly higher or lower velocities,
acting to broaden spectral features and increase the observed velocity
dispersion. Secondly, if the blueshifted components are brighter than
the redshifted components (or vice versa) the intrinsic velocity of
the pixel will be shifted slightly. Globally, this results in the
rotation curve appearing flatter than it may be intrinsically.

Understanding the kinematics of our sample is central to achieving the
key science goals of KROSS, e.g. investigating the origins of disk
turbulence and studying angular momentum. It is therefore essential
that we calibrate for the effects of beam smearing. Here we
investigate the systematic effects of beam smearing by creating a
series of mock KMOS observations. This will allow us to constrain the
biases introduced and derive an efficient method of correcting for
them.

\subsection{Methods}
To explore the impact of beam-smearing on our observations we create a
catalogue of $\sim$10$^5$ model galaxies, with properties to uniformly
sample the KROSS parameter space. For each galaxy we create two sets
of mock IFU observations. First, we model what the ionised gas
dynamics would look like in the absence of atmospheric turbulence
(i.e. KMOS sampling the intrinsic properties of the galaxy). Second,
we generate the same dynamical maps for observations made under
seeing-limited conditions. Differences between the two datasets will
allow us to understand how beam smearing affects measurements of the
rotation velocity ($v$) and intrinsic velocity dispersion ($\sigma_0$)
and learn how to correct for it.

\subsubsection{Intrinsic Properties of the Model Galaxies}
In the local Universe, galaxy dynamics can be described by the
contribution of a rotating disk of gas and stars plus a dark matter
halo, with the velocities added in quadrature as $v^2$\,=\,$v^2_{\rm
  h}$\,+\,$v^2_{\rm d}$. To create model galaxies we apply the same
principle, making some simple assumptions about each component,
following \citet{swinbank2017}. Firstly, we assume that the baryonic
surface density follows an exponential profile \citep{freeman1970}
characterised by a disk mass ($M_{\rm d}$) and scale length ($R_{\rm
  d}$):

\begin{equation}
\Sigma_{\rm d}(r)\,=\,\frac{M_{\rm d}}{2 \pi\,R^2_{\rm d}}e^{-r/R_{\rm d}}.
\end{equation}
The contribution of this disk to the circular velocity of the galaxy is

\begin{equation}
v^2_{\rm d}(x)\,=\,\frac{1}{2}\frac{GM_{\rm d}}{R_{\rm d}}(3.2x)^2(I_0K_0\,-\,I_1K_1),
\end{equation}
where $x\,=\,R/R_{\rm d}$ and $I_{\rm n}$, $K_{\rm n}$ are the
modified Bessel functions computed at 1.6\,$x$. For the halo we assume
$v_{\rm h}^2 = GM_{\rm h}(<r)/r$ with a dark matter density profile
described by a core radius ($r_{\rm c}$) and effective core density
($\rho_{\rm dm}$):

\begin{equation}
\rho(r)\,=\,\frac{\rho_{\rm dm}r_{\rm c}^3}{(r\,+\,r_{\rm c})(r^2\,+\,r_{\rm c}^2)},
\end{equation}
\citep{persic1988,burkert1995,salucci2000}. This results in a velocity profile of the form

\begin{multline}
v^2_{\rm h}(r)\,=\,\frac{6.4G\rho_{\rm dm}r_{\rm c}^3}{r} \,\,\, \times \\ \left\lbrace\,{\rm ln}\,\left(1\,+\,\frac{r}{r_{\rm c}}\right)\,-\,{\rm tan}^{-1}\left(\frac{r}{r_{\rm c}}\right)\,+\,\frac{1}{2}\,{\rm ln}\left(1\,+\,\left(\frac{r}{r_{\rm c}}\right)^2\right)\right\rbrace.
\end{multline}

The dark matter fraction of a galaxy ($f_{\rm dm}$) greatly influences
the shape of its rotation curve, hence it is important that the dark
matter properties of our model galaxies closely match those of the
KROSS sample. To satisfy this, we exploit results of the ``Evolution
and Assembly of GaLaxies and their Environments'' cosmological
simulation suite ({\sc eagle}; \citealt{schaye2015,crain2015}). These
are a set of hydrodynamical simulations, including sub-grid modelling
of star formation and stellar feedback, as well as feedback from
supermassive black hole accretion. The {\sc eagle} simulations produce
galaxies which closely match the observed Universe and so provide an
ideal way to estimate $f_{\rm dm}$ for our $z$\,$\sim$\,1
sample. Considering star-forming galaxies of a similar mass
(10$^9$\,$<$\,$M_{\rm d}$\,$<$\,10$^{11}$) and redshift
(0.8\,$<$\,$z$\,$<$\,1.0), we find a median and 1$\sigma$ range of
$f_{\rm dm}$\,=\,0.75\,$\pm$\,0.10 within the central 10\,kpc
\citep{schaller2015}. From this we can infer suitable values for
$\rho_{\rm dm}$.

To complete our galaxy model, we assume that the intrinsic velocity
dispersion of ionised gas ($\sigma_0$) is uniform across the disk and
that the distribution of H$\alpha$ (the emission line from which we
measure the kinematics) is exponential. Following the results of
\citet{nelson2016} we assume that the H$\alpha$ emission is more
extended than the stellar continuum, with $R_{\rm
  H\alpha}$\,$\sim$\,1.1\,$R_{\rm d}$. How we define the distribution
of light is significant, since beam smearing effects are luminosity
weighted. Star forming galaxies at $z$\,$\sim$\,1 often appear
irregular or ``clumpy'' in H$\alpha$ and in \S\ref{intensity_maps} we
explore how this may impact our results.

From this simple prescription we create a series of intensity maps,
velocity maps and velocity dispersion maps for model galaxies with
similar properties to those in the KROSS sample. We vary the disk
mass, disk scale length, inclination, dark matter fraction and
intrinsic velocity dispersion as follows:
\vspace{0.2cm}

$\bullet$ 9.0 $\leq$ log($M_{\rm d}$ / M$_\odot$) $\leq$ 11.2; steps of 0.15
\vspace{0.15cm}

$\bullet$ 0.5 $\leq$ $R_{\rm d}$ $\leq$ 2.5\,kpc; steps of 0.25\,kpc
\vspace{0.15cm}

$\bullet$ 20 $\leq$ $\theta$ $\leq$ 70\,$\deg$; steps of 5\,$\deg$
\vspace{0.15cm}

$\bullet$ 0.65 $\leq$ $f_{\rm dm}$ $\leq$ 0.85; steps of 0.10
\vspace{0.15cm}

$\bullet$ 20 $\leq$ $\sigma_0$ $\leq$ 80\,km\,s$^{-1}$; steps of 10\,km\,s$^{-1}$

\vspace{0.15cm}
\subsubsection{Mock IFU Observations}
\label{mock_observations}
After defining the intrinsic properties of a given galaxy, we wish to
understand how these same dynamical maps may look under ground-based
seeing conditions. To model this, we generate a mock observation of
the galaxy, forming a KMOS data cube which we can then convolve with
the seeing PSF. While pixel scale of this cube is set to match the
spatial resolution of our observations (0.1 arcsec), we choose to
retain a high \textit{spectral} resolution ($R\,\sim$\,10,000) and
omit instrument noise. This allows us to attribute any difference
between the model and ``observed'' data to beam smearing alone. The
$x$\,--\,$y$ footprint of the array is initially also larger than the
2.8\,$\times$\,2.8 arcsec size of the KMOS IFU, to allow for light
outside of this region which may be introduced to the IFU pixels via
beam smearing.

At each pixel we create a spectrum consisting of the H$\alpha$
emission line and [N{\sc ii}] doublet, assuming that each line is
described by a Gaussian profile with a linewidth set by the dispersion
and redshift set by the rotation velocity at that position. For
simplicity we adopt a fixed H$\alpha$/[N{\sc ii}] ratio and set the
flux ratio of the [N{\sc ii}]$\lambda\lambda$6548, 6583 doublet to be
3.06 \citep{osterbrock2006}. To simulate the effects of beam smearing
we then convolve each wavelength slice with the spatial PSF. We model
a range of atmospheric conditions, with FWHM$_{\rm
  seeing}$\,=\,0.5\,-\,0.9 arcsec in increments of 0.1 arcsec, and
assume a Gaussian profile each time. The median for our KROSS
observations is 0.7 arcsec with a standard deviation of 0.17 arcsec,
so this range encompasses the data well. Finally, we extract a
subsection of the array to match the size of the KMOS IFU. This is the
``observed'' data cube on which we perform our analysis.

To generate dynamical maps from the beam-smeared cube we fit the
emission lines at each pixel using the same $\chi^2$ minimisation
method as in the data. We require that all lines are Gaussian profiles
and share the same linewidth, with the relative positions of the lines
and [N{\sc ii}] flux ratio fixed to their model values. These
constraints leave the H$\alpha$ and [N{\sc ii}] intensity, centroid
and line width free to vary. Since our model does not include noise,
spatial binning is not necessary, however we explore how this process
may affect results in \S\ref{adaptive_binning}. We also extract the
rotation curve and one-dimensional dispersion profile of each
galaxy. To do so we take the median value of pixels along a 0.7 arcsec
``slit'' defined by the major kinematic axis.

\subsubsection{Kinematic Measurements}
\label{measurements}
In order to quantify the effects of beam smearing, we measure the
kinematics in the same way as for the KROSS sample
\citep{harrison2017}. Each rotation curve is fit by an exponential
disk model of the form

\begin{equation}
\label{eq1}
v(r)^2 = \frac{r^2\pi G \mu_0}{R_{\rm d,fit}}(I_0 K_0 - I_1 K_1) + v_{\rm off},
\end{equation}
where $r$ is the radial distance, $\mu_{\rm 0}$ is the peak mass
surface density, $R_{\rm d,fit}$ is the disk radii, $v_{\rm off}$ is
the velocity at the kinematic centre, and $I_n K_n$ are Bessel
functions evaulated at $\frac{1}{2}$\,$r$/$R_{\rm d,fit}$. We use this
model to interpolate through the data and measure the velocity at a
given radius. Other kinematic surveys define the characteristic
rotation velocity of a galaxy in various ways. We therefore wish to
understand how beam smearing may affect our results as a function of
radius. Using the input value of $R_{\rm d}$ for each model we
therefore measure velocities at 2.2$R_{\rm d}$, 3.4$R_{\rm d}$ and the
same again but for radii convolved with the seeing ($R_{\rm d,conv}$).

To characterise the velocity dispersion we record the median of the
profile outside 3.4$R_{\rm d}$ and also the median of all pixels
within the map. Although our simulated galaxies are constructed such
that it is possible to make both measurements, for 52\% of KROSS
galaxies it is only possible to make the latter (due to a large disk
scale length or poor signal-to-noise). Hence it is important to
understand both parameters. Since beam smearing is expected to be
strongest towards the dynamical centre, the median dispersion will
likely depend on the maximum radii of detected pixels. We therefore
measure the velocity dispersion within both a 2$R_{\rm d}$ and
3$R_{\rm d}$ aperture.

\subsection{Results}
\subsubsection{Dynamical Maps}
\label{beam_smearing_maps}

\begin{figure*}
\begin{center}
\includegraphics[width=5in]{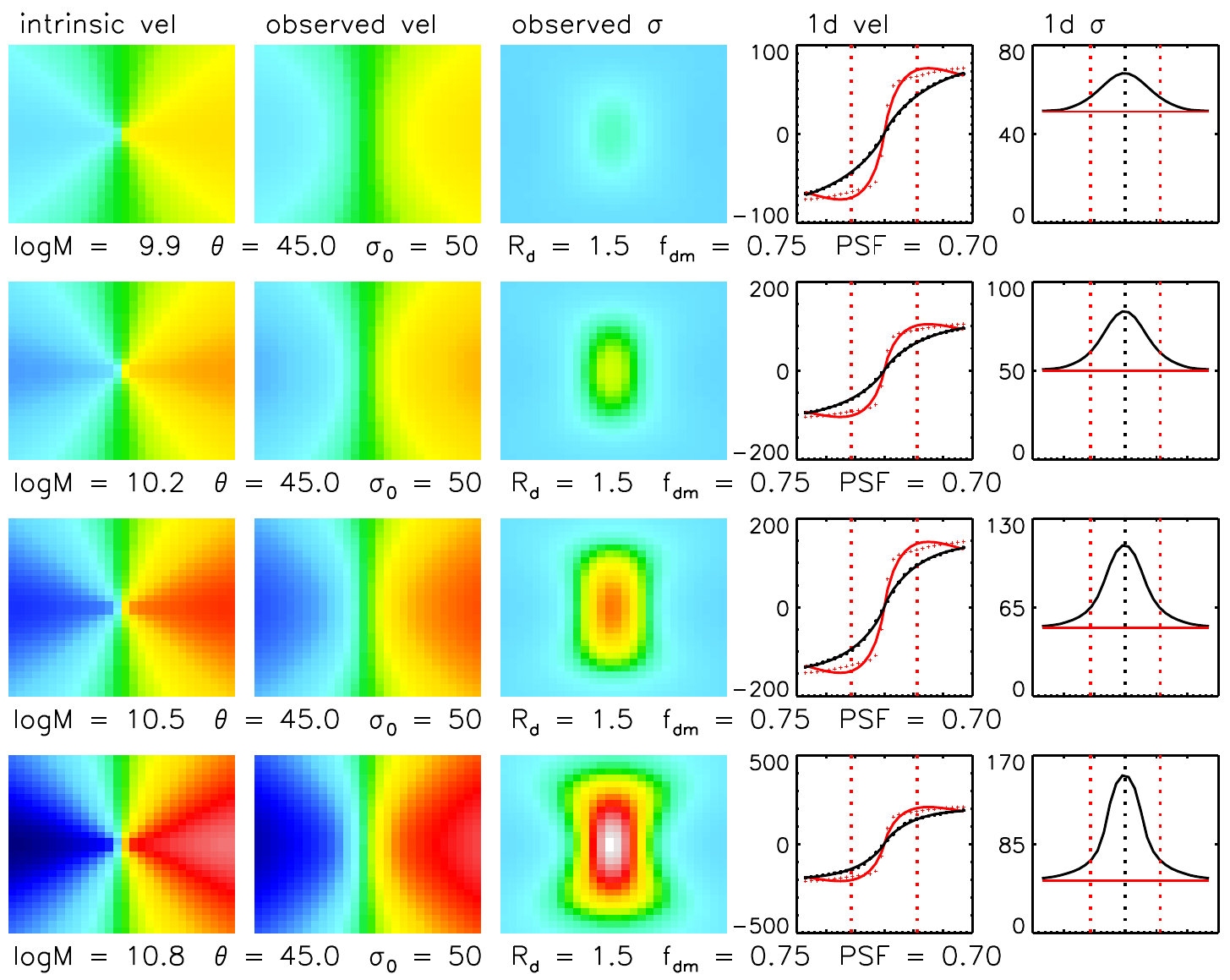}
\end{center}
\caption[Example dynamical maps for models of increasing disk
  mass]{Dynamical maps for a subset of $\sim$\,10$^5$ model galaxies
  created to explore the impacts of beam smearing. We investigate how
  closely we can recover the intrinsic velocity and velocity
  dispersion of a galaxy as a function of disk mass. In successive
  rows we increase the mass of the model while keeping all other
  parameters fixed. An increase in mass results in a steeper velocity
  gradient across the disk. This leads to a stronger beam smearing
  effect, with a larger peak in the observed velocity dispersion. Left
  to right we show the intrinsic velocity map, ``observed'' velocity
  map and velocity dispersion map, along with the rotation curve and
  line-of-sight dispersion profile extracted along the primary
  kinematic axis before (solid red line) and after (black) convolution
  with the seeing. Dashed red lines represent a radius of 3.4$R_{\rm
    d}$. On each row we describe the model input parameters where $M$
  is disk mass (M$_{\odot}$), $\theta$ inclination (deg), $\sigma_0$
  intrinsic velocity dispersion (km\,s$^{-1}$), $R_{\rm d}$ disk
  radius (kpc), $f_{\rm dm}$ dark matter fraction within 10\,kpc, and
  PSF the full width half maximum of seeing (arcsec). Each velocity
  map is scaled between $-$250 and 250\,km\,s$^{-1}$, and each
  dispersion map between 0 and ($\sigma_0$+100)\,km\,s$^{-1}$.}
\label{fig:mass_maps}
\end{figure*}

\begin{figure*}
\begin{center}
\includegraphics[width=5in]{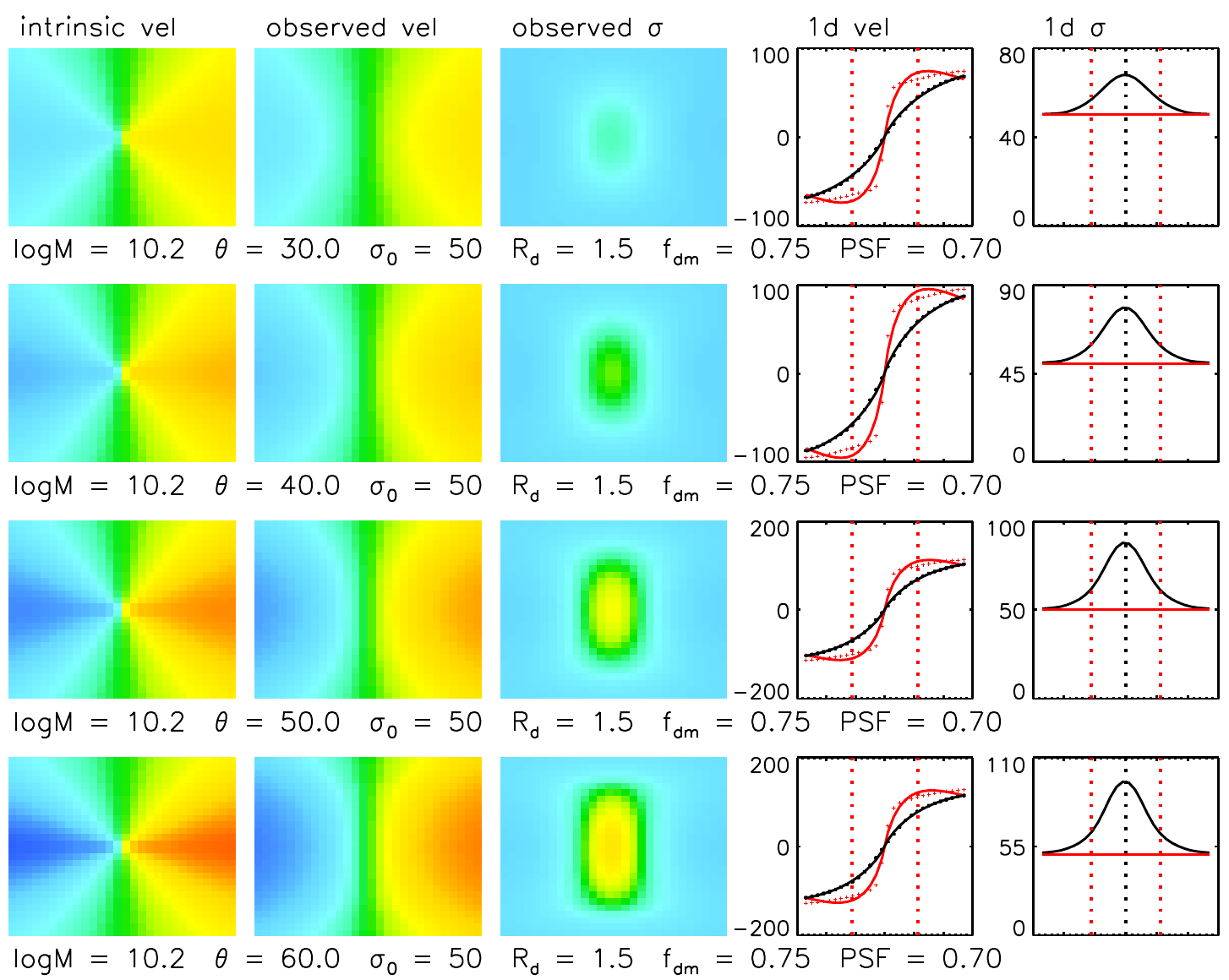}
\end{center}
\caption[Example dynamical maps for models of increasing
  inclination]{Example dynamical maps and velocity profiles with
  properties as described in Fig.\,\ref{fig:mass_maps}. Left to right
  we show the intrinsic velocity map, ``observed'' velocity map and
  velocity dispersion map, rotation curve and line-of-sight dispersion
  profile. In successive rows we increase the inclination of the model
  while all other input parameters remain fixed. As the disk is tilted
  towards edge-on, the maximum velocity of the rotation curve is
  increased and contours of the velocity map are pushed closer
  together (with the characteristic ``spider diagram'' shape). A
  steeper velocity gradient results in a stronger beam smearing
  effect. The more highly inclined the disk, the larger the peak in
  the observed velocity dispersion profile.}
\label{fig:incl_maps}
\end{figure*}

\begin{figure*}
\begin{center}
\includegraphics[width=5in]{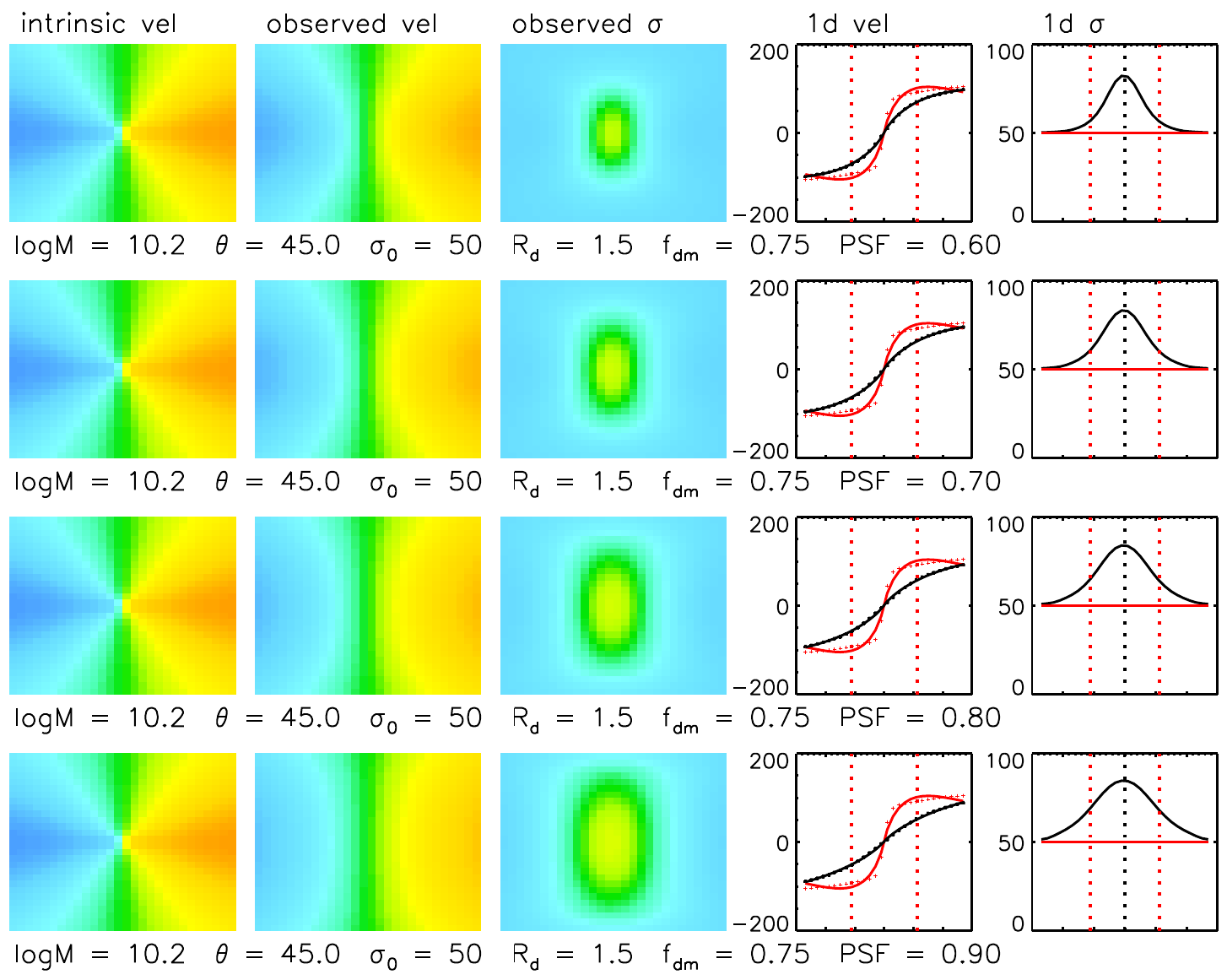}
\end{center}
\caption[Example dynamical maps for models of increasing spatial
  PSF]{Dynamical maps and velocity profiles with properties as
  described in Fig.\,\ref{fig:mass_maps}. Left to right we show the
  intrinsic velocity map, ``observed'' velocity map and velocity
  dispersion map, rotation curve and line-of-sight dispersion
  profile. In successive rows we increase the spatial PSF (the seeing)
  of the model while all other input parameters remain fixed. Poorer
  atmospheric conditions result in a more severe beam smearing
  effect. As the seeing is increased, the observed velocity gradient
  becomes shallower, structure in the velocity map is less visible and
  the peak in the observed velocity dispersion becomes broader. While
  at 0.6 arcsec the dispersion measured in the outskirts of the galaxy
  (beyond 3.4$R_{\rm d}$ -- red dashed lines) is close to the
  intrinsic value, at 0.9 arcsec this is a gross overestimate.}
\label{fig:psf_maps}
\end{figure*}

Before we perform a more rigorous analysis, quantifying the effects of
beam smearing using the variables defined in \S\ref{measurements}, we
note several trends in the dynamical maps. In
Fig.\,\ref{fig:mass_maps}--\ref{fig:psf_maps} we show the dynamical
maps, rotation curve and velocity dispersion profile of 12 model
galaxies where all parameters are kept fixed except for mass
(Fig.~\ref{fig:mass_maps}), inclination angle
(Fig.~\ref{fig:incl_maps}) or seeing (Fig.~\ref{fig:psf_maps}). We
compare the intrinsic kinematics to those recovered after the data
cube has been convolved with the spatial PSF. While the extent of the
beam smearing depends on the input parameters, the effects are broadly
similar in each case. The observed velocity map appears smoother and
the observed rotation curve (in black) is flatter than the intrinsic
(in red), particularly close to the dynamical centre. The beam smeared
rotation curve also peaks at a lower maximum velocity. Finally, the
observed dispersion map is no longer uniform, and we now see a
characteristic rise in the region of the steepest velocity gradient.

Fig.\,\ref{fig:mass_maps} explores the relationship between disk mass
and beam smearing, and we show four models for which mass is the only
parameter allowed to vary. We increase the disk mass over the range
log($M_{\rm d}$ / M$_\odot$)\,=\,9.9\,--\,10.8 and find that the beam
smearing effect becomes more apparent at each interval. Since the
effect of the mass is to increase the steepness of the inner rotation
curve, the peak of the observed velocity dispersion profile increases
from $\sim$\,70\,km\,s$^{-1}$ in the low mass galaxy, to
$\sim$\,160\,km\,s$^{-1}$ in the high mass example. As the disk mass
increases the velocity gradient across the disk becomes larger, hence
the components combined by the PSF have a greater velocity difference.

In Fig.\,\ref{fig:incl_maps} we use the same fiducial model as in
Fig.\,\ref{fig:mass_maps}, however this time fix the mass as
log($M_{\rm d}$ / M$_\odot$)\,=\,10.2 and vary the inclination from 30
to 60 degrees. This figure shows that the more inclined the disk, the
greater the beam smearing effect. As the disk approaches edge-on the
iso-velocity contours of the map are ``pinched'' together more
closely, an effect similar to increasing the disk mass.

Finally, Fig.\,\ref{fig:psf_maps} demonstrates the effect of
broadening the spatial PSF. We increase the seeing from 0.6 to 0.9
arcsec and study the impact this has on the observed velocity and
velocity dispersion. As with an increase in disk mass or greater disk
inclination, poorer atmospheric conditions result in a rotation curve
which is shallower than it would be intrinsically. However, while the
most noticeable effect of increasing the inclination or disk mass is
to increase the peak of the dispersion profile, the same is not true
of the seeing. An increase in the seeing instead acts to broaden the
dispersion profile. At a seeing of 0.6 arcsec the intrinsic dispersion
can be recovered at a radius of $\sim$\,3\,$R_{\rm d}$, but for a
seeing of 0.9 arcsec the required radius is approximately double.

\subsubsection{Impact of Model Parameters}
\label{effect_of_params}

As discussed in \S\ref{beam_smearing_maps}, from visual inspection of
the dynamical maps it is already possible to identify several trends
between model input parameters and the impact of beam
smearing. However we now wish to quantify these effects such that we
can apply corrections to our KROSS data.

\begin{figure*}
\begin{center}
\includegraphics[width=5.5in]{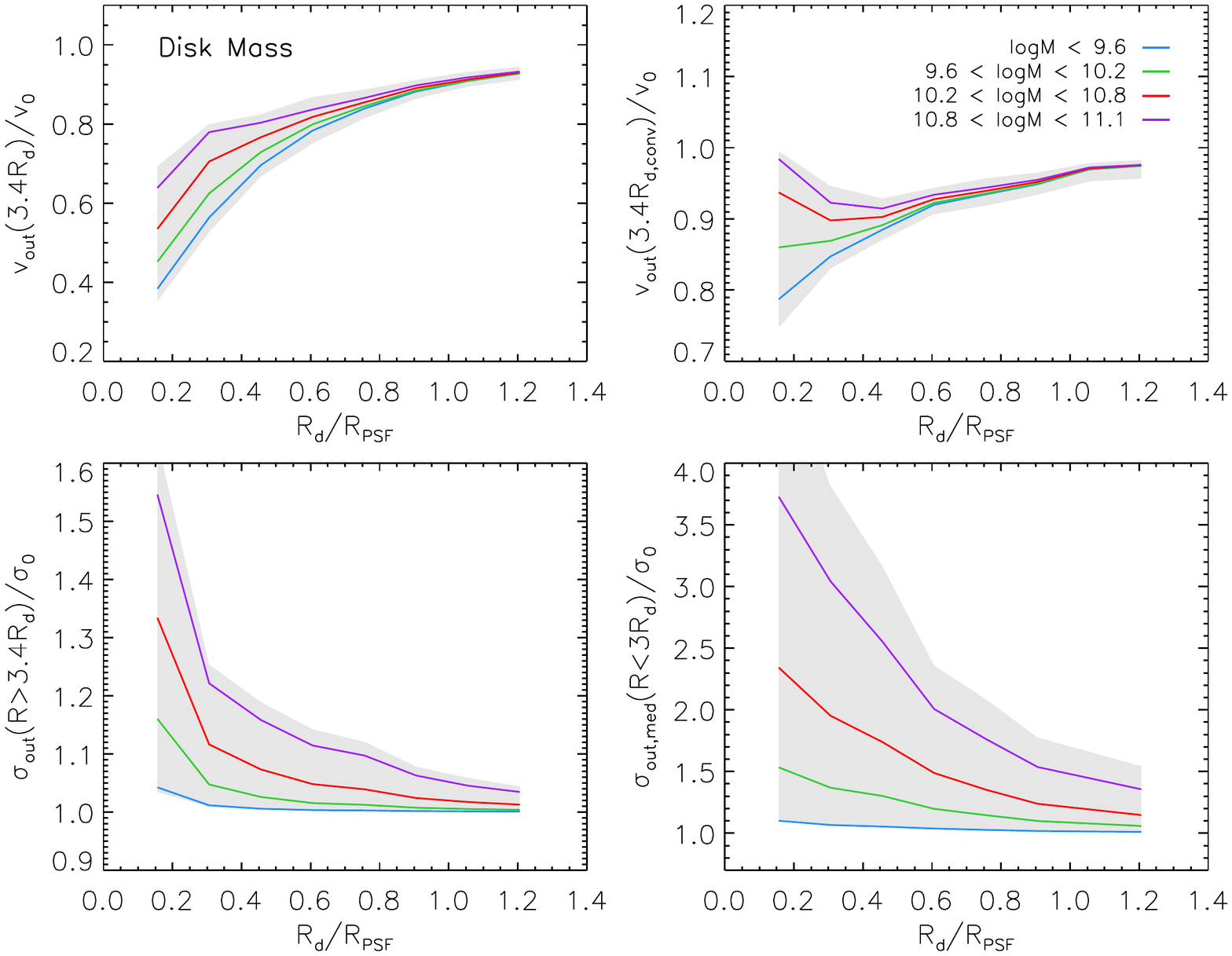}
\end{center}
\caption[Impact of beam smearing on measurements of rotation velocity
  and velocity dispersion, as a function of disk mass]{ The impact of
  beam smearing on measurements of rotation velocity and velocity
  dispersion, as a function of disk mass. $R_{\rm d}/R_{\rm PSF}$ is
  the ratio between the disk radius and the half width half maximum of
  the PSF, which determines which velocity components are combined by
  the seeing, and thus the extent of the beam smearing. We split
  models into four mass bins and plot a running median for each (solid
  lines). Shaded regions represent the 1$\sigma$ scatter of all
  models, showing that disk mass accounts for most, if not all, of
  this variation. \textit{Top Left:} Model velocity at 3.4$R_{\rm d}$
  as a fraction of the intrinsic value ($v_{0}$). The lower the
  $R_{\rm d}/R_{\rm PSF}$ and the smaller the disk mass, the more we
  underestimate the true velocity. \textit{Top Right:} Velocity at the
  same radius convolved with the seeing. This is a better estimate of
  $v_0$(3.4$R_{\rm d}$), with at most a 20\%
  difference. \textit{Bottom Left:} Median of the velocity dispersion
  profile beyond 3.4$R_{\rm d}$ as a fraction of the intrinsic
  ($\sigma_0$). The lower the $R_{\rm d}/R_{\rm PSF}$ and larger the
  disk mass, the more we overestimate the dispersion, with up to a
  50\% difference. \textit{Bottom Right:} The median dispersion within
  an aperture of 3$R_{\rm d}$. This measurement is more susceptible to
  beam smearing, with overestimates of up to a factor of four.}
\label{fig:mass_trend}
\end{figure*}

\begin{figure*}
\begin{center}
\includegraphics[width=5.5in]{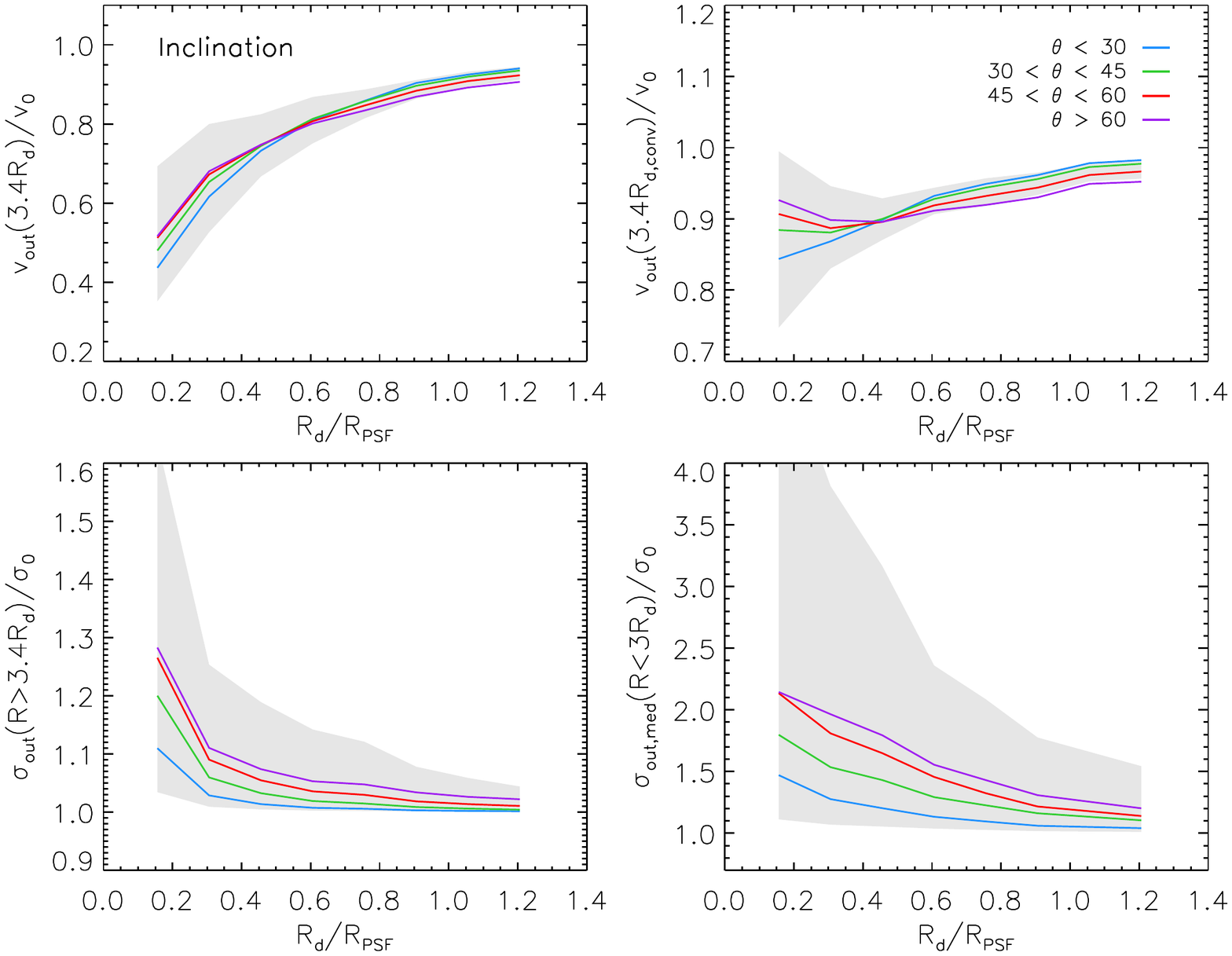}
\end{center}
\caption[Impact of beam smearing on measurements of rotation velocity
  and velocity dispersion, as a function of disk inclination]{The
  impact of beam smearing on measurements of rotation velocity and
  velocity dispersion, as a function of disk inclination. \textit{Top
    Left:} The lower the $R_{\rm d}/R_{\rm PSF}$, the more we
  underestimate the intrinsic rotation velocity ($v_0$). The extent of
  this difference is very similar for models of different inclinations
  (typically within 5\%). \textit{Top Right:} We extract the observed
  rotation velocity at the required radius convolved with the
  seeing. This results in a better estimate, but adjusting the
  inclination appears to have little influence. \textit{Bottom Left:}
  The lower the $R_{\rm d}/R_{\rm PSF}$ and the more inclined the
  disk, the more we overestimate the intrinsic velocity
  dispersion. The difference between a low inclination model ($<$\,30
  degrees) and a high inclination model ($>$\,30 degrees) is still
  relatively minor ($\sim$\,10\%) and cannot account for the full
  1$\sigma$ scatter of the data (shaded region). The trend between
  disk mass and beam smearing appears to be more
  dominant. \textit{Bottom Right:} If we estimate the velocity
  dispersion from a median of the map, as opposed to the outskirts of
  the dispersion profile, we overestimate $\sigma_0$ by almost twice
  as much.}
\label{fig:incl_trend}
\end{figure*}

\begin{figure*}
\begin{center}
\includegraphics[width=5.5in]{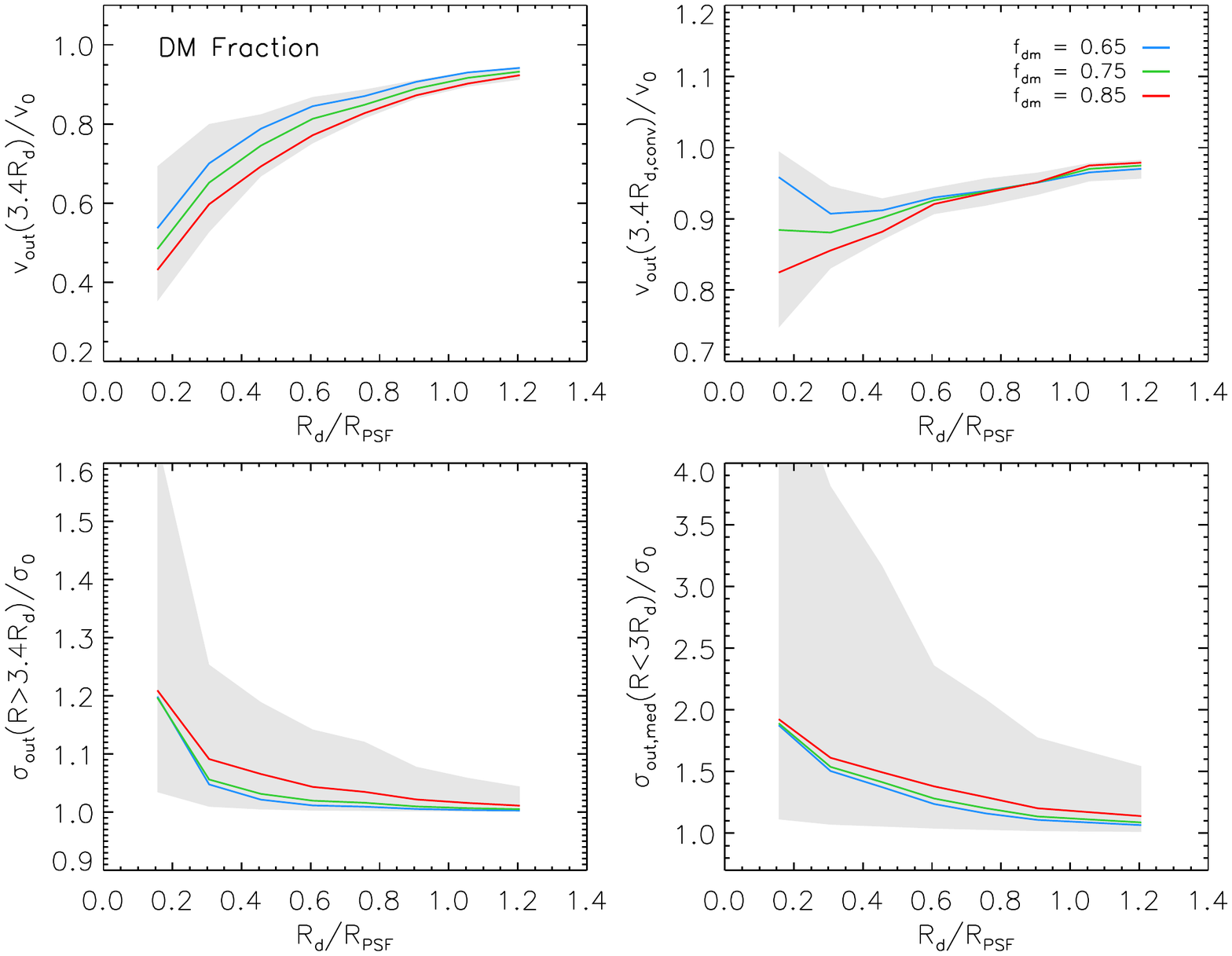}
\end{center}
\caption[Impact of beam smearing on measurements of rotation velocity
  and velocity dispersion, as a function of the dark matter fraction
  within 10\,kpc]{The impact of beam smearing on measurements of
  rotation velocity and velocity dispersion, as a function of the dark
  matter fraction within a radius of 10\,kpc. \textit{Top Left:} The
  lower the $R_{\rm d}/R_{\rm PSF}$ and the larger the dark matter
  fraction, the more we underestimate the intrinsic rotation velocity
  ($v_0$). There is a $\sim$\,10\% difference between $f_{\rm
    dm}$\,=\,0.65\,--\,0.85. Adjusting the dark matter fraction
  changes the shape of the rotation curve in the outer regions (more
  or less of a ``turn-over''), hence the velocity components
  ``merged'' by the seeing will be slightly different. \textit{Top
    Right:} We extract the observed rotation velocity at the required
  radius convolved with the seeing. This results in a better
  estimate. For low $R_{\rm d}/R_{\rm PSF}$ the difference in $f_{\rm
    dm}$ models is approximately the same, but for $R_{\rm d}/R_{\rm
    PSF}$\,$>$\,0.6 the models converge. \textit{Bottom Left:} The
  lower the $R_{\rm d}/R_{\rm PSF}$, the more we overestimate the
  intrinsic velocity dispersion. The difference between models of
  $f_{\rm dm}$\,=\,0.65 and models of $f_{\rm dm}$\,=\,0.85 is
  extremely small (a few per cent). \textit{Bottom Right:} If we
  estimate the velocity dispersion from a median of the map, as
  opposed to the outskirts of the dispersion profile, we overestimate
  $\sigma_0$ by almost twice as much. Again, the dark matter fraction
  appears to have little effect on this aspect of beam smearing.}
\label{fig:fdm_trend}
\end{figure*}

\begin{figure*}
\begin{center}
\includegraphics[width=5.5in]{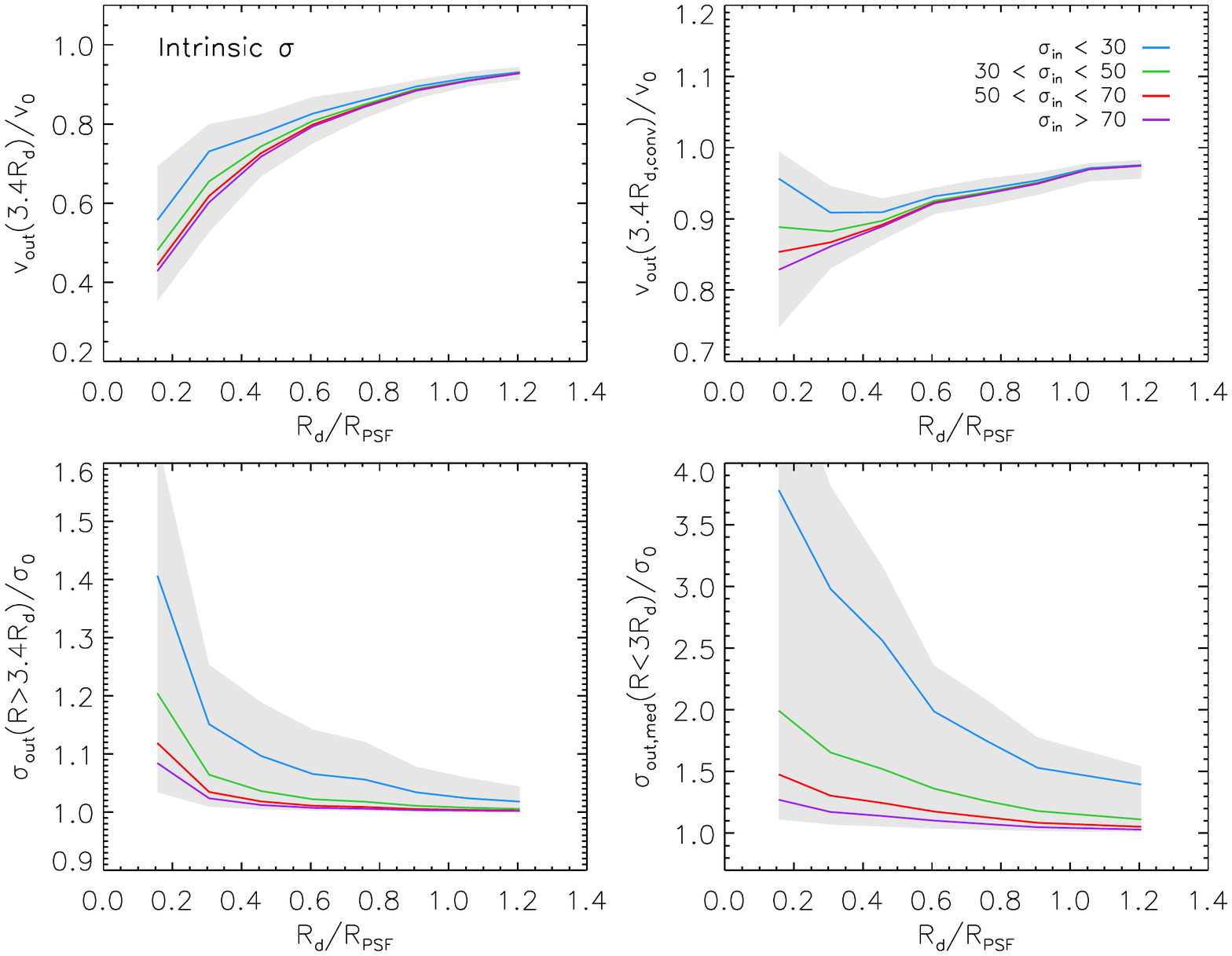}
\end{center}
\caption[Impact of beam smearing on measurements of rotation velocity
  and velocity dispersion, as a function of input velocity
  dispersion]{The impact of beam smearing on measurements of rotation
  velocity and velocity dispersion, as a function of input velocity
  dispersion. \textit{Top Left:} The lower the $R_{\rm d}/R_{\rm PSF}$
  and the greater the input dispersion of the model, the more we
  underestimate the intrinsic rotation velocity ($v_0$). The
  difference between model galaxies of $\sigma_0$\,=\,30\,km\,s$^{-1}$
  and $\sigma_0$\,=\,70\,km\,s$^{-1}$ is $\sim$\,10\% at low $R_{\rm
    d}/R_{\rm PSF}$, but the models converge as $R_{\rm d}/R_{\rm
    PSF}$ increases. \textit{Top Right:} We extract the observed
  rotation velocity at the required radius convolved with the
  seeing. This results in a better estimate. Again, results for the
  binned data converge beyond $R_{\rm d}/R_{\rm
    PSF}$\,=\,0.6. \textit{Bottom Left:} The lower the $R_{\rm
    d}/R_{\rm PSF}$, the more we overestimate the intrinsic velocity
  dispersion. The difference between model galaxies of
  $\sigma_0$\,=\,30\,km\,s$^{-1}$ and $\sigma_0$\,=\,70\,km\,s$^{-1}$
  is as much as $\sim$\,30\%. How well we can recover the intrinsic
  velocity dispersion appears to be strongly dependent on what its
  value was to begin with. \textit{Bottom Right:} If we estimate the
  velocity dispersion from a median of the map, as opposed to the
  outskirts of the dispersion profile, it is more difficult to recover
  $\sigma_0$. For very low dispersions ($\sigma_{\rm
    in}$\,$<$\,30\,km\,s$^{-1}$) the beam smearing effect is as strong
  as for very high mass galaxies
  (10.8\,$<$\,log($M$/M$_{\odot}$)\,$<$\,11.1), with $\sigma_0$
  overestimated by a factor of four at low $R_{\rm d}/R_{\rm PSF}$.}
\label{fig:sigma_trend}
\end{figure*}

Galaxies which are small in comparison to the PSF are more affected by
beam smearing (Fig.\,\ref{fig:psf_maps}), and the shape of the
rotation curve and ability to recover $\sigma_0$ in the outskirts of
the galaxy deteriorate rapidly as the seeing is increased. Since this
is perhaps the strongest universal trend, we choose to study how our
measurements of rotation velocity and dispersion are affected as a
function of $R_{\rm d}$/$R_{\rm PSF}$ where $R_{\rm PSF}$ is the half
width half maximum of the spatial PSF. To assess the impact of other
variables relative to this, we then bin the data in terms of disk mass
(Fig.\,\ref{fig:mass_trend}), inclination
(Fig.\,\ref{fig:incl_trend}), dark matter fraction
(Fig.\,\ref{fig:fdm_trend}) and intrinsic dispersion
(Fig.\,\ref{fig:sigma_trend}) in turn.

In each figure we demonstrate how four measurements are impacted by
beam smearing: the rotation velocity ($v_{\rm out}$) at 3.4$R_{\rm
  d}$, the rotation velocity at the same radius convolved with the
seeing, the median of the velocity dispersion profile at radii greater
than 3.4$R_{\rm d}$ ($\sigma_{\rm out}$), and the median of the
velocity dispersion map within a 3$R_{\rm d}$ aperture ($\sigma_{\rm
  out,med}$). Measurements taken at smaller radii result in the same
trends but with a systematic offset. We will discuss this further in
\S\ref{velocity_corrections}.

The tracks in Fig.\,\ref{fig:mass_trend}--\ref{fig:sigma_trend}
confirm many of our conclusions in \S\ref{beam_smearing_maps}. That
is, for low values of $R_{\rm d}$/$R_{\rm PSF}$ (i.e. galaxies which
are small compared to the spatial PSF), the rotation velocity we
recover is an underestimate of the intrinsic value. Nevertheless, as
the model galaxy is increased in size (or the seeing is decreased) we
approach $v_{\rm out}$/$v_{0}$\,=\,1. When extracting measurements
from the rotation curve at 3.4$R_{\rm d}$, a ratio of $R_{\rm
  d}$/$R_{\rm PSF}$\,=\,0.2 results in an average underestimate of a
factor of two. However, this effect is less significant when we
measure at the radius we require convolved with the seeing. Here the
rotation velocity is only underestimated by $\sim$\,10\,\% at $R_{\rm
  d}$/$R_{\rm PSF}$\,=\,0.2. In
Fig.\,\ref{fig:mass_maps}--\ref{fig:psf_maps} we can see why this may
be so; outer regions of the galaxy's rotation curve are less affected
by beam smearing.

In the lower two panels of
Fig.\,\ref{fig:mass_trend}--\ref{fig:sigma_trend} we can see that beam
smearing affects our ability to recover the intrinsic velocity
dispersion even more strongly. The lower the $R_{\rm d}$/$R_{\rm PSF}$
ratio, the more we overestimate the intrinsic velocity
dispersion. Most noticeably, measuring $\sigma$ in the outskirts of
the velocity dispersion profile is a much better estimate than a
median of the dynamical map. This is because the beam smearing effects
are largest in regions of steep velocity gradients (i.e. towards the
dynamical centre of a uniformly rotating disk). We see a range of
$\sigma_{\rm out}$/$\sigma_0$\,=\,1.0\,--\,1.5 compared to
$\sigma_{\rm out,med}$/$\sigma_0$\,=\,1.0\,--\,4.0 estimated using the
map.

Coloured tracks in these figures show the results for models with one
particular input parameter fixed and all others allowed to vary. The
shaded region illustrates the 1$\sigma$ range for all 10$^5$
models. In Fig.\,\ref{fig:mass_trend} we see that higher mass galaxies
result in estimates of rotation velocity closer to the intrinsic
value, since their rotation curve peaks more quickly, but the
systematic offset in $\sigma_0$ is larger due to the steeper velocity
gradient. The tracks of fixed disk mass cover the 1$\sigma$ range of
the data, suggesting that this is an important parameter.

As discussed in \S\ref{beam_smearing_maps}, galaxy models of a higher
inclination are more susceptible to beam smearing
(Fig.\,\ref{fig:incl_trend}). However the difference between the track
for $<$\,30 degree and $>$\,60 degree inclinations is small,
suggesting this effect is secondary to that caused by increasing the
disk mass. The same is true for models of varying dark matter fraction
(Fig.\,\ref{fig:fdm_trend}). Nonetheless, it is interesting to note
that galaxies with a greater dark matter fraction suffer more beam
smearing. We suggest that this is because the dark matter fraction
determines the degree of turn-over in the rotation curve, which in
turn affects the velocity gradient across pixels in the outer regions.

Finally, we explore the impact of adjusting the intrinsic velocity
dispersion of the model (Fig.\,\ref{fig:beam_smearing_tracks}). We
find that the lower the input dispersion, the more successful we are
in recovering the true rotation velocity, but that the opposite is
true of the velocity dispersion itself. For very low dispersions
($\sigma_{0}$\,$<$\,30\,km\,s$^{-1}$) the beam smearing effect is as
strong as for very high mass galaxies (10.8\,$<$\,log($M_{\rm
  d}$/M$_{\odot}$)\,$<$\,11.1), simply because the ratio of
$\Delta$\,$\sigma$/$\sigma_0$ is larger.

\subsection{Beam Smearing Corrections}
In \S\ref{effect_of_params} we found that adjusting the input
parameters our model galaxies can lead to a stronger or weaker beam
smearing effect. Now that we understand these systematics, we wish to
derive a series of corrections which can be applied to kinematic
measurements of the KROSS sample. In this section we discuss how best
this can be achieved.

\subsubsection{Measurements of Rotation Velocity}
\label{velocity_corrections}
In Fig.\,\ref{fig:mass_trend}--\ref{fig:sigma_trend} (top left panel)
we plot the relationship between $v_{\rm out}$/$v_{\rm 0}$ and $R_{\rm
  d}$/$R_{\rm PSF}$. The systematic offset between the input and
output rotation velocity is strongly correlated with how large the
galaxy is compared to the seeing. Values range between $v_{\rm
  out}$/$v_{\rm 0}$\,=\,0.5\,--\,0.9. However if we make measurements
at the same radii convolved with the seeing (top right panel) this
relation is less steep, with a range of only $v_{\rm out}$/$v_{\rm
  0}$\,=\,0.85\,--\,0.95. This is therefore the method we decide to
use for the KROSS sample. We note that if we had instead measured the
velocity at 2.2$R_{\rm d}$ (a radius commonly used by other kinematic
surveys) the results follow a similar trend, with a small shift
towards lower $v_{\rm out}$/$v_{\rm 0}$, but the effect is
$\lesssim$\,5\%.

While varying model parameters such as disk mass
(Fig.\,\ref{fig:mass_trend}) and inclination
(Fig.\,\ref{fig:incl_trend}) introduces scatter in $v_{\rm
  out}$/$v_{\rm 0}$, at the median $R_{\rm d}$/$R_{\rm PSF}$ of the
KROSS sample ($\sim$\,0.75) the difference is only a few per
cent. Moreover, one of the most dominant influences on the ratio of
$v_{\rm out}$/$v_{\rm 0}$ is the dark matter fraction, $f_{\rm dm}$,
which we are unable to constrain from our observations. In order to
correct the KROSS rotation velocities for beam smearing we therefore
consolidate the information from our models into a single relation for
each of $v_{\rm out}(3.4R_{\rm d})$ and $v_{\rm out}(2.2R_{\rm
  d})$. We define each correction track as the median outcome of all
models, with uncertainties to reflect the 1$\sigma$ scatter. Data
points are fit by an exponential of the form

\begin{equation}
\label{eq2}
1/\xi_v = v_{\rm out}/v_{\rm 0} = 1 - A e^{-B(R_{\rm d}/R_{\rm PSF})^C},
\end{equation}
where $A$, $B$ and $C$ are constants defined in Table
\ref{beam_smearing_params} and $\xi_v$ is the velocity correction
factor. We show these final tracks for beam smearing corrections to
the rotation velocity in Fig.\,\ref{fig:vel_tracks}.

\subsubsection{Measurements of Velocity Dispersion}

The effects of beam smearing on measurements of the intrinsic velocity
dispersion ($\sigma_{\rm out}$) are generally more significant than
for the rotation velocity. In
Fig.\,\ref{fig:mass_trend}--\ref{fig:sigma_trend} (bottom left, bottom
right) we see that for galaxies small in comparison to the spatial PSF
(i.e. for low $R_{\rm d}$/$R_{\rm PSF}$) the dispersion in the
outskirts of the disk can be a factor of $\sim$\,1.5 higher than the
intrinsic value. Estimates made using the median of the map may even
reach $\sigma_{\rm out}$/$\sigma_{\rm 0}$\,=\,5. However for larger
galaxies (or a smaller spatial PSF) $\sigma_{\rm out}$/$\sigma_{\rm
  0}$ appears to decrease exponentially.

Binning the data in Fig.\,\ref{fig:mass_trend}--\ref{fig:sigma_trend}
by disk mass, inclination, dark matter fraction and intrinsic
dispersion reveal that the input parameters of the model have a
significant impact on how we measure $\sigma_0$. At the median $R_{\rm
  d}$/$R_{\rm PSF}$ of our KROSS sample, the difference between high
mass and low mass models (for measurements made in the outskirts of
the dispersion profile) is $\sigma_{\rm out}$/$\sigma_{\rm
  0}$\,$\sim$\,0.1. The difference for disks close to edge-on or
face-on is $\sigma_{\rm out}$/$\sigma_{\rm 0}$\,$\sim$\,0.05. Changes
to $f_{\rm dm}$ or the input $\sigma_0$ itself have less of an impact
(unless $\sigma_0$\, is very small i.e. $<$\,30\,km\,s$^{-1}$), with
an average difference of only a few per cent.

Given the strong variation seen in our models, it is clear that we
cannot reduce the beam smearing correction to a single track for each
of the dispersion measurements (outer disk and median
values). Instead, we choose to make corrections as a function of
$v_{\rm out}(3.4R_{\rm d,conv})$, the rotation velocity measured at a
radius of 3.4$R_{\rm d}$ convolved with the seeing (referred to
hereafter as $v_{\rm out}$; not inclination corrected). This combines
the effects of the two most dominant parameters, disk mass and
inclination. For each method, we split the data into 50\,km\,s$^{-1}$
bins of $v_{\rm out}$ and calculate a series of running medians.

Models run with $\sigma_0$\,$<$\,30\,km\,s$^{-1}$ exhibit as much beam
smearing as high mass models, however we are unable to make
corrections as a function of $\sigma_0$ (it is what we are trying to
measure!). To account for the effect the intrinsic dispersion has on
the beam smearing, we instead resample the model data such that the
distribution of $\sigma_{\rm out}$ matches that of KROSS, and refit
the correction tracks.

\begin{figure}
\includegraphics[width=0.45\textwidth]{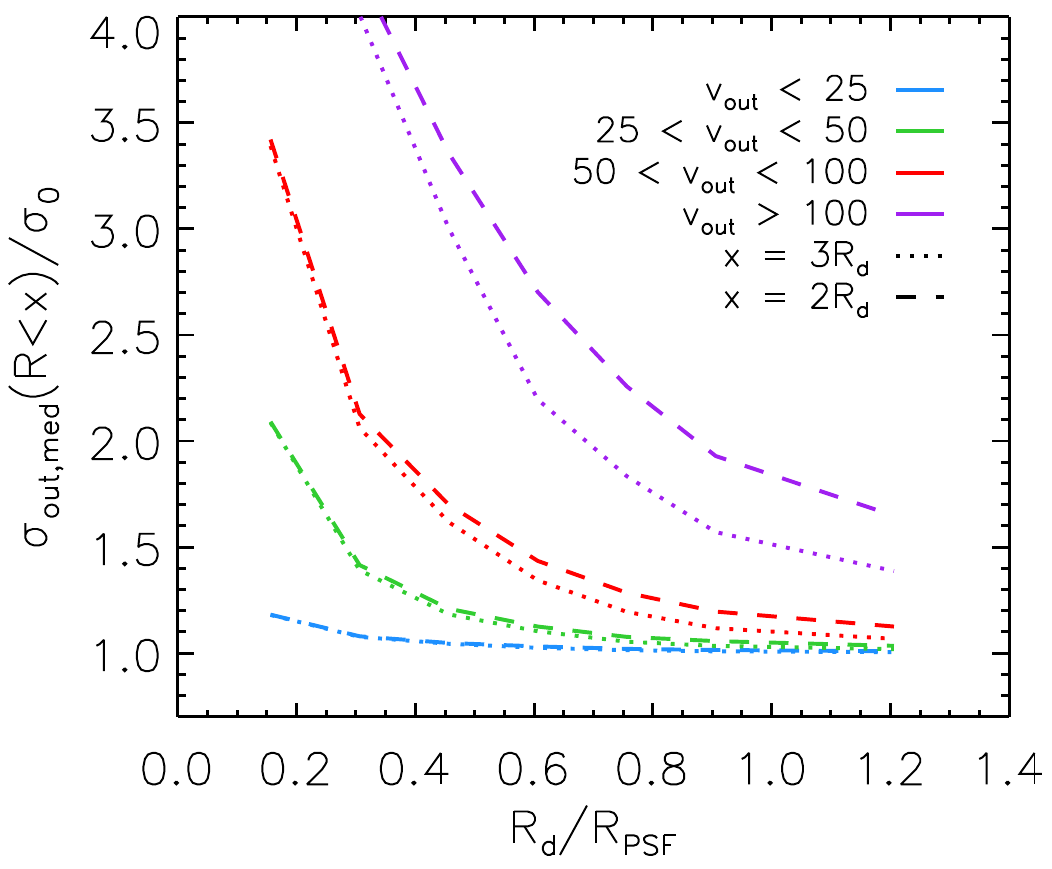}
\caption[Systematic offset in the velocity dispersion measured as a
  pixel-by-pixel median, as a function of aperture size]{Factor by
  which the intrinsic velocity dispersion is overestimated when
  measuring the pixel-by-pixel median within a 3$R_{\rm d}$ (dotted
  line) or 2$R_{\rm d}$ (dashed line) aperture. If the observed
  rotation velocity of the galaxy (at 3.4$R_{\rm d,conv}$) is small
  then the two estimates are almost identical (within 10\%). For
  larger velocities ($v_{\rm out}$\,$>$\,100\,km\,s$^{-1}$)
  corrections for the larger and smaller apertures (at the median
  $R_{\rm d}$/$R_{\rm PSF}$ of KROSS) are $\sigma_{\rm
    out,med}$/$\sigma_0$\,=\,1.8 and 2.2 respectively, however fewer
  than 25\% of our sample satisfy this criteria. We therefore create
  the final correction tracks (Fig.\,\ref{fig:beam_smearing_tracks})
  using only the values for 3$R_{\rm d}$.}
\label{fig:median_radii}
\end{figure}

As discussed in \S\ref{measurements}, we measured the median velocity
dispersion of each model within two apertures (2$R_{\rm d}$ and
3$R_{\rm d}$) since the size of the galaxy compared to the IFU, or the
surface brightness of the galaxy (hence signal-to-noise) will affect
the number of available pixels. Fig.\,\ref{fig:median_radii} compares
these two sets of results. If the observed rotation velocity is small
($v_{\rm out}$\,$<$\,100\,km\,s$^{-1}$) corrections at the median
$R_{\rm d}$/$R_{\rm PSF}$ of KROSS range between $\sigma_{\rm
  out,med}$/$\sigma_0$\,=\,1.0\,--\,1.3 and the difference between
results for the two apertures is typically $\sigma_{\rm
  out,med}$/$\sigma_0$\,$<$\,0.1. If the rotation velocity exceeds
this then corrections for the larger and smaller apertures are
$\sigma_{\rm out,med}$/$\sigma_0$\,=\,1.8 and 2.2 respectively. Since
the results are very similar, we therefore combine the results into a
single set of (velocity binned) tracks. Final tracks for beam smearing
corrections to the velocity dispersion are presented in
Fig.\,\ref{fig:beam_smearing_tracks}. The correction $\sigma_{\rm
  out}$ to $\sigma_{\rm 0}$ as a function of $R_{\rm d}$/$R_{\rm PSF}$
is defined by,

\begin{equation}
\label{eq3}
1/\xi_\sigma = \sigma_{\rm out}/\sigma_{\rm 0} = 1 + A e^{-B(R_{\rm d}/R_{\rm PSF})^C},
\end{equation}
where the constants $A$, $B$ and $C$ are defined in Table \ref{beam_smearing_params}.

\subsection{Additional Tests}
\subsubsection{Adaptive Binning}
\label{adaptive_binning}

To construct dynamical maps for each of the KROSS galaxies we employed
an ``adaptive binning'' technique. In fitting the spectrum of each
spaxel (\S\ref{mock_observations}) we required that the H$\alpha$
emission line was detected with a signal-to-noise ratio of $>$\,5. If
the line was too weak then we binned the spectra of neighbouring
pixels, increasing the size of the region until either the criterion
was met, or we reached an area of 0.7\,$\times$\,0.7 arcsec (the
typical seeing of our observations). To explore how this process may
affect measurement of the kinematics, we analyse our model data a
second time. When fitting the spectrum of each pixel we now include
all data within a 0.5\,$\times$\,0.5 arcsec region.

\begin{figure*}
\begin{center}
\includegraphics[width=0.9\textwidth]{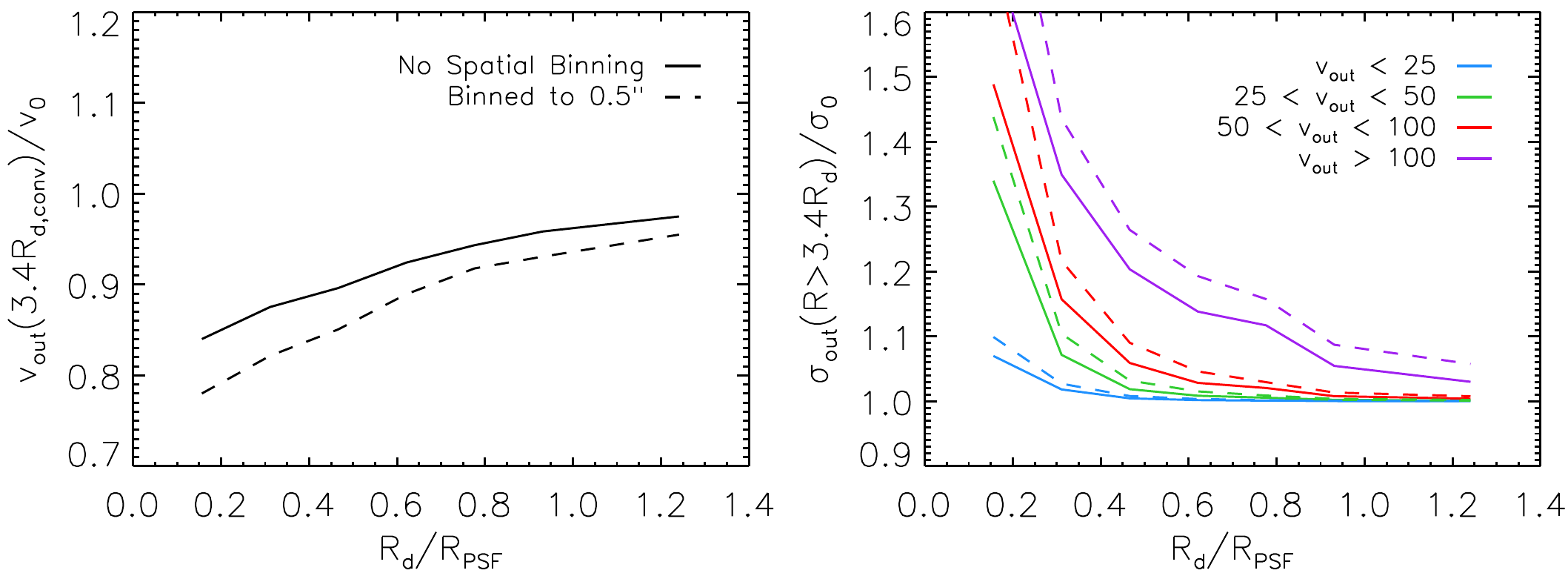}
\end{center}
\caption[Effects of spatially binning the model data]{Effects of
  spatial binning on measurements of the rotation velocity and
  velocity dispersion. Results are generated from our mock
  observations when the spectrum of each pixel is fit individually
  (solid line) and when the signal has been binned within a
  0.5\,$\times$\,0.5 arcsec region (dashed line). In the instances
  where data has been binned, the rotation velocity (\textit{left}) is
  underestimated by an additional $\sim$\,5\,--\,10\% and the
  dispersion (\textit{right}) overestimated by an additional
  2\,--\,3\% ($\sim$\,5\% for large $v_{\rm out}$). }
\label{fig:adaptive_binning}
\end{figure*}

Fig.\,\ref{fig:adaptive_binning} shows that binning acts to magnify
the effects of beam smearing, resulting in lower rotation velocities
and greater velocity dispersions. In the instances where data has been
binned, the rotation velocity is underestimated by an additional
$\sim$\,5\,--\,10\% and the dispersion overestimated by an additional
2\,--\,3\% ($\sim$\,5\% for large $v_{\rm out}$). This is a rather
exaggerated picture, since in our models have been uniformly binned
regardless of the surface brightness profile. In reality, outer
regions are more likely to have been binned, and some galaxies may not
have been binned at all. While this is an important effect to note we
do not attempt to correct for it, since details of the process are
unique to each KROSS galaxy.

\renewcommand{\arraystretch}{1.1}
\begin{table*}
\begin{center}{
\caption{Parametrisation of beam smearing correction tracks}
\label{beam_smearing_params}
\vspace{2mm}
\begin{tabular*}{\textwidth}{@{\extracolsep{\fill}} l | ccccc}
\hline

Correction Track & $v_{\rm min}$ & $v_{\rm max}$ & A & B & C \\
& (km\,s$^{-1}$) & (km\,s$^{-1}$) & & & \\ 
\hline
\hline
Velocity (3.4$R_{\rm d}$)  & - & - & 0.18 & 1.48 & 1.00 \\
Velocity (2.2$R_{\rm d}$)  & - & - & 0.18 & 1.26 & 0.88 \\
Dispersion (outskirts) & 0 & 50 & 0.53 & 8.22 & 0.94 \\
& 50 & 100 & 6.98 & 7.07 & 0.52 \\
& 100 & 150 & 3.27 & 4.96 & 0.59 \\
& 150 & -  & 2.06 & 3.67 & 0.70 \\
Dispersion (median) & 0 & 50 & 11.50 & 4.65 & 0.20 \\
& 50 & 100 & 52.85 & 5.55 & 0.34 \\
& 100 & 150 & 8.74 & 3.15 & 0.77 \\
& 150 & - & 14.15 & 3.05 & 0.69 \\
\hline

\end{tabular*}
\label{table:tracks}
}\end{center}
\vspace{-2mm} {\sc notes:} Constants $A$, $B$, $C$ for the beam
smearing correction tracks in Fig.\,\ref{fig:vel_tracks} and
\ref{fig:beam_smearing_tracks}, as defined by equations \ref{eq2} and
\ref{eq3}. For the velocity dispersion, $v_{\rm min}$ and $v_{\rm
  max}$ define the range of observed rotation velocities (uncorrected
for inclination) that each track covers.
\end{table*}

\begin{figure*}
\begin{center}
\includegraphics[width=0.9\textwidth]{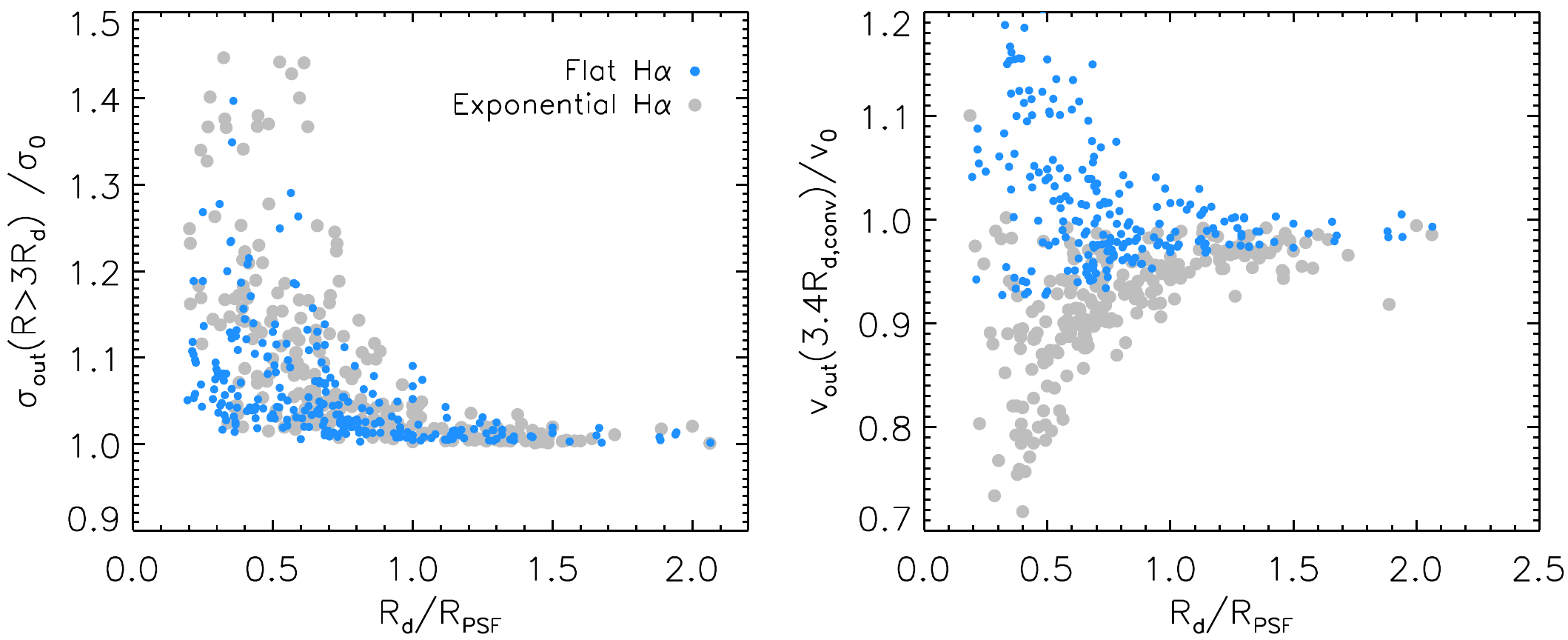}
\end{center}
\caption[Effects of adjusting the H$\alpha$ surface brightness profile
  of the model]{Beam smearing correction in $v$ and $\sigma$ as a
  function of the surface brightness profile. These results are
  generated from mock observations when the H$\alpha$ intensity
  follows an exponential profile (black points) or is uniform across
  the IFU (blue points). \textit{Left:} For a ``flat'' H$\alpha$
  profile effects of beam smearing on the velocity dispersion are
  weaker by $\sigma_{\rm out}$/$\sigma_{0}$\,$\sim$\,0.1 at low
  $R_{\rm d}$/$R_{\rm PSF}$. \textit{Right:} Results for the two flux
  distributions diverge significantly for $R_{\rm d}$/$R_{\rm
    PSF}$\,$<$\,0.7. When the H$\alpha$ follows a uniform distribution
  the recovered rotation curve is close to the intrinsic, hence if
  there is a turnover within the data using the convolved radius may
  actually result in an overestimate of the velocity. We see that $v$
  may be overestimated by as much as 20\%.}
\label{fig:exp_flat}
\end{figure*}

\subsubsection{Intensity Maps}
\label{intensity_maps}

For each model galaxy we have assumed that the stellar mass and light
follow an exponential profile, and this was propagated through to the
construction of model H$\alpha$ intensity maps. However, observations
suggest that the H$\alpha$ morphology of $z$\,$\sim$\,1 galaxies is
often irregular, with the presence of $\sim$\,kpc scale star-forming
``clumps''
\citep{genzel2011,wisnioski2012,swinbank2012,livermore2012,livermore2015}. These
deviations from an exponential profile may affect the beam smearing,
since within each pixel it will affect the relative contribution of
each new velocity component introduced (i.e. beam smearing is
luminosity weighted).

In Fig.\,\ref{fig:exp_flat} we compare the results of modelling
galaxies with an exponential or a uniform H$\alpha$ intensity
profile. The presence of bright, star-forming clumps may act to
flatten the H$\alpha$ profile, so this is suitable test of how (in the
most extreme case) this may affect the beam smearing. Measurements of
$\sigma_{\rm out}$ are less affected by beam smearing in the case of a
uniform flux distribution, with a difference of $\sigma_{\rm
  out}$/$\sigma_{0}$\,$\sim$\,0.1 on average. Effects on the shape of
the rotation curve are also less severe. Pixels in the outskirts of
the galaxy are still contamined by light from central regions, however
these regions are no longer as bright and contribute less
flux. Therefore pixels in the outskirts do not become as skewed
towards lower velocities. In the right hand panel of
Fig.\,\ref{fig:exp_flat} we see that the rotation velocity at
3.4$R_{\rm d,conv}$ is now an overestimate by as much as 20\% at low
$R_{\rm d}$/$R_{\rm PSF}$. However for $R_{\rm d}$/$R_{\rm
  PSF}$\,$>$\,0.7 the required corrections are within a few per cent
of those for an exponential profile.


\FloatBarrier

\label{lastpage}
\end{document}